\documentclass{jfm}
\usepackage{color}
\usepackage{amstext}
\usepackage{amssymb}
\usepackage{graphicx}
\usepackage{amsmath}

\newcommand\const{\mathrm{const}}
\newcommand\Div{\mathrm{div}\,}

\newcommand\vA{\boldsymbol{A}}
\newcommand\vB{\boldsymbol{B}}

\newcommand\vX{\boldsymbol{X}}
\newcommand\vP{\boldsymbol{P}}

\newcommand\vU{\boldsymbol{U}}
\newcommand\vV{\boldsymbol{V}}
\newcommand\va{\boldsymbol{a}}
\newcommand\vb{\boldsymbol{b}}
\newcommand\vc{\boldsymbol{c}}

\newcommand\vh{\boldsymbol{h}}

\newcommand\vk{\boldsymbol{k}}
\newcommand\vK{\boldsymbol{K}}
\newcommand\vl{\boldsymbol{l}}

\newcommand\vn{\boldsymbol{n}}
\newcommand\vu{\boldsymbol{u}}
\newcommand\vv{\boldsymbol{v}}
\newcommand\vw{\boldsymbol{w}}
\newcommand\vx{\boldsymbol{x}}
\newcommand\vy{\boldsymbol{y}}
\newcommand\vr{\boldsymbol{r}}

\newcommand\vp{\boldsymbol{p}}
\newcommand\vq{\boldsymbol{q}}
\newcommand\vQ{\boldsymbol{Q}}
\newcommand\vg{\boldsymbol{g}}

\newcommand\veta{\boldsymbol{\eta}}
\newcommand\vxi{\boldsymbol{\xi}}
\newcommand\vzeta{\boldsymbol{\zeta}}

\newcommand\vkappa{\boldsymbol{\kappa}}
\newcommand\vnu{\boldsymbol{\nu}}

\begin{document}

 {\title[Admixture and Drift in Oscillating Fluid Flows] {Admixture and Drift in Oscillating Fluid Flows}}

\author[V. A. Vladimirov]{V.\ns A.\ns V\ls l\ls a\ls d\ls i\ls m\ls i\ls r\ls o\ls v}

\affiliation{Dept of Mathematics, University of York, Heslington, York, YO10 5DD, UK}

\pubyear{2010} \volume{xx} \pagerange{xx-xx}
\date{September 21st, 2010}

\setcounter{page}{1}\maketitle \thispagestyle{empty}

\begin{abstract}
The motions of a passive scalar $\widehat{a}$ in a general high-frequency oscillating flow are studied. Our aim is
threefold: (i) to obtain different classes of general solutions; (ii) to identify, classify, and develop related
asymptotic procedures; and (iii) to study the notion of drift motion and the limits of its applicability. The used
mathematical approach combines a version of the two-timing method, the Eulerian averaging procedure, and several novel
elements. Our main results are: (i) the scaling procedure produces two independent dimensionless scaling parameters:
inverse frequency $1/\omega$ and displacement amplitude $\delta$; (ii) we propose the \emph{inspection procedure} that
allows to find the natural functional forms of asymptotic solutions for $1/\omega\to 0, \delta\to 0$ and leads to the
key notions of \emph{critical, subcritical, and supercritical asymptotic families} of solutions; (iii) we solve the
asymptotic problems for an arbitrary given oscillating flow and any initial data for $\widehat{a}$; (iv) these
solutions show that there are at least three different drift velocities which correspond to different asymptotic paths
on the plane $(1/\omega,\delta)$; each velocity has dimensionless magnitude ${O}(1)$; (v) the obtained solutions also
show that the averaged motion of a scalar represents a pure drift for the zeroth and first approximations and a drift
combined with
\emph{pseudo-diffusion} for the second approximation; (vi) we have shown how the changing of a time-scale produces new
classes of solutions; (vii) we develop the two-timing theories of a drift based on both the \emph{GLM}-theory and the
dynamical systems approach; (viii) examples illustrating different options of drifts and pseudo-diffusion are
presented.

\end{abstract}

\section{Introduction \label{sect01}}

The transport of a scalar field $\widehat{a}$ by an oscillating velocity field $\widehat{\vu}$ represents a classical
problem of fluid dynamics. One can find a number of models, physical effects, and applications in
\cite{Taylor, Moffatt3, Pedley1, Frisch, Moffatt1, Pedley, Leal}. The transport of a scalar represents a
paradigm problem for the use of the dynamical systems approach in fluid dynamics (see
\cite{Aref, Ottino, Solomon, Wiggins}) and for the studies of turbulence (see \cite{Sree, Warhaft, Balk}).
It is well-known that even in purely oscillating flows the motion of a material particle consists of two parts:
oscillating and non-oscillating. The latter is known as a \emph{drift}, see
\cite{Stokes, Maxwell, Lamb, LH, Darwin, Hunt, Batchelor, Schlichting, Lighthill0, McIntyre, Craik, Grimshaw, Craik0,
Benjamin, Eames, Hunt1}. Different notions of a drift are used in physics, see \emph{e.g.} \cite{Vergassola1,
Chierchia}.

In our paper we consider a high-frequency asymptotic problem of a scalar transport by a given oscillatory velocity
field.  A characteristic length $L$ for both $\widehat{\vu}$ and $\widehat{a}$ is arbitrary. We use a simple and
straightforward formulation of the problem that allows to study the general properties of solutions without the
consideration of particular applications. The evolution of $\widehat{a}$ is exploited for the study of the concept of a
drift motion and the limits of its applicability; this approach allows a great degree of flexibility and has not been
used before. We employ the mathematical method which comprises the two-timing approach (\cite{Nayfeh}), the fast-time
averaging operation, and several novel elements; we call it
\emph{two-timing-and-averaging method (TTAM)}. The interplay of the two-timing form of solutions and a
fast-time-averaging operation was exploited by many authors. The well-known applications in classical mechanics belong
to \cite{Kapitza1, Kapitza2, Landau}. A simplified version of the two-timing method has been extensively used in
mechanical engineering and is known as \emph{vibrational mechanics}, see \emph{e.g.} \cite{Blekhman1, Blekhman2}.
However
\cite{Kapitza1,Kapitza2, Landau,Blekhman1, Blekhman2} did not use (and even deliberately denied) the underlying
asymptotic nature of the method. This nature has been emphasised and exploited by \cite{Vladimirov2, Yudovich,
Vladimirov1} in a particular form of \emph{TTAM} used in this paper.

Let us outline here the content and the main results:

In \emph{Sect.2} we formulate the problem, introduce two-timing and dimensionless versions of the governing PDE. The
two independent dimensionless scaling parameters are inverse frequency $1/\omega$ and displacement amplitude $\delta$
of oscillations. Our asymptotic procedure implies both $1/\omega\to 0$ and $\delta\to 0$. It produces the multiplicity
of available asymptotic paths on the plane $(1/\omega, \delta)$; we parameterize these paths with a single small
parameter $\varepsilon_\alpha=1/\omega^\alpha$, $\alpha=\const>0$. This multiplicity gives us new opportunities for the
study of a drift. At the end of the section we consider an important case when $\widehat{\vu}$ does not contain any
slow time-scale $T$: in this case we show the presence of an infinite number of solutions with a range of different
$T$.

In \emph{Sect.3} we develop \emph{the inspection procedure} for the governing equations. It leads to classification of
different asymptotic solutions and to the key notions of
\emph{critical, sub-critical, and super-critical} asymptotic families of solutions. For a critical asymptotic family:
(i) the small parameter is $\varepsilon_{1/2}$ (or $\alpha=1/2$); and  (ii) the correlation $\langle\widetilde{\vu}
\widetilde{a}\rangle$ between the oscillatory parts
$\widetilde{\vu}$ and $\widetilde{a}$ of $\widehat{\vu}$ and $\widehat{a}$ enters the ${O}(1)$ approximation of the
averaged governing equation. For a super-critical family $\alpha<1/2$ and again $\langle\widetilde{\vu}
\widetilde{a}\rangle={O}(1)$, while some constrains on $\widehat{\vu}$ must be imposed. For a sub-critical asymptotic
family $\langle\widetilde{\vu} \widetilde{a}\rangle$ appears only in higher than ${O}(1)$ approximations.

In \emph{Sect.4} and \emph{Appendix A} we obtain the general solution of an asymptotic problem for the critical
asymptotic family ($\alpha=1/2$) with an arbitrary purely oscillating velocity $\widetilde{\vu}$ and arbitrary initial
conditions for $\widehat{a}$. Each solution $\widehat{a}=\overline{a}+\widetilde{a}$ consists of explicit formulae for
its oscillating part $\widetilde{a}$ and an advection-diffusion-type equation for the averaged part $\overline{a}$. Our
detailed analytical calculations include five steps of successive approximations. The obtained solutions satisfy the
governing equations with small explicitly calculated residual right-hand-sides. It appears that evolution of the
averaged field $\overline{a}$ includes a drift with the velocity
$\overline{\vV}=\overline{\vV}_0+\varepsilon\overline{\vV}_1+\varepsilon^2\overline{\vV}_2 +{O}(\varepsilon^3)$ and
pseudo-diffusion with a matrix of coefficients
$\overline{\kappa}_{ik}=\varepsilon^2\overline{\chi}_{ik}+{O}(\varepsilon^3)$, where
$\varepsilon\equiv\varepsilon_{1/2}$ and $\overline{\vV}_0, \overline{\vV}_1,
\overline{\vV}_2,\overline{\chi}_{ik}$ have magnitudes ${O}(1)$.
Our term {\emph{pseudo-diffusion}} means that the evolution of $\overline{a}$ is described by an
advection-diffusion-type equation where the diffusion-type term appears not in the leading approximation.

In \emph{Sect.5} and \emph{Appendix B} we present the asymptotic solutions for supercritical asymptotic families with
$\alpha=1/3$ and $1/4$ and degenerated velocity field $\widetilde{\vu}$. For $\alpha=1/3$ the required degeneration is
$\overline{\vV}_0\equiv 0$ which leads to
$\overline{\vV}=\overline{\vV}_1+\varepsilon\overline{\vV}_2+{O}(\varepsilon^2)$ and $\overline{\kappa}_{ik}\equiv
0+{O}(\varepsilon^2)$, where $\varepsilon\equiv\varepsilon_{1/3}$. For $\alpha=1/4$ the required degeneration is
$\overline{\vV}_0\equiv 0$ and $\overline{\vV}_1\equiv 0$ which leads to
$\overline{\vV}=\overline{\vV}_2+{O}(\varepsilon)$ and $\overline{\kappa}_{ik}\equiv 0+{O}(\varepsilon)$, where
$\varepsilon\equiv\varepsilon_{1/4}$. Hence, the vanishing  of the main term of a drift velocity
$\overline{\vV}_0\equiv 0$ (or the first two terms  $\overline{\vV}_0\equiv 0$ and $\overline{\vV}_1\equiv 0$) does not
lead to the smallness of a dimensionless drift. Instead, a different expression $\overline{\vV}={O}(1)$ appears from a
different asymptotic procedure. At the same time any supercritical oscillations do not produce pseudo-diffusion (within
the precision of our calculations).

In \emph{Sect.6} we generalise the results of \emph{Sects.4} and \emph{5} considering more general $\widehat{\vu}$,
which represents a power series of a small parameter $\varepsilon$ and contains both oscillating and non-oscillating
parts. The expressions for drift velocities appear to be much more general, while pseudo-diffusion stays the same as in
\emph{Sect.4}.

\emph{Sect.7} contains eleven examples of drift velocities and/or pseudo-diffusivity calculated for various flows.
These examples are: (1) the velocity field $\widetilde{\vu}$ consisting of two arbitrary modulated fields of the same
frequency; (2) $\widetilde{\vu}$ that represents the Fourier series of modulated fields; (3) the Stokes drift, which
allows us to make comparison (in \emph{Sect.10}) with classical results; (4) $\widetilde{\vu}$ is a spherical
`acoustic' wave; (5) $\overline{\vV}_1$-drift (\emph{Sect.5}), which takes place for $\overline{\vV}_0\equiv 0$; (6)
$\widetilde{\vu}$ is a plane travelling wave of a general shape or the superposition of two such waves; (7) the
polynomial velocity $\widehat{\vu}$, which has a nonzero average and represents polynomial of a small parameter; (8)
$\widetilde{\vu}$ as the Bjorknes configuration of two pulsating point sources; (9) a simple $\widetilde{\vu}$ that
produces a drift $\overline{\vV}_0$ leading to chaotic dynamics of material particles; (10) the rigid-body-type
oscillations of a fluid where a pseudo-diffusion term in the equations appears naturally; this example explains both
mathematical and physical nature of pseudo-diffusion; (11) we consider a sub-critical asymptotic procedure and show
that the averaged equations for $\overline{{a}}$ contain \emph{pseudo-drift}  and pseudo-diffusion; it emphasizes the
significance of the critical and super-critical procedures.

The \emph{TTAM} used in \emph{Sects.3-7} combines five components: (i) a version of the two-timing method; (ii) the
averaging over fast-time; (iii) the inspection procedure with the multiplicity of asymptotic procedures; (iv) the
obtaining of general asymptotic solutions for these procedures; and (v) the estimating of RHS-residuals (which has been
only mentioned in
\emph{Sects.4-6} and is actually given in \emph{Appendices A,B}).

In \emph{Sect.8} we study the links between the \emph{TTAM}-results of \emph{Sects.2-7} and other theories: (i) the
dynamical systems approach, see \cite{Aref, Ottino, Wiggins}); (ii) \emph{GLM}-theory by \cite{McIntyre}; (iii) the
classical drift theory by \cite{Stokes, LH, Batchelor}; (iv) the Krylov-Bogoliubov averaging method, see
\cite{Bog,Krylov,Verhulst}; (v) the homogenization method, see \cite{Lions, Berdichevsky};
(vi) the theory with nonzero molecular diffusion; and (vii) the MHD-kinematics in oscillating flows, see
\cite{Moffatt}. The links between our \emph{TTAM}-results of \emph{Sects.2-7} and the theories (i), (ii) can be
established only if we adapt and develop last two theories. Therefore, in
\emph{Sect.8.1} we calculate a drift velocity by the averaging of characteristics for transport equations
and in \emph{Sect.8.2} we develop a two-timing version of \emph{GLM}-kinematics. It is important that the
\emph{TTAM}-theory of \emph{Sects.2-7} uses the Eulerian averaging while the theories
(i) and (ii) operate with the Lagrangian averaging. These different averaging operations lead to different results for
a drift and pseudo-diffusion.

Finally, in \emph{Sect.9} we discuss the assumptions used, the results obtained, and the problems to be addressed. We
also present short discussion/remarks at the end of each section and subsection.

\section{Basic Equations and Operations \label{sect02}}

\subsection{Exact problem}

The equation for a scalar field $a=a(\vx^*,{s}^*)$ in Cartesian coordinates $\vx^*=(x_1^*,x_2^*,x_3^*)$ and time
${s}^*$ is
\begin{eqnarray}
\left(\frac{\partial}{\partial {s}^*}+(\vu^*\cdot\nabla^*)\right) a(\vx^*,{s}^*)=0
\label{exact-1}
\end{eqnarray}
where $\vu^*=\vu^*(\vx^*,{s}^*)$ is \emph{a given velocity field}, the asterisks mark dimensional variables. All
functions used in this paper are considered to be sufficiently smooth. Equation (\ref{exact-1}) describes the motion of
a Lagrangian marker in either an incompressible or compressible fluid or the advection of a passive scalar admixture
with concentration $a(\vx^*,{s}^*)$ in an incompressible fluid. For a compressible fluid it also describes the
advection of a passive scalar admixture, where $\widehat{a}$ represents the ratio of concentration of admixture to
density of a fluid. We study the motion of a Lagrangian marker (\ref{exact-1}) due to our primary interest in the
motions of material particles.

The velocity field $\vu^*$ has the functional form of a \emph{hat-function}
\begin{eqnarray}
&& \vu^*=\widehat{\vu}^*(\vx^*, t^*, \tau);\quad t^*={s}^*,\ \tau=\omega^* {s}^*\label{exact-2}
\end{eqnarray}
where $t^*$ and $\tau$ are two mutually dependent time variables, which are introduced as two different time-scales
$k_1{s}^*$ and $k_2{s}^*$ with $k_1=1$ and $k_2=\omega^*$ ($\omega^*$ is frequency). The $\tau$-dependence in
(\ref{exact-2}) is always $2\pi$-periodic. The functional form (\ref{exact-2}) is aimed to describe modulated
oscillations of high-frequency. Following established terminology we call $t^*$ \emph{slow time} and $\tau$
\emph{fast time}. We do not require (\ref{exact-2}) to satisfy any equations of motion; such a general setting
is often accepted, for example in the kinematic MHD-dynamo theory (\cite{Moffatt}).

The functional structure of $\widehat{\vu}^*$ (\ref{exact-2}) underpins the idea that the solution of (\ref{exact-1})
also represents a hat-function:
\begin{eqnarray}
&& a=\widehat{a}(\vx^*, t^*, \tau)\label{exact-3}
\end{eqnarray}
Then after the use of the chain rule the equation (\ref{exact-1}) is
\begin{eqnarray}
&&\mathfrak{D}^*\widehat{a}\equiv\left(\omega^*\frac{\partial}{\partial\tau}+\frac{\partial}{\partial
t^*}+(\widehat{\vu}^*\cdot\nabla^*)\right)\widehat{a}= 0\label{exact-6}\\
&&\frac{\partial}{\partial{s}^*}=\omega^*\left.\frac{\partial}{\partial\tau}\right\vert_{\vx^*,t^*}+
\left.\frac{\partial}{\partial
t^*}\right\vert_{\vx^*,\tau}\equiv\omega^*\frac{\partial}{\partial\tau}+\frac{\partial}{\partial t^*}\label{chain}
\end{eqnarray}
We also denote partial derivatives by subscripts, \emph{e.g.} $a_\tau\equiv\partial a/\partial\tau$.

We obtain the dimensionless form of (\ref{exact-6}) by introducing the characteristic scales of slow time $T$, length
$L$, and velocity $U$. An example of a velocity field is
\begin{eqnarray}
\widehat{\vu}^*=U (1+\sin\tau)(1+C\sin(t^*/T))\,\vg(x^*/L,y^*/L,y^*/L)\label{u-sample}
\end{eqnarray}
where $\vg$ is an arbitrary vector-function of three scalar variables (one can also take $T=L/U$, but it narrows the
class of velocity fields). Dimensionless variables and parameters are not asteriated:
\begin{eqnarray}
&& t=t^*/T ,\ \vx=\vx^*/L,\ \widehat{\vu}=\widehat{\vu}^*/U, \ \omega=\omega^*T \label{scales}
\end{eqnarray}
According to Buckingham's $\pi$-theorem the problem (\ref{exact-6})-(\ref{scales}) possesses two independent
dimensionless scaling parameters
\begin{eqnarray}
&&\varepsilon_1\equiv 1/\omega^* T=1/\omega,\ \mathcal{\delta}\equiv U/(\omega^* L)\label{exact-6aa}
\end{eqnarray}
where $\delta$ can be interpreted as either of two physical parameters: (i) characteristic dimensionless displacement
 $\delta=\delta^*/L$ (for displacement $\delta^*\equiv U/\omega^*$);
 or (ii) Strouhal number $\delta=\Omega/\omega^*$
(for frequency $\Omega\equiv U/L$). The parameter $\varepsilon_1$ represents another Strouhal number and follows a more
general notation:
\begin{eqnarray}
\varepsilon_\alpha\equiv 1/\omega^\alpha,\quad\text{with}\quad  \alpha=\const>0.\label{epsilon}
\end{eqnarray}
The dimensionless versions of (\ref{chain}), (\ref{exact-6}) are
\begin{eqnarray}
&&\mathfrak{D}\widehat{a}\equiv\left(\omega\frac{\partial}{\partial\tau}+\frac{\partial}{\partial
t}+\omega\delta{\widehat{\vu}}\cdot\nabla\right)\widehat{a}= 0\label{exact-6-L}\\
&&\frac{\partial}{\partial{s}}=\omega\left.\frac{\partial}{\partial\tau}\right\vert_{\vx,t}+
\left.\frac{\partial}{\partial
t}\right\vert_{\vx,\tau}\equiv\omega\frac{\partial}{\partial\tau}+\frac{\partial}{\partial t}\label{chain-L}
\end{eqnarray}
Eqn. (\ref{exact-6-L}) can also be written in the form containing only small parameters:
\begin{eqnarray}
&&\mathfrak{D} \widehat{a}/\omega=
\widehat{a}_\tau+\varepsilon_1\widehat{a}_{t}+\delta \widehat{\vu}\cdot\nabla \widehat{a}= 0
\label{exact-6a}
\end{eqnarray}
One can see that we operate on the plane $(\varepsilon_{\alpha}, \delta)$ of two independent dimensionless scaling
parameters. To use a rigorous asymptotic procedure one has to choose a one-parametric asymptotic path on this plane. In
order to form such a path one might prescribe the dependence of each characteristic parameter in (\ref{scales}) on the
chosen single parameter $\varepsilon_{\alpha}$. The simplest choice is
\begin{equation}\label{path-simple}
 T=\const,\quad L=\const,\quad UT/L=O(\varepsilon_{\alpha})
\end{equation}
We will exploit different options for paths, which go to the same limit
\begin{equation}\label{exact-7}
(\varepsilon_{\alpha},\delta(\varepsilon_{\alpha}))\to 0\quad\text{as}\quad \varepsilon_{\alpha}\to 0
\end{equation}
\underline{Remark:} More details on the scaling procedure are given in
\emph{Sect.2.3} and \emph{Sect.7.3}.

\subsection{ The classes of $\mathbb{H}, \mathbb{B}, \mathbb{T}$, and $\mathbb{O}(1)$-functions}

\underline{\emph{Notations and definitions:}}

\underline{\emph{Definition 1.}} The class $\mathbb{H}$ of \emph{hat-functions}  is defined as
\begin{eqnarray}
\widehat{f}\in \mathbb{H}:\quad
\widehat{f}(\vx, t, \tau)=\widehat{f}(\vx,t,\tau+2\pi)\label{tilde-func-def}
\end{eqnarray}
where $t=s$ and $\tau\equiv\omega s$ are two mutually dependent time variables ($\omega^*$ is dimensionless frequency);
the $\tau$-dependence is always $2\pi$-periodic; the dependencies on $\vx$ and $t$ are not specified.

\noindent\underline{\emph{Comments:}} (A) We have already accepted that $\vu^*\in \mathbb{H}$ (\ref{exact-2})
and $a\in \mathbb{H}$ (\ref{exact-3}). (B) For any $\widehat f\in \mathbb{H}$ a partial time-derivative can be
expressed
\emph{via} the chain rule as
\begin{eqnarray}
&&\frac{\partial \widehat{f}}{\partial {s}}=\left(\frac{\partial}{\partial t}+ \omega
\frac{\partial }{\partial \tau}\right) \widehat f(\vx, t, \tau)\equiv\widehat f_t+\omega\widehat f_{\tau},
\quad \text{where}\ t={s}, \ \tau=\omega {s}
\label{exact-4}
\end{eqnarray}
(C) In any version of the two-timing method the variables $t$ and $\tau$  are considered to be mutually independent
after the chain rule (\ref{exact-4}) has been applied. The justification of this auxiliary assumption is given
\emph{a posteriori} (see
\emph{Appendix A}).

\underline{\emph{Definition 2.}}
For an arbitrary $\widehat{f}\in \mathbb{H}$ the \emph{Eulerian averaging operation} is
\begin{eqnarray}
\langle {\widehat{f}}\rangle \equiv \frac{1}{2\pi}\int_{\tau_0}^{\tau_0+2\pi}
\widehat{f}(\vx, t, \tau)\,d\tau,\qquad\forall\ \tau_0\label{oper-1}
\end{eqnarray}
where  during the integration $t=\const$ and $\langle {\widehat{f}}\rangle$ does not depend on $\tau_0$.

\underline{\emph{Definition 3.}} The class $\mathbb{T}$ of \emph{tilde-functions} is such that
\begin{eqnarray}
\widetilde f\in \mathbb{T}:\quad
\widetilde f(\vx, t, \tau)=\widetilde f(\vx,t,\tau+2\pi),\quad\text{with}\quad
\langle \widetilde f \rangle =0,\label{oper-2}
\end{eqnarray}
\underline{\emph{Comments:}} (A) One can see that a \emph{$\mathbb{T}$-function} represents a special case of
$\mathbb{H}$-function (\ref{tilde-func-def}) with zero average (\ref{oper-1}). (B) In the text the tilde-functions are
also called the purely oscillating functions.

\underline{\emph{Definition 4.}} The class $\mathbb{B}$ of \emph{bar-functions} is defined as
\begin{eqnarray}
\overline{f}\in \mathbb{B}:\quad  \overline{f}_{\tau}\equiv 0,\quad
\overline{f}(\vx, t)=\langle\overline f(\vx,t)\rangle
 \label{oper-3}
\end{eqnarray}
\underline{\emph{Comments:}} (A) Any \emph{$\mathbb{B}$-function} depends on $\vx$ and $t$ and does not depend on $\tau$;
in particular, $\mathbb{B}$-function can appear from $\mathbb{H}$-function after its averaging. (B) With the use of the
average (\ref{oper-1}) any $\mathbb{H}$-function can be uniquely separated into its $\mathbb{B}$- and $\mathbb{T}$-
parts:
\begin{eqnarray}
\widehat{f}(\vx,t,\tau)=\overline{f}(\vx, t) +\widetilde{f}(\vx,t,\tau)\quad
\text{with}\quad \langle \widehat{f}(\vx,t,\tau)\rangle\equiv\overline{f}(\vx,t)\label{oper-4}
\end{eqnarray}
\underline{\emph{Definition 5.}}
\emph{The $\mathbb{T}$-integration}: For a given $\widetilde{f}$ we introduce a new function $\widetilde{f}^{\tau}$
called the $\mathbb{T}$-integral of $\widetilde{f}$:
\begin{eqnarray}
&&\widetilde{f}^{\tau}\equiv\int_0^\tau \widetilde{f}(\vx,t,\sigma)d\sigma -\frac{1}{2\pi}\int_0^{2\pi}\Bigl(\int_0^\mu
\widetilde{f}(\vx,t,\sigma)d\sigma\Bigr)d\mu.\label{oper-7}
\end{eqnarray}
which represents the unique solution of the partial differential equation $\partial
\widetilde{f}^{\tau}/\partial\tau=\widetilde{f}$ with the condition
$\langle \widetilde f\, \rangle=\langle \widetilde f^\tau \rangle =0$ (\ref{oper-2}).

\underline{\emph{Comments:}} (A) The $\tau$-derivative of $\mathbb{T}$-function always represents
$\mathbb{T}$-function. However the $\tau$-integration of $\mathbb{T}$-function can produce an $\mathbb{H}$-function. An
example: $\widetilde{f}=\overline{f}_1\sin\tau$ where $\overline{f}_1$ is an arbitrary function of $\vx, t$: one can
see that $\langle\widetilde{f}\rangle\equiv 0$, however $\langle\int_0^\mu\widetilde{f}d\tau\rangle=\overline{f}_1\neq
0$, unless $\overline{f}_1\equiv 0$. Formula (\ref{oper-7}) keeps the result of integration inside the
$\mathbb{T}$-class. (B) The $\mathbb{T}$-integration is inverse to the $\tau$-differentiation
\begin{eqnarray}
&&(\widetilde{f}^{\tau})_{\tau}=(\widetilde{f}_{\tau})^{\tau}=\widetilde{f}.\label{oper-8}
\end{eqnarray}
The proof is omitted.

\underline{\emph{Definition 6.}}
A dimensionless function $f=f(\vx,t,\tau,\varepsilon)$ ($\varepsilon$ is a small parameter) belongs to the class
$\mathbb{O}(1)$ if $f={O}(1)$ and all partial $\vx$-, $t$-, and $\tau$-derivatives of $f$ (required for further
consideration) are also ${O}(1)$.

\noindent\underline{\emph{Comments:}}  In all text below: (A) The highest spatial derivatives will be of
the fourth order, while all $t$- and $\tau$-derivatives will be of the first order; (B) All large or small parameters
(in all this paper) are represented by various degrees of $\omega$ only; these parameters appear as explicit
multipliers in all formulae containing tilde- and bar-functions; these functions always belong to
$\mathbb{O}(1)$-class.

\noindent\underline{\emph{Definition 7.}}
The commutator of two vector fields $\va(\vx)$ and $\vb(\vx)$ is
\begin{eqnarray}
&& [\va,\vb]\equiv(\vb\cdot\nabla)\va-(\va\cdot\nabla)\vb. \label{oper-11}
\end{eqnarray}
\noindent\underline{\emph{Comment:}} The commutator is antisymmetric and always satisfies Jacobi's identity
for three vector fields $\va(\vx)$, $\vb(\vx)$, and $\vc(\vx)$:
\begin{eqnarray}
&&[\va,\vb]=-[\vb,\va],\quad [\va,[\vb,\vc]]+ [\vc,[\va,\vb]]+[\vb,[\vc,\va]]=0 \label{oper-13}
\end{eqnarray}
Two more useful properties of commutators are:
\begin{eqnarray}
&& \Div\va=0,\quad \Div\vb=0\quad \Rightarrow\quad \Div[\va,\vb]=0\label{oper-11a}\\
&&\va\cdot\vn=0,\quad \vb\cdot\vn=0\quad \Rightarrow\quad [\va,\vb]\cdot\vn=0\label{oper-11b}
\end{eqnarray}
where $\vn$ is a normal vector to a surface $\partial\Omega$.

\noindent\underline{\emph{Properties of $\tau$-differentiation and $\mathbb{T}$-integration:}}

For $\tau$-derivatives it is clear that
\begin{eqnarray}
\widehat{f}_\tau=\overline{{f}}_\tau+\widetilde{f}_\tau=\widetilde{f}_\tau, \quad
\langle\widehat{f}_\tau\rangle=\langle\widetilde{f}_\tau\rangle=0 \label{oper-6}
\end{eqnarray}
The product of two $\mathbb{T}$-functions $\widetilde{f}$ and $\widetilde{g}$ forms a $\mathbb{H}$-function:
$\widetilde{f}\widetilde{g}\equiv\widehat{F}$, say. Separating $\mathbb{T}$-part $\widetilde{F}$ from $\widehat{F}$ we
write
\begin{eqnarray}
&&\widetilde{F}=\widehat{F}-\langle\widehat{F}\rangle
=\widetilde{f}\widetilde{g}-\langle\widetilde{f}\widetilde{g}\rangle=
\widetilde{\widetilde{f}\widetilde{g}}\equiv\{\widetilde{f}\widetilde{g}\}
\label{oper-5}
\end{eqnarray}
Since we do not use a two-level tilde notation for the $\mathbb{T}$-parts of long expressions, we introduce braces
instead. As the average operation (\ref{oper-1}) is proportional to the integration over $\tau$, for products
containing functions $\widetilde{f},\widetilde{g},\widetilde{h}$ and their derivatives we have
\begin{eqnarray}
&&\langle\widetilde{f}\widetilde{g}_\tau\rangle=\langle(\widetilde{f}\widetilde{g})_\tau\rangle-
\langle\widetilde{f}_\tau\widetilde{g}\rangle=-\langle\widetilde{f}_\tau \widetilde{g}\rangle=-
\langle\widetilde{f}_\tau \widehat{g}\rangle
\label{oper-9}\\
&&\langle\widetilde{f}_\tau\widetilde{g}\widetilde{h}\rangle+\langle\widetilde{f}\widetilde{g}_\tau\widetilde{h}\rangle+
\langle\widetilde{f}\widetilde{g}\widetilde{h}_\tau\rangle=0
\label{oper-9a}\\
&&\langle\widetilde{f}\widetilde{g}^\tau\rangle=\langle(\widetilde{f}^\tau\widetilde{g}^\tau)_\tau\rangle-
\langle\widetilde{f}^\tau\widetilde{g}\rangle=-\langle\widetilde{f}^\tau \widetilde{g}\rangle=-
\langle\widetilde{f}^\tau \widehat{g}\rangle
\label{oper-10}
\end{eqnarray}
which represent different versions of the integration by parts. Similarly
\begin{eqnarray}
&&\langle[\widetilde{\va},\widetilde{\vb}_\tau]\rangle=-\langle[\widetilde{\va}_\tau,\widetilde{\vb}]\rangle=-
\langle[\widetilde{\va}_\tau, \widehat{\vb}]\rangle,\
\langle[\widetilde{\va},\widetilde{\vb}^\tau]\rangle=-\langle[\widetilde{\va}^\tau,\widetilde{\vb}]\rangle=-
\langle[\widetilde{\va}^\tau, \widehat{\vb}]\rangle
\label{oper-15}
\end{eqnarray}
\underline{\emph{The equation to be solved:}}

Let us accept that a dimensionless velocity in (\ref{exact-6-L}), (\ref{exact-6a}) is
\begin{eqnarray}
&& \widehat{\vu}(\vx,t,\tau)\in\mathbb{H}\cap\mathbb{O}(1), \quad\text{and}\quad
\delta(\omega)=\delta_0\omega^{\beta-1},\
\beta=\const<1\label{insp-1}
\end{eqnarray}
where $\widehat{\vu}$ is a given function; indefinite constant $\beta$ allows us to consider different asymptotic
paths; one can take $\delta_0\equiv 1$ since any other $\delta_0$ can be incorporated into $\widehat{\vu}$. Eqn.
(\ref{exact-6-L}) can be rewritten as
\begin{eqnarray}
&&\mathfrak{D}
\widehat{a}=\omega\widehat{a}_\tau+\widehat{a}_t+
\omega^\beta(\widehat{\vu}\cdot\nabla)\widehat{a}=0.\label{insp-6}
\end{eqnarray}
In \emph{Sects.3-7} we will study and solve this equation.

\underline{Remark:} The $\tau$-periodicity of the given velocity (\ref{exact-2}),(\ref{insp-1}) represents a restriction which is artificially
imposed to simplify further calculations. At the same time we will show that a $\tau$-periodic asymptotic solution
(\ref{exact-3}) can be build for any initial data; hence we are not forced to go outside the $\mathbb{H}$-class
(\ref{tilde-func-def}).

\subsection{Choice of scale $T$ for $t$-independent velocity}

A $t$-independent velocity
\begin{eqnarray}
\widehat{\vu}=\widehat{\vu}(\vx,\tau)\label{t-indep}
\end{eqnarray}
is important theoretically. This case represents a degeneration, where the presence of slow time $t$ in
eqns.(\ref{exact-6-L}),(\ref{insp-6}) is provided only by the additional agreement $\widehat{a}_t\in
\mathbb{O}(1)$ leading to two mutually dependent time-scales $(t, \tau) =(s, \omega s)$.
It is clear that for (\ref{t-indep}) this agreement represents only one available option and an infinite number of
different slow-time-scales can be introduced. To make it possible one can write
\begin{eqnarray}\label{lambda}
\tau=\omega^{1-\lambda}(\omega^\lambda s)=\omega_\lambda t_\lambda\equiv\tau_\lambda\ \text{where}\
\omega_\lambda\equiv\omega^{1-\lambda},\ t_\lambda\equiv\omega^\lambda s,\ \lambda=\const<1-\lambda_0
\end{eqnarray}
with a small constant $\lambda_0>0$ which provides $\omega_\lambda\to\infty$ when $\omega\to\infty$. These
manipulations allow to introduce time-scales $(t_\lambda,
\tau_\lambda) =(\omega^\lambda s, \omega s)$ and a new version of eqn.(\ref{insp-6})
\begin{eqnarray}
\left(\frac{\partial}{\partial {s}}+\omega^{\beta}\widehat{\vu}_\lambda\cdot\nabla\right) \widehat{a}=0,\quad
\frac{\partial}{\partial {s}}=\omega^\lambda\frac{\partial}{\partial t_\lambda}+\omega\frac{\partial}{\partial\tau_\lambda}
\label{eqn-lambda}
\end{eqnarray}
with $\widehat{\vu}_\lambda\equiv\widehat{\vu}(\vx,\tau_\lambda)$. Further transformations of (\ref{eqn-lambda}) yield
\begin{eqnarray}
\left(\frac{\partial}{\partial t_\lambda}+\omega_\lambda\frac{\partial}{\partial\tau_\lambda}\right)\widehat{a}+
\omega_\lambda^{\beta_\lambda}(\widehat{\vu}_\lambda\cdot\nabla)\widehat{a} =0,\quad
\label{eqn-lambda-1}
\end{eqnarray}
where $\beta_\lambda\equiv(\beta-\lambda)/(1-\lambda)$ and $\widehat{a}=\widehat{a}(\vx,t_\lambda,\tau_\lambda)$.
Finally, we replace the old scaling agreement $\widehat{a}_{t}\in
\mathbb{O}(1)$ with a new one $\widehat{a}_{t_\lambda}\in
\mathbb{O}(1)$ ($\widehat{a}_{\tau_\lambda}\in
\mathbb{O}(1)$ remains valid since $\tau_\lambda=\tau$). Now one can see that the problem
(\ref{eqn-lambda-1}) is mathematically identical to (\ref{insp-6}), which means that for any solution in variables
$(t,\tau)$ we have an additional solution in variables $(t_\lambda,\tau_\lambda)$ (and
\emph{vice versa}); the varying of $\lambda$ gives us an infinite number of such additional solutions.
Hence, an infinite number of new solutions to (\ref{eqn-lambda-1}) can be obtained by rescaling of the known solutions
to (\ref{insp-6}), which will be considered in \emph{Sects.3-7}. The existence of such additional solutions means that
motions with any slow-time-scale $\omega^\lambda t$ (\ref{lambda}) are possible. Indeed, it sounds physically
reasonable: if a slow-time-scale is not enforced externally (does not appear in the given velocity) then any time-scale
that is longer than imposed oscillations can be considered as a slow one. For brevity, in the text below we keep
$\widehat{a}_t\in\mathbb{O}(1)$ and two time-scales $(t, \tau) =(s, \omega s)$.

\underline{Remark:} Similar to (\ref{t-indep}),(\ref{eqn-lambda-1}), the multiplicity of
slow-time-scales takes place, for instance, in dynamics of a classical pendulum with a harmonically vibrating pivot,
when an external gravity field is absent.

\section{The Inspection Procedure \label{sect03}}

For simplicity we consider now eqns.(\ref{insp-1}),(\ref{insp-6}) with purely oscillatory velocity
$\widehat{\vu}=\widetilde{\vu}\in\mathbb{T}\cap\mathbb{O}(1)$. In our
\emph{inspection procedure} we use a test solution to (\ref{insp-6})
\begin{eqnarray}
&&\widehat{a}(\vx,t,\tau)=\overline{a}(\vx,t)+\frac{1}{\omega^\alpha}\,\widetilde{b}(\vx,t,\tau);
\quad \alpha=\const>0\label{insp-7}
\end{eqnarray}
where $\overline{a}\in \mathbb{B}\cap\mathbb{O}(1),\  \widetilde{b}\in \mathbb{T}\cap\mathbb{O}(1)$ and the amplitude
of an oscillating part is given by the small parameter $\varepsilon_\alpha=1/\omega^\alpha$ (\ref{epsilon}). The
substitution of (\ref{insp-7}) into (\ref{insp-6}) yields
\begin{eqnarray}\label{insp-8}
\overline{a}_t+\frac{1}{\omega^\alpha}\left(\omega
\widetilde{b}_\tau+\widetilde{b}_t\right)+
\omega^\beta\left((\widetilde{\vu}\cdot\nabla) \overline{a}+
\frac{1}{\omega^{\alpha}}(\widetilde{\vu}\cdot\nabla) \widetilde{b} \right)=0
\end{eqnarray}
The \emph{$\mathbb{B}$-part} of (\ref{insp-8}) is
\begin{eqnarray}\label{insp-9}
\overline{a}_t+\omega^{\beta-\alpha}\langle(\widetilde{\vu}\cdot\nabla) \widetilde{b}\rangle=0
\end{eqnarray}
and the subtraction of (\ref{insp-9}) from (\ref{insp-8}) produces the \emph{ $\mathbb{T}$-part} of (\ref{insp-8})
\begin{eqnarray}
&&\underline{\omega^{1-\alpha}
\widetilde{b}_\tau+\omega^{-\alpha}\widetilde{b}_t}+
\underline{\omega^\beta(\widetilde{\vu}\cdot\nabla) \overline{a}+
\omega^{\beta-\alpha}\{(\widetilde{\vu}\cdot\nabla) \widetilde{b}\}}=0\label{insp-10}
\end{eqnarray}
with the notation $\{\cdot\}$ (\ref{oper-5}) used.  In each underlined group in (\ref{insp-10}) the first term
dominates over the second one (due to $\alpha>0$ (\ref{insp-7})). Therefore, to make this equation meaningful we need
to accept that the dominating terms are of the same order:
\begin{equation}
1-\alpha=\beta\quad\text{or} \quad   \alpha+\beta=1.\label{insp-10a}
\end{equation}
Then the imposed restrictions $\alpha>0$ (\ref{insp-7}) and $\beta<1$ (\ref{insp-1}) are mutually compatible and
(\ref{insp-9}), (\ref{insp-10}) can be rewritten as
\begin{eqnarray}
&&\overline{a}_t+
\omega^\gamma\langle(\widetilde{\vu}\cdot\nabla) \widetilde{b}\rangle=0 \quad \text{where}\quad
\gamma\equiv2\beta-1=1-2\alpha \label{insp-9a}\\
&&\omega\left(\widetilde{b}_\tau+(\widetilde{\vu}\cdot\nabla) \overline{a}\right)+\widetilde{b}_t
+\omega^{\beta}\{(\widetilde{\vu}\cdot\nabla) \widetilde{b}\}=0\label{insp-9b}
\end{eqnarray}
The second term in eqn.(\ref{insp-9a}) represents the changes of $\overline{a}$ due to the averaged nonlinear effects
of oscillations. Taking the different values of $\gamma$ we introduce three qualitatively different asymptotic families
of solutions:
\begin{eqnarray}
&\gamma>0\ (1/2<\beta<1,\ 0<\alpha<1/2) &- \text{\emph{super-critical oscillations}}
\label{insp-11a}\\
&\gamma=0\ (\beta=1/2, \ \alpha=1/2) &- \text{\emph{critical oscillations}}
\label{insp-11b}\\
&\gamma<0\ (-\infty<\beta<1/2, \ 1/2<\alpha<\infty) &- \text{\emph{sub-critical oscillations}}\label{insp-11c}
\end{eqnarray}
One can notice that \emph{super-critical, critical, and sub-critical oscillations} have been introduced as the
asymptotic families of solutions, not as particular solutions (recall that any particular solution can be treated as a
member of several different asymptotic families).

For the family of \emph{critical oscillations} the averaged oscillatory term in (\ref{insp-9a}) has the same order
${O}(1)$ as a mean solution. Then the leading terms (for large $\omega$) in (\ref{insp-9a}), (\ref{insp-9b}) are:
\begin{eqnarray}
&& \overline{a}_t+\langle(\widetilde{\vu}\cdot\nabla)\, \widetilde{b}\rangle=0;\label{insp-12}\\
&&\widetilde{b}_\tau+(\widetilde{\vu}\cdot\nabla)\, \overline{a}=0.\label{insp-12a}
\end{eqnarray}
This system gives a single equation for $\overline{a}$
\begin{eqnarray}
&& \overline{a}_t-\langle(\widetilde{\vu}\cdot\nabla)(\widetilde{\vxi}\cdot\nabla)\rangle\,\overline{a}=0,\quad
\widetilde{\vxi}\equiv\widetilde{\vu}^\tau\label{insp-12b}
\end{eqnarray}
which is obtained by $\mathbb{T}$-integration (\ref{oper-7}) of (\ref{insp-12a}) and the following substitution of
$\widetilde{b}(\vx,t,\tau)=-(\widetilde{\vxi}\cdot\nabla)\,\overline{a}$ into (\ref{insp-12}).

The critical family of oscillations (\ref{insp-11b}) separates the larger families of sub-critical and super-critical
oscillations from each other. For super-critical oscillations $\gamma>0$  and the correlation in (\ref{insp-9a}) can
not be balanced by $\overline{a}_t={O}(1)$. Then the condition of solvability requires this correlation to vanish,
hence the leading order of (\ref{insp-12}), (\ref{insp-12b}) is replaced with $\langle(\widetilde{\vu}\cdot\nabla)\,
\widetilde{b}\rangle=0$. It means that the imposed velocity must degenerate (satisfy
certain additional restrictions).

\underline{Remarks:}

1. The inspection procedure is aimed to find and classify all possible asymptotic solutions (of the functional form
(\ref{insp-7})) of a given PDE. This procedure also allows to choose a natural small parameter (\ref{epsilon}) with
different values of $\alpha$ for different classes of solutions.

2. The critical and super-critical oscillations are most interesting both theoretically and practically. We justify
this statement with three arguments: (A) \emph{Sects.4-7} below are completely devoted to these two oscillatory
families. The obtained results are  mathematically satisfactory and look physically convincing. (B)  These families of
oscillations are the least studied, probably due to the reason that for them  $U\to\infty$ as $\omega\to\infty$ in
(\ref{path-simple}); however it is well-known that such an increase of $U$ does not diminish the validity of asymptotic
procedures. (C) In
\emph{Sect.7.11} we will consider the sub-critical family (\ref{insp-11c}) with $\beta=0$ in (\ref{insp-1}),(\ref{insp-6}).
We will show that for this family a drift is replaced with a \emph{pseudo-drift} and the general solution is diverging.

\section{Family of Critical Oscillations. Purely Oscillatory Velocity.  \label{sect04}}

\subsection{Critical asymptotic procedure with  $\alpha={1/2}$}

Following (\ref{insp-11b}) we accept that (\ref{insp-1}) belongs to a critical asymptotic family:
\begin{eqnarray}
&& \widehat{\vu}(\vx, t, \tau)=\widetilde{\vu}(\vx,t,\tau),\quad \delta=\omega^{-1/2},
\label{basic-1}
\end{eqnarray}
Then (\ref{insp-6}), (\ref{basic-1}) yield:
\begin{eqnarray}
&& \mathfrak{D} \widehat{a}=\omega\widehat{a}_\tau+\widehat{a}_t+\sqrt\omega(\widetilde{\vu}\cdot\nabla)\,\widehat{a}=
0
\label{basic-2}
\end{eqnarray}
The small parameter $\varepsilon_{1/2}= 1/\sqrt\omega$ (\ref{epsilon}) allows to rewrite it as
\begin{eqnarray}
&& \mathfrak{D}_2
\widehat{a}\equiv\widehat{a}_{\tau}+\varepsilon(\widetilde{\vu}\cdot\nabla)\,\widehat{a}+\varepsilon^2\widehat{a}_t=0, \quad
\mathfrak{D}_2\equiv\varepsilon^2\mathfrak{D}=\mathfrak{D}/\omega\label{basic-3}
\end{eqnarray}
where the subscript in $\varepsilon_{1/2}$ has been dropped. We are looking for the solution of (\ref{basic-3}) in the
form of regular series
\begin{eqnarray}
&&\widehat{a}(\vx,t,\tau)=\sum_{k=0}^\infty\varepsilon^k\widehat{a}_k(\vx,t,\tau),\quad \widehat{a}_k\in
\mathbb{H}\cap\mathbb{O}(1),
\quad k=0,1,2,\dots
\label{basic-4aa}
\end{eqnarray}
The substitution of (\ref{basic-4aa}) into (\ref{basic-3}) produces the equations of successive approximations
\begin{eqnarray}
&&\widehat{a}_{0\tau}=0 \label{basic-5}\\
&&\widehat{a}_{1\tau}=-(\widetilde{\vu}\cdot\nabla)\, \widehat{a}_0\label{basic-6}\\
&&\widehat{a}_{n\tau}=-(\widetilde{\vu}\cdot\nabla)\,
\widehat{a}_{n-1}-\partial_t\,\widehat{a}_{n-2},\quad\partial_t\equiv\partial/\partial t,\quad n=2,3,4,\dots
\label{basic-7}\\
&&\widehat{a}_{k}=\overline{a}_k(\vx,t)+\widetilde{a}_k(\vx,t,\tau),\ \overline{a}_k\in \mathbb{B}\cap \mathbb{O}(1),\
\widetilde{a}_k\in \mathbb{T}\cap \mathbb{O}(1), \ k=0,1,2,\dots\nonumber
\end{eqnarray}
where (\ref{basic-6}) represents a systematically derived version of (\ref{insp-12a}). Let $\widehat{a}^{[n]}$ be a
truncated solution (\ref{basic-4aa}) of order $n$:
\begin{eqnarray}
&&\widehat{a}^{[n]}(\vx,t,\tau)\equiv\sum_{k=0}^n\varepsilon^k\widehat{a}_k(\vx,t,\tau)
\label{basic-4a}
\end{eqnarray}
Its substitution into (\ref{basic-3}) produces the RHS-residual $\mathrm{Res}{[n]}$:
\begin{eqnarray}
&& \mathfrak{D}_2 \widehat{a}^{[n]} =\mathrm{Res}{[n]} \label{basic-3res}
\end{eqnarray}

\subsection{General solution for the first five approximations}

The detailed solving of (\ref{basic-5})-(\ref{basic-7}) for $n=0,1,2,3,4$ is given in \emph{Appendix A}. The truncated
general solution $\widehat{a}^{[4]}$ is
\begin{eqnarray}
&&\widehat{a}^{[4]}=\widehat{a}_0+\varepsilon\widehat{a}_1+\varepsilon^2\widehat{a}_2+\varepsilon^3\widehat{a}_3+
\varepsilon^4\widehat{a}_4
\label{4.10}
\end{eqnarray}
with the tilde-parts $\widetilde{a}_k$ given by explicit recurrent expressions
\begin{eqnarray}
&&\widetilde{a}_0\equiv 0,\label{4.11a}\\
&& \widetilde{a}_1= -(\widetilde{\vxi}\cdot\nabla)\,\overline{a}_0,\label{4.11}\\
&&\widetilde{a}_{2}=-(\widetilde{\vxi}\cdot\nabla)\, \overline{a}_1
-\{(\widetilde{\vu}\cdot\nabla)\,\widetilde{a}_1\}^\tau, \label{4.12}\\
&&\widetilde{a}_{3}=-(\widetilde{\vxi}\cdot\nabla)\,\overline{a}_2
-\{(\widetilde{\vu}\cdot\nabla)\,\widetilde{a}_2\}^\tau-\widetilde{a}_{1t}^\tau, \label{4.13}\\
&&\widetilde{a}_{4}=-(\widetilde{\vxi}\cdot\nabla)\,\overline{a}_3-
\{(\widetilde{\vu}\cdot\nabla)\,\widetilde{a}_3\}^\tau-\widetilde{a}_{2t}^\tau, \label{4.14}\\
&&\widetilde{\vxi}\equiv\widetilde{\vu}^\tau=\int_0^\tau \widetilde{\vu}(\vx,t,\sigma)d\sigma
-\frac{1}{2\pi}\int_0^{2\pi}\left(\int_0^\mu
\widetilde{\vu}(\vx,t,\sigma)d\sigma\right)d\mu\label{01-appr-5}
\end{eqnarray}
and the bar-parts $\overline{a}_k$ satisfy the equations
\begin{eqnarray}
&&\left(\partial_t+ \overline{\vV}_0\cdot\nabla\right)\overline{a}_{0}=0\label{4.15}\\
&&\left(\partial_t+ \overline{\vV}_0\cdot\nabla\right)
\overline{a}_{1}+(\overline{\vV}_1\cdot\nabla)\overline{a}_0=0\label{4.16}\\
&&\left(\partial_t+ \overline{\vV}_0\cdot\nabla\right)\overline{a}_{2}+
(\overline{\vV}_1\cdot\nabla)\overline{a}_1+(\overline{\vV}_2\cdot\nabla)\overline{a}_0=
\frac{\partial}{\partial x_i}\left(
\overline{\chi}_{ik}\frac{\partial\overline{a}_0}{\partial x_k}\right),
\label{4.17}\\
&&\widehat{\vV}_0\equiv \frac{1}{2}[\widetilde{\vu},\widetilde{\vxi}],\quad
\overline{\vV}_1\equiv\frac{1}{3}\langle[[\widetilde{\vu},\widetilde{\vxi}],\widetilde{\vxi}]\rangle,
\label{4.18}\\
&&\overline{\vV}_2\equiv\frac{1}{4}\langle[[\widehat{\vV}_0,\widetilde{\vxi}],\widetilde{\vxi}]\rangle +
\frac{1}{2}\langle[\widetilde{\vV}_0,\widetilde{\vV}_0^\tau]\rangle+
\frac{1}{2}\langle[\widetilde{\vxi},\widetilde{\vxi}_t]\rangle
+\frac{1}{2}\langle\widetilde{\vxi}\Div\widetilde{\vu}'+ \widetilde{\vu}'\Div\widetilde{\vxi}\rangle,\label{4.19}
\\
&&\widetilde{\vu}'\equiv\widetilde{\vxi}_t-[\overline{\vV}_0,\widetilde{\vxi}],
\label{4.20a}\\
&&2\overline{\chi}_{ik}\equiv\langle\widetilde{u'}_i\widetilde{\xi}_k+\widetilde{u'}_k\widetilde{\xi}_i\rangle=
\mathfrak{L}_{\overline{\vV}_0}\langle\widetilde{\xi}_i\widetilde{\xi}_k\rangle,\label{4.20}\\
&&\mathfrak{L}_{\overline{\vV}_0}\overline{f}_{ik}\equiv
\left(\partial_t+
\overline{\vV}_0\cdot\nabla\right)\overline{f}_{ik}-\frac{\partial\overline{V}_{0k}}{\partial x_m}\overline{f}_{im}-
\frac{\partial\overline{V}_{0i}}{\partial x_m}\overline{f}_{km}
\label{4.20b}
\end{eqnarray}
where the operator $\mathfrak{L}_{\overline{\vV}_0}$ is such that $\mathfrak{L}_{\overline{\vV}_0}
\overline{f}_{ik}=0$ represents the condition for tensorial field $\overline{f}_{ik}(\vx,t)$
to be `frozen' into velocity field $\overline{\vV}_0(\vx,t)$. Three equations (\ref{4.15})-(\ref{4.17}) can be written
as a single advection-pseudo-diffusion equation (with the error ${O}(\varepsilon^3)$)
\begin{eqnarray}
&&\left(\partial_t+ \overline{\vV}\cdot\nabla\right)\overline{a} =
\frac{\partial}{\partial x_i}\left(\overline{\kappa}_{ik}\frac{\partial\overline{a}}{\partial x_k}\right)
\label{4.21}\\
&&\overline{\vV}=\overline{\vV}^{[2]}=\overline{\vV}_0+\varepsilon\overline{\vV}_1+\varepsilon^2
\overline{\vV}_2,\label{4.22}\\
&&\overline{\kappa}_{ik}=\overline{\chi}_{ik}^{[2]}=\varepsilon^2\overline{\chi}_{ik}\label{4.23}\\
&&\overline{a}=\overline{a}^{[2]}=\overline{a}_0+\varepsilon\overline{a}_1+\varepsilon^2\overline{a}_2
\label{4.22aa}
\end{eqnarray}
where all coefficients are given in (\ref{4.18})-(\ref{4.20b}). Eqn. (\ref{4.21}) shows that the averaged motion of
$\widehat{a}$ represents a drift with velocity $\overline{\vV}=\overline{\vV}^{[2]}+{O}(\varepsilon^3)$ and
\emph{pseudo-diffusion} with the matrix coefficient
$\overline{\kappa}_{ik}=\varepsilon^2\overline{\chi}_{ik}+{O}(\varepsilon^3)$.

\underline{\emph{Definition:}} The term {\emph{pseudo-diffusion (PD)}} means that:
(i) the evolution of $\overline{a}$ is described by an advection-diffusion-type equation (\ref{4.21}); (ii) the
diffusion-type term appears in (\ref{4.21}) not in the leading approximation (see (\ref{4.17}), where it represents a
known source-type-term in the second approximation); and (ii) the equation (\ref{4.21}) is valid only for regular
asymptotic expansions (\ref{4.22})-(\ref{4.22aa}).

In $\widehat{a}^{[4]}$ (\ref{4.10}) the expressions for $\overline{a}_0$, $\overline{a}_1$, $\overline{a}_2$,
$\widetilde{a}_{1}$, $\widetilde{a}_{2}$, $\widetilde{a}_{3}$, and $\widetilde{a}_{4}$ are given by
(\ref{4.11a})-(\ref{4.20b}), while $\overline{a}_3$ and $\overline{a}_4$ are to be found from higher approximations.
The substitution of $\widehat{a}^{[4]}$ into (\ref{basic-3}) produces the RHS-residual of order $\varepsilon^5$
\begin{eqnarray}
\mathfrak{D}_2\widehat{a}^{[4]}=\mathrm{Res}{[4]}={O}(\varepsilon^5)={O}(1/\omega^{5/2})\label{4-appr-7}
\end{eqnarray}
One can notice that the residual of order ${O}(1/\omega^{5/2})$ for $\mathfrak{D}_2$ leads to the residual
${O}(1/\omega^{3/2})$ for the original operator $\mathfrak{D}$(\ref{basic-2}).

\underline{Remarks:}

1. The statement that (\ref{4.11a})-(\ref{4.20b}) describe the solution of the problem with arbitrary initial data is
based on the fact that initial data for (\ref{4.15})-(\ref{4.17}) can be chosen arbitrarily.

2. We have explicitly solved the equations of the first five approximations. Such persistence does not occur often in
the use of asymptotic methods. However, the well-known examples are the surface gravity wave, see
\cite{Stokes,Stoker,Debnath} and the low-$Re$-solutions for a sphere, see \cite{Chester}.
Our motivation for doing the cumbersome calculations of \emph{Appendix A} is based on two our results: (i) the first
and the second approximations of the critical asymptotic solutions appear as the main-order terms in super-critical
solutions (see \emph{Sect.5}); and (ii) \emph{PD} (which qualitatively complements a drift) appears only in the second
approximation.


3. The term \emph{PD} is different from \emph{super-diffusion}, which means the spreading of an admixture faster than
$\sqrt t$ (see \cite{Majda, Volpert}).


\section{Super-critical Oscillations\label{sect05}}

\subsection{Super-critical asymptotic procedure with  $\alpha={1/3}$}

For critical oscillations (\ref{basic-1}) the drift velocity (\ref{4.22}) is $\overline{\vV}\sim\overline{\vV}_0=
{O}(1)$ (\ref{4.18}). From (\ref{4.22}) one can expect that for the degenerated $\widetilde{\vu}$ such that
\begin{eqnarray}
&&\overline{\vV}_0= \frac{1}{2}\langle[\widetilde{\vu},\widetilde{\vxi}]\rangle  \equiv 0 \label{5.0}
\end{eqnarray}
the drift is $\overline{\vV}\sim\varepsilon\overline{\vV}_1={O}(1/\sqrt\omega)$. However, one can find that the drift
velocity is still ${O}(1)$ in a different asymptotic family. To show it, we accept (\ref{5.0}) and consider a
super-critical family of purely oscillating velocities
\begin{eqnarray}
&& \widehat{\vu}(\vx, t, \tau)=\widetilde{\vu}(\vx,t,\tau),\quad \delta=\omega^{-1/3}
\label{5.1}
\end{eqnarray}
where $\widetilde{\vu}(\vx,t,\tau)\in\mathbb{T}\cap\mathbb{O}(1)$ is a given function. Then eqns. (\ref{insp-6}),
(\ref{5.1}) yield:
\begin{eqnarray}
&& \mathfrak{D} \widehat{a}=\omega\widehat{a}_\tau+
\widehat{a}_t+\omega^{2/3}(\widetilde{\vu}\cdot\nabla)\,\widehat{a}= 0
\label{5.2}
\end{eqnarray}
A small parameter $\varepsilon_{1/3}=\omega^{-1/3}$ allows to rewrite (\ref{5.2}) as
\begin{eqnarray}
&& \mathfrak{D}_3
\widehat{a}\equiv\widehat{a}_{\tau}+\varepsilon(\widetilde{\vu}\cdot\nabla)\,\widehat{a}
+\varepsilon^3\,\widehat{a}_t=0, \quad
\mathfrak{D}_3\equiv\varepsilon^3\mathfrak{D}=\mathfrak{D}/\omega\label{5.3}
\end{eqnarray}
where the subscript in $\varepsilon_{1/3}$ is dropped. We are looking for a solution of (\ref{5.3}) in the form of
regular series (\ref{basic-4aa}) with the redefined $\varepsilon$. The substitution of (\ref{basic-4aa}) into
(\ref{5.3}) produces the equations for successive approximations (\emph{cf.} with (\ref{basic-5})-(\ref{basic-7}))
\begin{eqnarray}
&&\widehat{a}_{0\tau}=0 \label{5.5}\\
&&\widehat{a}_{1\tau}=-(\widetilde{\vu}\cdot\nabla)\, \widehat{a}_0\label{5.6}\\
&&\widehat{a}_{2\tau}=-(\widetilde{\vu}\cdot\nabla)\, \widehat{a}_1\label{5.7}\\
&&\widehat{a}_{n\tau}=-(\widetilde{\vu}\cdot\nabla)\,\widehat{a}_{n-1} -\partial_t\,
\widehat{a}_{n-3},\quad n=3,4,5,\dots\label{5.8}
\end{eqnarray}
These equations can be solved in the most general form (see \emph{Appendix B}). Here we present the truncated solution
$\widehat{a}^{[4]}$ (\ref{4.10}). For the tilde-part of each approximation one can obtain explicit recurrent formulae
\begin{eqnarray}
&&\widetilde{a}_0\equiv 0,\label{5.10a}\\
&& \widetilde{a}_1= -(\widetilde{\vxi}\cdot\nabla)\,\overline{a}_0,\label{5.10}\\
&&\widetilde{a}_{2}=-(\widetilde{\vxi}\cdot\nabla)\, \overline{a}_1
-\{(\widetilde{\vu}\cdot\nabla)\,\widetilde{a}_1\}^\tau \label{5.11}\\
&&\widetilde{a}_{3}=-(\widetilde{\vxi}\cdot\nabla)\,\overline{a}_2
-\{(\widetilde{\vu}\cdot\nabla)\,\widetilde{a}_2\}^\tau \label{5.12}\\
&&\widetilde{a}_{4}=-(\widetilde{\vxi}\cdot\nabla)\,\overline{a}_3-
\{(\widetilde{\vu}\cdot\nabla)\,\widetilde{a}_3\}^\tau-\widetilde{a}_{1t}^\tau, \forall\overline{a}_3 \label{5.13}
\end{eqnarray}
and the transport equation (which is valid with the error ${O}(\varepsilon^3)$) for averaged motion
\begin{eqnarray}
&&\left(\partial_t+ \overline{\vV}\cdot\nabla\right)\overline{a} =0,
\label{5.14}\\
&&\overline{\vV}=\overline{\vV}^{[2]}=\overline{\vV}_1+\varepsilon \overline{\vV}_2,\label{5.15}\\
&&\overline{a}=\overline{a}^{[2]}=\overline{a}_0+\varepsilon\overline{a}_1,\quad  \forall \overline{a}_2,
\overline{a}_3, \overline{a}_4\label{5.16}\\
&&\overline{\vV}_1= -\frac{1}{3}\langle[\widehat{\vK},\widetilde{\vxi}]\rangle,\
\overline{\vV}_2=-\frac{1}{8}(\overline{\vkappa}+\overline{\vK''}),
\label{5.17}\\
&&\widehat{\vK}\equiv[\widetilde{\vxi},\widetilde{\vu}],\,
\widehat{\vkappa}\equiv[\widetilde{\vK}^\tau,\widetilde{\vK}],\
\widehat{\vK}''\equiv[[\widehat{\vK},\widetilde{\vxi}],\widetilde{\vxi}]
\label{5.17a}
\end{eqnarray}
Eqn.(\ref{5.14}) means that the averaged motion of a passive admixture (up to the first two approximations) represents
a pure drift with velocity $\overline{\vV}^{[2]}$ (\ref{5.15}).

Hence, in $\widehat{a}^{[4]}$ (\ref{4.10}) functions $\overline{a}_0$, $\overline{a}_1$, $\widetilde{a}_{1}$,
$\widetilde{a}_{2}$, $\widetilde{a}_{3}$, and $\widetilde{a}_{4}$ are given by (\ref{5.14})-(\ref{5.17}) and
(\ref{5.10})-(\ref{5.13}), while $\overline{a}_2$, $\overline{a}_3$ and $\overline{a}_4$ are to be found from further
approximations. The substitution of $\widehat{a}^{[4]}$ into (\ref{5.3}) produces the RHS-residual of order
$\varepsilon^5$
\begin{eqnarray}
\mathfrak{D}_3\widehat{a}^{[4]}=\mathrm{Res}{[4]}={O}(\varepsilon^5)={O}(1/\omega^{5/3})\label{5.18}
\end{eqnarray}
The residual of order ${O}(1/\omega^{5/3})$ for the operator $\mathfrak{D}_3$ means that the residual for original
operators (\ref{exact-1}), (\ref{exact-6}), (\ref{insp-6}), and (\ref{5.2}) is ${O}(1/\omega^{2/3})$.

\subsection{Super-critical asymptotic procedure with  $\alpha={1/4}$}

Next we consider the oscillating velocity fields with first two terms of the drift velocity (\ref{4.22}) vanishing
\begin{eqnarray}
&&\overline{\vV}_0= \frac{1}{2}\langle[\widetilde{\vu},\widetilde{\vxi}]\rangle\equiv 0 \quad\text{and}\quad
\overline{\vV}_1= \frac{1}{3}\langle[[\widetilde{\vu},\widetilde{\vxi}],\widetilde{\vxi}]\rangle\equiv
0\label{5.19}
\end{eqnarray}
Here one can find that the drift of magnitude ${O}(1)$ appears within the asymptotic family with $\alpha={1/4}$. Let us
accept (\ref{5.19}) and introduce a family of super-critical oscillations
\begin{eqnarray}
&& \widehat{\vu}(\vx, t, \tau)=\widetilde{\vu}(\vx,t,\tau),\quad \delta=\omega^{-1/4}\label{5.20}
\end{eqnarray}
where $\widetilde{\vu}(\vx,t,\tau)\in\mathbb{T}\cap\mathbb{O}(1)$ is a given function. Eqns.
(\ref{insp-6}),(\ref{5.20}) give:
\begin{eqnarray}
&& \mathfrak{D} \widehat{a}=\omega\widehat{a}_\tau+
\widehat{a}_t+\omega^{3/4}(\widetilde{\vu}\cdot\nabla)\,\widehat{a}= 0
\label{5.21}
\end{eqnarray}
The  small parameter $\varepsilon_{1/4}=\omega^{-1/4}$ allows us to rewrite it as
\begin{eqnarray}
&& \mathfrak{D}_4
\widehat{a}\equiv\widehat{a}_{\tau}+\varepsilon(\widetilde{\vu}\cdot\nabla)\,\widehat{a}
+\varepsilon^4\,\widehat{a}_t=0, \quad
\mathfrak{D}_4\equiv\varepsilon^4\mathfrak{D}=\mathfrak{D}/\omega\label{5.22}
\end{eqnarray}
where the subscript in $\varepsilon_{1/4}$ is dropped. We are looking for a solution of (\ref{5.22}) in the form of
regular series (\ref{basic-4aa}) with the redefined $\varepsilon$. The substitution of (\ref{basic-4aa}) into
(\ref{5.22}) produces the equations for successive approximations (\emph{cf.} with (\ref{basic-5})-(\ref{basic-7}) and
(\ref{5.5})-(\ref{5.8}))
\begin{eqnarray}
&&\widehat{a}_{0\tau}=0 \label{5.24}\\
&&\widehat{a}_{1\tau}=-(\widetilde{\vu}\cdot\nabla)\, \widehat{a}_0\label{5.25}\\
&&\widehat{a}_{2\tau}=-(\widetilde{\vu}\cdot\nabla)\, \widehat{a}_1\label{5.26}\\
&&\widehat{a}_{3\tau}=-(\widetilde{\vu}\cdot\nabla)\, \widehat{a}_2\label{5.27}\\
&&\widehat{a}_{n\tau}=-(\widetilde{\vu}\cdot\nabla)\,\widehat{a}_{n-1} -\partial_t\,
\widehat{a}_{n-4},\quad n=4,5,\dots\label{5.28}
\end{eqnarray}
Again, these equations can be solved in the most general form (see \emph{Appendix B}). Here we present the truncated
solution $\widehat{a}^{[4]}$ (\ref{4.10}). For the tilde-part of each approximation one can obtain explicit recurrent
formulae
\begin{eqnarray}
&&\widetilde{a}_0\equiv 0,\label{5.30a}\\
&& \widetilde{a}_1= -(\widetilde{\vxi}\cdot\nabla)\,\overline{a}_0,\label{5.30}\\
&&\widetilde{a}_{2}=-(\widetilde{\vxi}\cdot\nabla)\, \overline{a}_1
-\{(\widetilde{\vu}\cdot\nabla)\,\widetilde{a}_1\}^\tau \label{5.31}\\
&&\widetilde{a}_{3}=-(\widetilde{\vxi}\cdot\nabla)\,\overline{a}_2
-\{(\widetilde{\vu}\cdot\nabla)\,\widetilde{a}_2\}^\tau \label{5.32}\\
&&\widetilde{a}_{4}=-(\widetilde{\vxi}\cdot\nabla)\,\overline{a}_3-
\{(\widetilde{\vu}\cdot\nabla)\,\widetilde{a}_3\}^\tau,\ \forall \overline{a}_3 \label{5.33}
\end{eqnarray}
and the transport equation for the averaged motion
\begin{eqnarray}
&&\left(\partial_t+ \overline{\vV}\cdot\nabla\right)\overline{a} =0,
\label{5.34}\\
&&\overline{\vV}=\overline{\vV}^{[2]}=\overline{\vV}_2,\ \overline{a}=\overline{a}^{[2]}=\overline{a}_0;\nonumber
\label{5.35}
\end{eqnarray}
with the same notations as in (\ref{5.17}), (\ref{5.17a}). Equation (\ref{5.34}) means that averaged motion represents
a pure drift with velocity $\overline{\vV}_2$.

Hence, $\widehat{a}^{[4]}$ can be written as (\ref{4.10}) where $\overline{a}_0$, $\widetilde{a}_{1}$,
$\widetilde{a}_{0}$, $\widetilde{a}_{2}$, $\widetilde{a}_{3}$, and $\widetilde{a}_{4}$ are given by
(\ref{5.30a})-(\ref{5.34}), while $\overline{a}_1$, $\overline{a}_2$, $\overline{a}_3$ and $\overline{a}_4$ are to be
found from further approximations. The substitution of $\widehat{a}^{[4]}$ into (\ref{5.22}) produces the RHS-residual
of order $\varepsilon^5$
\begin{eqnarray}
\mathfrak{D}_4\widehat{a}^{[4]}=\mathrm{Res}{[4]}={O}(\varepsilon^5)={O}(1/\omega^{5/4})\label{5.36}
\end{eqnarray}
The residual of order ${O}(1/\omega^{5/4})$ for the operator $\mathfrak{D}_4$ means that the residual for original
operators (\ref{exact-1}), (\ref{exact-6}), (\ref{insp-6}), and (\ref{5.21}) is ${O}(1/\omega^{1/4})$.

\underline{Remark:} One can see that the results for both the super-critical families are simpler than for the critical one
(\emph{Sect.4}) and they are build up from the same elements: if the conditions of degenerations (\ref{5.0}) or
(\ref{5.19}) are valid, then either $\overline{\vV}_1$ or $\overline{\vV}_2$ (\ref{4.22}) becomes the leading
$O(1)$-term in (\ref{5.16}) or (\ref{5.35}). At the same time pseudo-diffusion does not appear (within the considered
precision).

\section{Polynomial Velocity for Critical and Super-Critical Oscillations.  \label{sect07}}

Expressions (\ref{basic-1}), (\ref{5.1}), (\ref{5.20}) for velocity have a special form. One can expect that the adding
of higher-order terms to these expressions may potentially change both the drift and pseudo-diffusion.

\subsection{Critical asymptotic procedure with  $\alpha={1/2}$}

Let us consider a critical asymptotic family with a velocity in (\ref{exact-2}),(\ref{exact-6-L}),(\ref{insp-6}) given
as (\emph{cf.} with (\ref{basic-1}))
\begin{eqnarray}
&&\widehat{\vu}=\widehat{\vv}+\frac{1}{\sqrt\omega}\widehat{\vw}+\frac{1}{\omega}\widehat{\vr}, \quad
\delta=\omega^{-1/2},\label{7.1}
\end{eqnarray}
where $\widehat{\vv}, \widehat{\vw}, \widehat{\vr}\in\mathbb{H}\cap\mathbb{O}(1)$ are given functions. This formula
might be seen as the first three terms of an infinite series, truncated by the omitting of terms
${O}({1}/{\omega^{3/2}})$. Eqns. (\ref{insp-6}), (\ref{7.1}) yield:
\begin{eqnarray}
&& \mathfrak{D} \widehat{a}=\omega\widehat{a}_\tau+\sqrt\omega(\widehat{\vv}\cdot\nabla)\,\widehat{a}+
(\partial_t+\widehat{\vw}\cdot\nabla)\widehat{a}+\frac{1}{\sqrt\omega}(\widehat{\vr}\cdot\nabla)\,\widehat{a}= 0
\label{7.2}
\end{eqnarray}
The small parameter $\varepsilon_{1/2}={1}/{\sqrt\omega}$  allows to rewrite it as
\begin{eqnarray}
&& \mathfrak{D}_2
\widehat{a}\equiv\widehat{a}_{\tau}+\varepsilon(\widehat{\vv}\cdot\nabla)\,\widehat{a}+
\varepsilon^2(\partial_t+\widehat{\vw}\cdot\nabla)\widehat{a}+\varepsilon^3(\widehat{\vr}\cdot\nabla)\,\widehat{a}
=0
\label{7.3}
\end{eqnarray}
where $\mathfrak{D}_2\equiv\varepsilon^2\mathfrak{D}=\mathfrak{D}/\omega$, and the subscript in $\varepsilon_{1/2}$ is
dropped. We are looking for the solution of (\ref{7.3}) in the form of regular series (\ref{basic-4aa}). The
substitution of (\ref{basic-4aa}) into (\ref{7.3}) produces the equations for successive approximations
\begin{eqnarray}
&&\widehat{a}_{0\tau}=0 \label{7.5}\\
&&\widehat{a}_{1\tau}=-(\widehat{\vv}\cdot\nabla)\, \widehat{a}_0\label{7.6}\\
&&\widehat{a}_{2\tau}=-(\widehat{\vv}\cdot\nabla)\widehat{a}_1-(\partial_t+\widehat{\vw}\cdot\nabla)\widehat{a}_0\,
\label{7.6c}\\
&&\widehat{a}_{3\tau}=-(\widehat{\vv}\cdot\nabla)\widehat{a}_2-(\partial_t+\widehat{\vw}\cdot\nabla)\widehat{a}_1-
(\widehat{\vr}\cdot\nabla)\widehat{a}_0
\label{7.6a}
\end{eqnarray}
The solving of (\ref{7.5})-(\ref{7.6a}) follows the same steps as in \emph{Appendix A}. An important new element is:
the bar-part of eqn. (\ref{7.6}) is $\overline{\vv}\cdot\nabla\overline{a}_0=0$. It requires $\overline{\vv}\equiv 0$
as a solvability condition. Therefore the leading term in the prescribed velocity (\ref{7.1}) must always be purely
oscillatory.

Here we present only the truncated solution $\widehat{a}^{[3]}$ (\ref{4.10}). For the tilde-parts $\widetilde{{a}}_k$
with $k=0,1,2,3$ the equations (\ref{7.5})-(\ref{7.6a}) yield the recurrent expressions
\begin{eqnarray}
&&\widetilde{a}_0\equiv 0,\label{7.11a}\\
&& \widetilde{a}_1= -(\widetilde{\vxi}\cdot\nabla)\,\overline{a}_0,\label{7.11}\\
&&\widetilde{a}_{2}=-\{(\widetilde{\vv}\cdot\nabla)\,\widetilde{a}_1\}^\tau -(\widetilde{\vxi}\cdot\nabla)\,
\overline{a}_1 -(\widetilde{\veta}\cdot\nabla)\,\overline{a}_0
\label{7.12}\\
&&\widetilde{a}_{3}=-\{(\widetilde{\vv}\cdot\nabla)\,\widetilde{a}_2\}^\tau
-\{(\widetilde{\vw}\cdot\nabla)\,\widetilde{a}_1\}^\tau-\label{7.13}\\
&&-(\widetilde{\vxi}\cdot\nabla)\,\overline{a}_2-(\widetilde{\veta}\cdot\nabla)\,\overline{a}_1
-(\widetilde{\vzeta}\cdot\nabla)\,\overline{a}_0
-(\partial_t+\overline{\vw}\cdot\nabla)\widetilde{a}_{1}^\tau \nonumber\\
&&\widetilde{\vxi}\equiv\widetilde{\vv}^\tau,\quad \widetilde{\veta}\equiv\widetilde{\vw}^\tau,\quad
\widetilde{\vzeta}\equiv\widetilde{\vr}^\tau,\label{7.15}
\end{eqnarray}
and for the bar-parts $\overline{a}_k$ we obtain the transport equation
\begin{eqnarray}
&&\left(\partial_t+ (\overline{\vw}+\overline{\vV}_0)\cdot\nabla\right)\overline{a}_{0}=0\label{7.16}\\
&&\left(\partial_t+ (\overline{\vw}+\overline{\vV}_0)\cdot\nabla\right)
\overline{a}_{1}+((\overline{\vr}+\overline{\vV}_1+\overline{\vV}_{12})\cdot\nabla)\overline{a}_0=0\label{7.17}\\
&&\overline{\vV}_0\equiv \frac{1}{2}\langle[\widetilde{\vv},\widetilde{\vxi}]\rangle,\quad
\overline{\vV}_1\equiv\frac{1}{3}\langle[[\widetilde{\vv},\widetilde{\vxi}],\widetilde{\vxi}]\rangle,\quad
\overline{\vV}_{12}\equiv \langle[\widetilde{\vv},\widetilde{\veta}]\rangle,
\label{7.19}
\end{eqnarray}
which can be rewritten (with the error ${O}(\varepsilon^2)$)  as a single advection equation
\begin{eqnarray}
&&\left(\partial_t+ \overline{\vV}\cdot\nabla\right)\overline{a} =0
\label{7.23}\\
&&\overline{\vV}=(\overline{\vw}+\overline{\vV}_0)+
\varepsilon(\overline{\vr}+\overline{\vV}_1+\overline{\vV}_{12}),\quad
\overline{a}=\overline{a}_0+\varepsilon\overline{a}_1\label{7.24}
\end{eqnarray}
Notice that the solution given by (\ref{7.11a})-(\ref{7.24}) satisfies eqn. (\ref{7.3}) with
$\mathrm{Res}{[3]}={O}(\omega^{-2})$.

\underline{Remark:} In (\ref{7.5})-(\ref{7.24}) we present only four steps $(k=0,1,2,3)$ of successive approximations since that is
enough to give us an instructive correction $\overline{\vV}_{12}$ to the previously calculated drift (\ref{4.16}),
(\ref{4.18}). The fifth step $(k=4)$ has been also carried out but for brevity it is not shown here: it produces a
combination of a drift term $\overline{\vV}_2$ (which is more cumbersome than (\ref{4.19})) and pseudo-diffusion
\emph{with the same coefficients} (\ref{4.20}). The related calculations are similar to those of
\emph{Appendix A}.

\subsection{Super-critical asymptotic procedure with  $\alpha={1/3}$}

Next, we consider a super-critical asymptotic family with a velocity similar to (\ref{7.1})
\begin{eqnarray}
&&\widehat{\vu}=\widehat{\vv}+\omega^{-1/3}\widehat{\vw}+\omega^{-2/3}\,\widehat{\vr}, \quad
\delta=\omega^{-1/3}
\label{7.25}
\end{eqnarray}
where $\widehat{\vv}, \widehat{\vw}, \widehat{\vr}\in\mathbb{H}\cap\mathbb{O}(1)$ are given functions. Expression
(\ref{7.25}) might be seen as the first three terms in an infinite series, truncated by omitting the terms
${O}({1}/\omega)$. Eqns. (\ref{insp-6}), (\ref{7.25}) yield:
\begin{eqnarray}
&& \mathfrak{D}
\widehat{a}=\omega\widehat{a}_\tau+\omega^{2/3}(\widehat{\vv}\cdot\nabla)\,\widehat{a}
+\omega^{1/3}(\widehat{\vw}\cdot\nabla)\,\widehat{a}+ (\partial_t+\widehat{\vr}\cdot\nabla)\widehat{a}= 0
\label{7.26}
\end{eqnarray}
The small parameter $\varepsilon_{1/3}= \omega^{-1/3}$ (\ref{epsilon}) allows to rewrite (\ref{7.26}) as
\begin{eqnarray}
&& \mathfrak{D}_3
\widehat{a}\equiv\widehat{a}_{\tau}+\varepsilon(\widehat{\vv}\cdot\nabla)\,\widehat{a}+
\varepsilon^2(\widehat{\vw}\cdot\nabla)\,\widehat{a}+
\varepsilon^3(\partial_t+\widehat{\vr}\cdot\nabla)\widehat{a}
=0 \label{7.27}
\end{eqnarray}
where $\mathfrak{D}_3\equiv\varepsilon^3\mathfrak{D}=\mathfrak{D}/\omega$ and the subscript in $\varepsilon_{1/3}$ is
dropped. We are looking for the solution of (\ref{7.27}) in the form of regular series (\ref{basic-4aa}). The
substitution of (\ref{basic-4aa}) into (\ref{7.27}) produces the equations for successive approximations
\begin{eqnarray}
&&\widehat{a}_{0\tau}=0 \label{7.28}\\
&&\widehat{a}_{1\tau}=-(\widehat{\vv}\cdot\nabla)\, \widehat{a}_0\label{7.29}\\
&&\widehat{a}_{2\tau}=-(\widehat{\vv}\cdot\nabla)\widehat{a}_1-(\widehat{\vw}\cdot\nabla)\widehat{a}_0\,
\label{7.30}\\
&&\widehat{a}_{3\tau}=-(\widehat{\vv}\cdot\nabla)\widehat{a}_2-
(\widehat{\vw}\cdot\nabla)\widehat{a}_1-(\partial_t+\widehat{\vr}\cdot\nabla)\widehat{a}_0
\label{7.31}
\end{eqnarray}
The solving of (\ref{7.28})-(\ref{7.31}) follows the same steps as in \emph{Appendix B}. There are two solvability
conditions
\begin{equation}
    \overline{\vv}\equiv 0,\quad \overline{\vV}_0=-\overline{\vw}\label{7.31a}
\end{equation}
which are required for the existence of the bar-part solutions in (\ref{7.29}) and (\ref{7.30}). One might see the
consideration of \emph{Sect.5} as being physically incomplete. Indeed, one can expect that the results for small
$\overline{\vV}_0\neq 0$ are close to that for $\overline{\vV}_0=0$. This expectation has been met and fulfilled by a
more general degeneration condition $\overline{\vV}_0=-\overline{\vw}$ (\ref{7.31a}) which leads to a more general
(than (\ref{5.15})) expression for a drift.

For tilde-parts $\widetilde{{a}}_k$ one can obtain the explicit recurrent expressions which coincide with
(\ref{7.11})-(\ref{7.13}), while the formula (\ref{7.15}) for $\widetilde{a}_{3}$ should be replaced by
\begin{eqnarray}
&&\widetilde{a}_{3}=-\{(\widetilde{\vv}\cdot\nabla)\,\widetilde{a}_2\}^\tau
-\{(\widetilde{\vw}\cdot\nabla)\,\widetilde{a}_1\}^\tau-\label{7.35}\\
&&-(\widetilde{\vxi}\cdot\nabla)\,\overline{a}_2-(\widetilde{\veta}\cdot\nabla)\,\overline{a}_1
-(\widetilde{\vzeta}\cdot\nabla)\,\overline{a}_0 -(\overline{\vw}\cdot\nabla)\widetilde{a}_{1}^\tau \nonumber
\end{eqnarray}
For the bar-parts one can derive the equation
\begin{eqnarray}
&&(\partial_t+ \overline{\vV}\cdot\nabla)\overline{a}_{0}= 0,
\quad \overline{\vV}=\overline{\vr}+\overline{\vV}_1+\overline{\vV}_{12}
\label{7.37}
\end{eqnarray}
which is valid with the error ${O}(\varepsilon^2)$ and contains the same $\overline{\vV}_1$ and $\widehat{\vV}_{12}$ as
in (\ref{7.19}). Notice that this solution satisfies the equation (\ref{7.27}) with
$\mathrm{Res}{[3]}={O}(\omega^{-4/3})$.

\underline{Remarks:}

1. One might additionally consider a polynomial version of velocity for $\omega^{1/4}$-family (\ref{5.20}). The related
results could be predicted by the analogy between the material of \emph{Sect.6.2} and \emph{Sect.5.}

2. For the critical family (\ref{7.1}) slow motion represents the drift (\ref{7.24}) of magnitude ${O}(1)$ combined
with the same  pseudo-diffusion of order ${O}(1/\omega)$ as in (\ref{4.22aa}), hence the results are very similar to
those of \emph{Sect.4}.

3. All the results of \emph{Sects.2-6} can be generalized to an arbitrary flow domain $\Omega$ with fixed boundary
$\partial\Omega$. Then (\ref{oper-11b}) yields that
\begin{equation}
\overline{\vV}_0\cdot\vn=\overline{\vV}_1\cdot\vn=\overline{\vV}_2\cdot\vn
=\widetilde{\vu}'\cdot\vn=0\quad \text{on}\quad\partial\Omega\label{4-appr-8f}
\end{equation}
provided $\widehat{\vu}$ satisfies the no-leak condition. One can also see that if a fluid is incompressible then
(\ref{oper-11a}) yields that all drift velocities in (\ref{4.15})-(\ref{4.20b}) are also incompressible:
\begin{equation}
\Div\widetilde{\vu}=0\quad
\Rightarrow\quad\Div\overline{\vV}_0=\Div\overline{\vV}_1=\Div\overline{\vV}_2=0\label{4-appr-8s}
\end{equation}
The explicit expressions (\ref{4.15})-(\ref{4.20a}) show that due to incompressibility only the last term in
$\overline{\vV}_2$ (\ref{4.19}) vanishes. Similar (for (\ref{7.16})-(\ref{7.19})) the incompressibility yields
\begin{equation}
\Div\widetilde{\vv}=\Div\widehat{\vw}=\Div\widehat{\vr}=0\quad
\Rightarrow\quad\Div\overline{\vV}_0=\Div\overline{\vV}_1=\Div\overline{\vV}_{12}=0\label{incompr}
\end{equation}

4.  For the super-critical family (\ref{7.25}) with the degeneration (\ref{7.31a}) slow motion is a pure drift with the
velocity (\ref{7.37}) which is `upgraded' to ${O}(1)$ from ${O}(\varepsilon)$-term in (\ref{7.24}). The qualitatively
new addition to a drift (in comparison with a critical family)  is a cross-term $\overline{\vV}_{12}\equiv
\langle[\widetilde{\vv},\widetilde{\veta}]\rangle$. It represents an averaged commutator between two mutually
independent functions $\widetilde{\vv}$ and $\widetilde{\veta}$; hence $\overline{\vV}_{12}$ expands the class of
available analytical expressions for a drift. For example, for an incompressible fluid
$\langle[\widetilde{\vv},\widetilde{\veta}]\rangle=\nabla\times\langle\widetilde{\vv}\times\widetilde{\veta}\rangle$,
so one can see that $\langle\widetilde{\vv}\times\widetilde{\veta}\rangle$ represents an arbitrary vector-potential for
$\overline{\vV}_{12}$.

\section{Examples. \label{sect08}}

\subsection{Superposition of two modulated oscillatory fields of the same frequency}

The velocity field $\widetilde{\vu}$ (\ref{basic-1})  and $\widetilde{\vxi}\equiv\widetilde{\vu}^\tau$ (\ref{oper-7})
are
\begin{eqnarray}
&&\widetilde{\vu}(\vx,t,\tau)=\overline{\vp}(\vx,t)\sin\tau+
\overline{\vq}(\vx,t)\cos\tau  \label{Example-1-1}\\
&&\widetilde{\vxi}(\vx,t,\tau)=-\overline{\vp}(\vx,t)\cos\tau+
\overline{\vq}(\vx,t)\sin\tau\label{Example-1-2}
\end{eqnarray}
with arbitrary $\mathbb{B}$-functions  $\overline{\vp}$ and $\overline{\vq}$. The straightforward calculations yield
\begin{eqnarray}
&[\widetilde{\vu},\widetilde{\vxi}]=[\overline{\vp},\overline{\vq}]\label{Example-1-3}
\end{eqnarray}
so the commutator surprisingly is not oscillating. The drift velocities (\ref{4.18}), (\ref{4.19}) are
\begin{eqnarray}
&&\overline{\vV}_0=\frac{1}{2}\langle[\widetilde{\vu},\widetilde{\vxi}]\rangle=
\frac{1}{2}[\overline{\vp},\overline{\vq}],\quad
\overline{\vV}_1=\frac{1}{3}\langle[[\widetilde{\vu},\widetilde{\vxi}],\widetilde{\vxi}]\rangle\equiv 0,
\label{Example-1-4}\\
&&\overline{\vV}_2=\frac{1}{8}\left([\overline{\vP},\overline{\vp}]+[\overline{\vQ},\overline{\vq}]\right)
-\frac{1}{4}\left([\overline{\vp}_t,\overline{\vp}]+[\overline{\vq}_t,\overline{\vq}]\right)+\label{Example-1-4a}\\
&&+\frac{1}{4}\left(\overline{\vp}\,\Div\overline{\vP}'+\overline{\vq}\Div\overline{\vQ}'
+\overline{\vP}'\Div\overline{\vp}+\overline{\vQ}'\Div\overline{\vq}\right),\nonumber\\
&&\overline{\vP}\equiv[\overline{\vV}_0,\overline{\vp}],\quad
\overline{\vQ}\equiv[\overline{\vV}_0,\overline{\vq}],\quad
\overline{\vP}'\equiv\overline{\vp}_t-\overline{\vP},\quad
\overline{\vQ}'\equiv\overline{\vq}_t-\overline{\vQ},\nonumber\\
&&\langle\widetilde{\xi}_i\widetilde{\xi}_k\rangle=
\frac{1}{2}(\overline{p}_i \overline{p}_k + \overline{q}_i \overline{q}_k)\label{Example-1-5}
\end{eqnarray}
A pseudo-diffusion matrix $\overline{\kappa}_{ik}$, which follows after the substitution of (\ref{Example-1-5}) into
(\ref{4.20}).

\underline{Remarks:}

1. The degeneration $\overline{\vV}_0\equiv 0$ in an incompressible fluid  (\ref{4-appr-8s}),(\ref{incompr})
corresponds to $\overline{\vp}\times\overline{\vq}=\nabla\overline{\varphi}$ with an arbitrary potential
$\overline{\varphi}(\vx,t)$. For any such fields all results of \emph{Sect.5.1} are valid.

2. The velocity $\widetilde{\vu}$ (\ref{Example-1-1}) is general enough to produce any given function
$\overline{\vV}_0(\vx,t)$. To obtain $\overline{\vp}(\vx,t)$ and $\overline{\vq}(\vx,t)$ one has to solve a bilinear
first-order PDE
\begin{eqnarray}\label{bi-linear}
[\overline{\vp},\overline{\vq}]=(\overline{\vq}\cdot\nabla)\overline{\vp}-
(\overline{\vp}\cdot\nabla)\overline{\vq}=2\overline{\vV}_0(\vx,t)
\end{eqnarray}
which represents an underdetermined bi-linear PDE-problem for two unknown functions.

\subsection{Fourier series of modulated fields}
The previous example can be generalized by the consideration of a velocity
\begin{eqnarray}
&&\widetilde{\vu}(\vx,t,\tau)=\sum_{k=1}^{\infty}\overline{\vp}_k(\vx,t)\sin k\tau+
\overline{\vq}_k(\vx,t)\cos k\tau  \label{Example-2-1}\\
&&\widetilde{\vxi}=\sum_{k=1}^{\infty}-\frac{\overline{\vp}_k}{k}\cos k\tau+
\frac{\overline{\vq}_k}{k}\sin k\tau
\label{Example-2-2}
\end{eqnarray}
The calculations of $\overline{\vV}_0$ lead to an infinite sum of commutators
\begin{eqnarray}
&&\overline{\vV}_0=\frac{1}{2}\langle[\widetilde{\vu},\widetilde{\vxi}]\rangle=\sum_{k=1}^{\infty}
\frac{1}{2k}[\overline{\vp}_k,\overline{\vq}_k]\label{Example-2-3}
\end{eqnarray}
In particular, (\ref{Example-2-3}) gives an infinite number of fields $\widetilde{\vu}$ with $\overline{\vV}_0\equiv
0$. For example, if $\overline{\vp}_k\equiv 0,\ \forall k$ then
\begin{eqnarray}\label{Example-2-4}
\overline{\vV}_1=\frac{1}{3}\langle[[\widetilde{\vu},\widetilde{\vxi}],\widetilde{\vxi}]\rangle=
\frac{1}{12}\sum_{m\pm n\pm l=0}\frac{1}{ml}[[\overline{\vq}_n,\overline{\vq}_m],\overline{\vq}_l]
\end{eqnarray}
where the sum is taken over all sets of three positive integers $m,n,l$ such that $m\pm n\pm l=0$. It is clear that for
many combinations $\overline{\vV}_0\equiv 0$ and $\overline{\vV}_1\ne 0$ (see \emph{Sect.5.1}).

\subsection{The Stokes drift}

This most celebrated example of a drift we consider in some details. We start with the scaling (\ref{scales}),
(\ref{exact-6aa}) and the limit (\ref{exact-7}). The dimensional solution for a plane potential travelling gravity wave
(see
\cite{Stokes, Lamb, Debnath}) is
\begin{equation}\label{Stokes}
   \widehat{\vu}^*=U\widetilde{\vu},\quad U=\frac{k^*g^*h^*}{\omega^*},
   \quad\widetilde{\vu}=\exp(k^*y^*)\left(\begin{array} {c} \cos(k^*x^*-\tau)\\ \sin (k^*x^*-\tau)\end{array}\right)
\end{equation}
where $\omega^*$, $k^*$, $h^*$, and $g^*$ are dimensional frequency, wavenumber, spatial amplitude,  and gravity;
$(x^*,y^*)$ are Cartesian coordinates, and $\tau\equiv\omega^* t^*$. One can immediately notice that: (i) the
characteristic length is $L=1/k^*$; and (ii) the scale $U$ is apparent. Hence the dimensionless scaling parameters
(\ref{exact-6aa}), (\ref{exact-6a}) and the asymptotic limot  (\ref{exact-7}) appear as $k=1$ and
\begin{equation}\label{limits2}
\varepsilon_1=1/T\omega^*=1/\omega\to 0,\ \delta\equiv U/\omega^* L=g^*h^*/L^2\omega^{*2}=gh/\omega^2\to 0\
 \text{as}\ \omega\to \infty
\end{equation}
where one can choose the scale $T=T(\omega^*)$ in an arbitrary way, its only mission is to provide $\varepsilon_1\to 0$
as $\omega\to\infty$ (see \emph{Sect.2.3}). The simplest way to provide $\delta=1/\sqrt\omega$ (\ref{basic-1}) is to
choose an asymptotic family as
\begin{equation}
T=\const,\quad L=\const,\quad gh={O}(\omega^{3/2})\label{limits3}
\end{equation}

The dimensionless velocity field (\ref{Stokes}) and $\widetilde{\vxi}$ are
\begin{eqnarray}
\widetilde{\vu}=Ae^{ky}\left(\begin{array} {c} \cos(kx-\tau)\\ \sin (kx-\tau)\end{array}\right),\quad
\widetilde{\vxi}=Ae^{ky}\left(\begin{array}{c} -\sin(kx-\tau)\\ ~\cos (kx-\tau)\end{array}\right)
\label{Example-3-1}
\end{eqnarray}
where in the chosen system of units $A=1$ and $k=1$; however, we keep both $A$ and $k$ in the formulae for tracking its
physical meaning. The fields $\overline{\vp}(x,y)$, $\overline{\vq}(x,y)$ (\ref{Example-1-1}) are
\begin{eqnarray}
\overline{\vp}=Ae^{ky}\left(\begin{array}{c} \sin kx\\ -\cos kx\end{array}\right),\quad
\overline{\vq}=Ae^{ky}\left(\begin{array}{c} \cos kx\\ \sin kx\end{array}\right)
\label{Example-3-2}
\end{eqnarray}
The calculations (with the use of (\ref{Example-1-3}), (\ref{Example-1-4})) yield
\begin{eqnarray}
\overline{{\vV}}_0=k A^2 e^{2ky}\left(\begin{array}{c} 1\\ 0\end{array}\right), \quad  \overline{{\vV}}_1\equiv 0
\label{Example-3-3}
\end{eqnarray}
which gives $\overline{{\vV}}_0$ proportional to the classical Stokes drift and a zero value for the first correction
to it;  the explicit formula for $\overline{{\vV}}_2$  is not given here for brevity. At the same time
\begin{eqnarray}
\langle\widetilde{\xi}_i\widetilde{\xi}_k\rangle=\Xi(x,y,t)\delta_{ik}, \quad\text{with}\quad
\Xi=\frac{1}{2}A^2 e^{2ky}
\label{Example-3-4}
\end{eqnarray}
which can bee seen as `locally isotropic' oscillations. The matrix of pseudo-diffusion (\ref{4.20}) is
\begin{eqnarray}
&&2\overline{\chi}_{ik}=
\left\{\left(\partial_t+
\overline{\vV}_0\cdot\nabla\right)\delta_{ik}-2\overline{\chi}\, \overline{e}_{ik}\right\}\Xi,\quad
2\overline{e}_{ik}\equiv\frac{\partial\overline{V}_{0i}}{\partial x_k}+
\frac{\partial\overline{V}_{0k}}{\partial x_i}\nonumber
\nonumber
\end{eqnarray}
Further calculations show that
\begin{eqnarray}\nonumber
&&\overline{\chi}_{ik}=-\overline{\chi}
\begin{pmatrix}
0 & 1\\
1 & 0
\end{pmatrix},\quad\text{with}\quad \overline{\chi}\equiv\frac{1}{4}k^2 A^4 e^{3ky}
\end{eqnarray}
One can see that the eigenvalues $\overline{\chi}_1=-\overline{\chi}$ and $\overline{\chi}_2=\overline{\chi}$
correspond to ordinary diffusion in one direction and anti-diffusion in the perpendicular direction, so here we have a
mixed case of pseudo-diffusion. The averaged equation (\ref{4.21}) (with an error $O(\varepsilon^3)$) can be written as
\begin{eqnarray}
&&\overline{a}_{t}+(\overline{V}_0+\varepsilon^2 \overline{V}_2) \overline{a}_x=
    \varepsilon^2(\overline{\chi}_y \overline{a}_{x}+\overline{\chi}\, \overline{a}_{xy})\label{Example-3-5}\\
    &&\overline{a}=\overline{a}_0+\varepsilon\overline{a}_1+\varepsilon^2\overline{a}_2\nonumber
\end{eqnarray}
where $\overline{V}_0$ and $\overline{V}_2$ are the $x$-components of corresponding velocities (their $y$-components
vanish). This equation has an exact solution $\overline{a}=\overline{a}(y)$ (with an arbitrary function
$\overline{a}(y)$), which is not blurred by pseudo-diffusion.

\underline{Remarks:}

1. In order to avoid confusion one should recall that the choice of asymptotic family (\ref{limits3}) has nothing to do
with physical relations between parameters. Any asymptotic family represents only a formal one-dimensional
parametrization in the functional space of all possible functions $\widehat{\vu}^*$ (\ref{Stokes}). The only aim of
such a parametrization is to build a valid asymptotic procedure. Therefore any particular wave (\ref{Stokes}) with
given values $k^*=k^*_0$ and $\omega^*=\omega^*_0$  can be considered as a point on an asymptotic curve with $k=1$ and
with variable $\omega$ irrespectively to the presence of any dispersion relation between $k^*_0$ and $\omega^*_0$.

2. A parameter $gh\to\infty$ as $\omega\to\infty$  (\ref{limits3}) which is a common property for many asymptotic
procedures.

3. More general (than (\ref{limits3})) asymptotic families are considered in \emph{Sect.2.3} and in
(\ref{exact-1R}),(\ref{exact-1RR}).

4. Both fields (\ref{Example-3-1}) are unbounded as $y\to\infty$, but it is not essential for our purposes.

\subsection{Spherical `acoustic' wave}

A velocity potential for an outgoing spherical wave is
\begin{eqnarray}
&& \widetilde{\phi}=\frac{A}{r}\sin(kr-\tau)\label{Example-4-1}
\end{eqnarray}
where $A$, $k$, and $r$ are an amplitude, a wavenumber, and a radius in a spherical coordinate system. The velocity is
purely radial has a form (\ref{Example-1-1})
\begin{eqnarray}
&& \widetilde{u}=\overline{p}\sin\tau+\overline{q}\cos\tau,\\
&&\overline{p}=A\left(\frac{1}{r^2}\cos kr+\frac{k}{r}\sin kr\right),\quad\overline{q}=A\left(-\frac{1}{r^2}\sin
kr+\frac{k}{r}\cos kr\right)\label{Example-4-1a}
\end{eqnarray}
where $\widetilde{u}, \overline{p}$, and $\overline{q}$ are radial components of corresponding vector-fields. The
fields $\widetilde{\vxi}$ and $[\widetilde{\vu},\widetilde{\vxi}]$ are also purely radial; the radial component for the
commutator is
\begin{eqnarray}
&& \widetilde{\xi} \widetilde{u}_r-\widetilde{u}\widetilde{\xi}_r=A^2 k^3/r^2\label{Example-4-3}
\end{eqnarray}
where $\xi$ is radial component of $\widetilde{\vxi}$ and subscript $r$ stands for radial derivative. The drift
(\ref{Example-1-3}),(\ref{Example-1-4}) is purely radial with
\begin{eqnarray}
&&\overline{V}_0=\frac{A^2 k^3}{2\,r^2},\quad
\overline{V}_1=0,\quad\overline{V}_2=\frac{A^4k^5}{16r^4}\left(3k^2-\frac{5}{r^2}\right)
\label{Example-4-5}
\end{eqnarray}
It is interesting that $\overline{V}_0$ formally coincides with the velocity caused by a point source in an
incompressible fluid and for small $r$ the convergence of $\overline{\vV}$ is endangered, since $\overline{V}_2$
dominates over $\overline{V}_0$. Further calculations yield
\begin{eqnarray}
&&\langle\xi^2\rangle=\frac{A^2}{2r^2}(k^2+1/r^2),\quad \overline{\chi}=A^4k^5/4r^2>0
\label{Example-4-6}
\end{eqnarray}
where $\overline{\chi}$ stands for the only nonzero $rr$-component of $\overline{\chi}_{ik}$. One can see that in this
case pseudo-diffusion appears as ordinary diffusion.

\subsection{The $\overline{\vV}_1$-drift.}

If $\overline{\vV}_0\equiv 0$ then the drift of order ${O}(1)$ is given by (\ref{5.15}). Let the velocity field
(\ref{5.1}) be a superposition of two standing waves of frequencies $\omega$ and $2\omega$:
\begin{eqnarray}
&&\widetilde{\vu}(\vx,t,\tau)=\overline{\vp}(\vx,t)\sin\tau+\overline{\vq}(\vx,t)\sin 2\tau  \label{Example-5-1}\\
&&\widetilde{\vxi}(\vx,t,\tau)=-\overline{\vp}(\vx,t)\cos\tau-\frac{1}{2}\overline{\vq}(\vx,t)\cos
2\tau\label{Example-5-2}\\
&&[\widetilde{\vu},\widetilde{\vxi}]=\frac{1}{2}[\overline{\vp},\overline{\vq}](2\cos\tau\sin 2\tau-\cos
2\tau\sin\tau)\label{Example-5-3}
\end{eqnarray}
Hence (\ref{4.18}) yields
\begin{eqnarray}\label{Example-5-4}
\overline{\vV}_0=\frac{1}{2}\langle[\widetilde{\vu},\widetilde{\vxi}]\rangle\equiv 0,\quad
\overline{\vV}_1=\frac{1}{3}\langle[[\widetilde{\vu},\widetilde{\vxi}],\widetilde{\vxi}]\rangle=
\frac{1}{8}[[\overline{\vp},\overline{\vq}],\overline{\vp}]
\end{eqnarray}
These expressions produce infinitely many examples of flows with ${O}(1)$-drift in a super-critical asymptotic family
(\ref{5.15}).

\subsection{A plane travelling  wave of a general shape}

The velocity field (\ref{Example-1-1}) for a plane wave in three-dimensional space is
\begin{eqnarray}
\widetilde{\vu}=\vA \widetilde{f}'(\vk\cdot\vx-\tau)\label{Example-6-1}
\end{eqnarray}
where $\vA$ and $\vk$ are two constant vectors, $\widetilde{f}$ is a $\mathbb{T}$-function of a scalar variable, its
$\tau$-periodicity automatically leads to periodicity w.r.t. $\vk\cdot\vx$, primes stand for ordinary derivatives.
Calculations yield
\begin{eqnarray}
\overline{{\vV}}_0=\vA(\vA\cdot\vk)\langle \widetilde{f}'^2\rangle,\quad
\overline{{\vV}}_1=\vA(\vA\cdot\vk)^2\langle \widetilde{f}'^3\rangle\label{Example-6-2}
\end{eqnarray}
It shows that the only case of a zero drift corresponds to a transversal wave $\vA\bot\vk$ (an incompressible fluid)
and the maximal drift takes place for a longitudinal wave $\vA\|\vk$. A pseudo-diffusion matrix is
\begin{eqnarray}
\overline{\chi}_{ik}\equiv 0\quad\text{for}\quad
\langle\widetilde{\xi}_i \widetilde{\xi}_k\rangle=A_i A_k \langle \widetilde{f}^2\rangle,
\label{Example-6-3}
\end{eqnarray}
Next step is to consider the superposition of two travelling plane  waves (\ref{Example-6-1}):
\begin{eqnarray}
&&\widetilde{\vu}=\vA \widetilde{f}'(\vk\cdot\vx-\tau)+\vB \widetilde{g}'(\vl\cdot\vx-\tau)
=-\vA \widetilde{f}_\tau(\vk\cdot\vx-\tau)-\vB \widetilde{g}_\tau(\vl\cdot\vx-\tau)\nonumber\\
&&\widetilde{\vxi}=-\vA\widetilde{f}_\tau(\vk\cdot\vx-\tau)-\vB\widetilde{g}_\tau(\vl\cdot\vx-\tau)\label{Example-7-1}
\end{eqnarray}
with constant vectors $\vA$, $\vB$, $\vk$, $\vl$. Calculations yield
\begin{eqnarray}
\overline{{\vV}}_0=(\vA\cdot\vk)\vA\langle \widetilde{f}'^2\rangle+
(\vB\cdot\vk)\vB\langle \widetilde{g}'^2\rangle+\left((\vA\cdot\vl)\vB+(\vB\cdot\vk)\vA\right)\langle
\widetilde{f}'\widetilde{g}'\rangle\label{Example-7-2}
\end{eqnarray}
which exhibits the interference (third) term. If $f'=\sin(\vk\cdot\vx-\tau)$ and $g'=\sin(\vl\cdot\vx-\tau)$ then
(\ref{Example-7-2}) gives
\begin{eqnarray}
2\overline{{\vV}}_0=(\vA\cdot\vk)\vA+
(\vB\cdot\vk)\vB+\left((\vA\cdot\vl)\vB+(\vB\cdot\vk)\vA\right)\cos(\vk-\vl)\vx\label{Example-7-3}
\end{eqnarray}
while $\overline{\vV}_1$ is too cumbersome to be presented here. Also
\begin{eqnarray}
2\langle\xi_i \xi_k\rangle=A_i A_k +B_i B_k +(A_i B_k+A_k B_i)\cos(\vk-\vl)\vx\nonumber
\end{eqnarray}
For two mutually perpendicular longitudinal waves $(\vA\|\vk)\bot(\vB\|\vl)$ one can obtain
\begin{eqnarray}
\overline{\chi}_{ik}=-\frac{1}{2}(A_i B_k+A_k B_i)\left[(\vA\cdot\vk)^2-(\vB\cdot\vl)^2\right]\sin(\vk-\vl)\vx
\end{eqnarray}
which for the plane flow $\vA=(A,0), \vB=(0,B)$ gives
\begin{eqnarray}
\overline{\chi}_{ik}=-\frac{1}{8}AB
\left[(\vA\cdot\vk)^2-(\vB\cdot\vl)^2\right]
\begin{pmatrix}
0 & 1\\
1 & 0
\end{pmatrix}
\sin(\vk-\vl)\vx
\end{eqnarray}
Here we have the eigenvalues $\overline{\chi}_1=-\overline{\chi}_2$ of opposite signs, they are also oscillating in
space. Once again (as in \emph{Sect.7.3}), we deal with the sign-indefinite case of pseudo-diffusion. This example can
be linked to plane sound waves; the drift (\ref{Example-6-2}) can give an addition to an acoustic streaming (see
\emph{e.g.} \cite{Lighthill1}).

\subsection{The drift for the polynomial velocity (\ref{7.25})}

The expression (\ref{7.37}) for the ${O}(1)$-drift within a super-critical family contains the term given in
(\ref{7.19})
\begin{eqnarray}
\overline{\vV}_{12}\equiv \langle[\widetilde{\vv},\widetilde{\veta}]\rangle
\label{7.19a}
\end{eqnarray}
which is built by two mutually independent oscillating functions. Such a functional freedom allows to construct broad
functional classes of drifts. However, in this case one should keep in mind that the degeneration condition
$\overline{\vV}_0=-\overline{\vw}$ (\ref{7.31a}) imposes a restriction on the coefficients of (\ref{7.25}). For
example, one can chose $\widetilde{\vv}$ such that $\overline{\vV}_0=-\overline{\vw}=0$. In this case the function
$\widetilde{\vv}$ in (\ref{7.19a}) is not arbitrary. Alternatively, one can consider $\overline{\vV}_0=-\overline{\vw}$
as the definition of $\overline{\vw}$. In this case the functions $\widetilde{\vv}$ and $\widetilde{\veta}$ in
(\ref{7.19a}) are indeed mutually independent.

\subsection{The Bjorknes configuration of two pulsating point sources}

This example is aimed to clarify the global structure of drift motion in an oscillating flow related to \cite{Cooke,
Hicks1, Hicks2} and to show that interesting oscillating flow, which are different from waves, do exist. An
incompressible velocity from the class of flows (\ref{Example-1-1}) in Cartesian coordinates $\vx$ is given by
\begin{eqnarray}
\overline{\vq}(\vx)=\nabla\frac{1}{|\vx|},\quad \overline{\vp}(\vx)=\nabla\frac{1}{|\vy|},\quad
\vy=\vx-\vl,\quad\vl=\const
\end{eqnarray}
which represents a superposition of two oscillating point sources. Calculations yield a rotationally symmetric w.r.t.
the $\vl$-axis field:
\begin{eqnarray}
\overline{\vV}_0=\frac{2(\vx\cdot\vy)}{(|\vx|^2|\vy|^2)^2}(|\vy|^2\vx-|\vx|^2\vy), \quad
\overline{\vV}_0\cdot(\vx\times\vy)=0
\end{eqnarray}
Let us introduce cylindrical coordinates $(\rho,\phi,z)$ with an origin at $\vx=\vl/2$. The related ODEs (\ref{Cauchy})
are
\begin{eqnarray}
&&\dot\rho=M\rho z,\quad \dot\phi=0,\quad \dot z=-\frac{M}{2}\left(\frac{l^2}{4}-z^2+\rho^2\right)\label{Bjorknes}\\
&&l\equiv|\vl|,\quad M\equiv-36\frac{\rho^2+z^2-l^2/4}{|\vx|^2|\vy|^2}\nonumber\\
&&|\vx|^2=\rho^2+(z-l/2)^2,\quad |\vy|^2=\rho^2+(z+l/2)^2\nonumber
\end{eqnarray}
The qualitative dynamics of particles described by this system is: (i) two singularities at the points $\vx=\pm\vl/2$;
(ii) any point of the sphere $\rho^2+z^2=l^2/4$ represents an equilibrium; and (iii) the directions of particle motions
inside and outside of this sphere are topologically opposite to each other.

\subsection{The complexity of particle dynamics for $\overline{\vV}_0$}

Now we illustrate the complexity of particle dynamics (\ref{Cauchy}) for the zero-order drift velocity
$\overline{\vV}_0$ (\ref{4.18}) (see \cite{Aref, Ottino, Wiggins}). Let an incompressible velocity (\ref{Example-1-1})
be
\begin{eqnarray}
\overline{\vp}=\left(\begin{array}{c}
  \cos y \\
  0 \\
  \sin y
\end{array}\right), \qquad
\overline{\vq}=\left(\begin{array}{c}
  a\sin z \\
  b\sin x+a\cos z \\
  b\cos x
\end{array}\right)
\end{eqnarray}
where  $(x,y,z)$ are  Cartesian coordinates, $a, b$ are constants. Either of this fields, taken separately, produces
simple integrable dynamics of particles. Calculations yield
\begin{eqnarray}
\overline{\vV}_0=\left(\begin{array}{c}
  -a\sin y\sin x-2b\sin y\cos z \\
  b\sin z\sin y-a\cos x\cos y \\
  b\cos z\cos y+2a\sin x\cos y
\end{array}\right),
\end{eqnarray}
Straightforward computations for this steady flow exhibit chaotic dynamics of particles. In particular, positive
Lyapunov exponents have been observed. This example shows, that the drift created by simple oscillatory field can
produce complex dynamics. Since the averaged dynamics is chaotic then related results by
\cite{Arnold,Aref,Ottino,Wiggins,Chierchia} can be applied to it. This example brings up new questions: (i) what is
the relationship between chaotic motions for the original dynamical system and the averaged one? (ii) can the
oscillatory part of a solution also cause  chaotic dynamics? (iii) can a chaotic drift and the pseudo-diffusion of
\emph{Sect.4} complement each other? (iv) how a chaotic drift can be used in the theory of mixing? This example has
been constructed and computed by A.B.Morgulis (private communications) for the use in this paper.

\subsection{The appearance of pseudo-diffusion (PD) due to slow-time-modulations}

Let us consider a simple example that clarifies the meaning of pseudo-diffusion.

An infinite fluid oscillates as a rigid body with the small displacement of any material particle
$\widetilde{\vx}=\widetilde{\vx}(t,\tau)$. The Eulerian coordinate of a particle is
\begin{equation}\label{osil-rigid}
\vx=\overline{\vx}+\widetilde{\vx}, \quad\langle\widetilde{\vx}\rangle\equiv 0,
\quad\widetilde{\vx}=\varepsilon\widetilde{\vx}_1, \ \widetilde{\vx}_1\in \mathbb{O}(1)
\end{equation}
where $\langle\cdot\rangle$ is the $\tau$-average (\ref{oper-1}); the small parameter $\varepsilon$ and the function
$\widetilde{\vx}_1(t,\tau)$ can be either chosen by us or taken from (\ref{ODE-crit1})-(\ref{ODE-Appr-1}). A drift for
a rigid-body oscillations is absent $\overline{\vV}\equiv 0$, hence the mean coordinate $\overline{\vx}$ of a particle
is not changing with time.  For simplicity we also accept that the mean coordinate $\overline{\vx}$ coincides with the
initial (Lagrangian) coordinate $\vX$ of a particle, hence $\widetilde{\vx}(0,0)=0$.

A distribution of a Lagrangian tracer $\widehat{a}$ is given as $\widehat{a}=A(\vX)$. Then
\begin{equation}\label{mm1}
\widehat{a}(\vx,t,\tau)=A(\vx-\varepsilon\widetilde{\vx}_1)
\end{equation}
One can expand both sides of (\ref{mm1}) as
\begin{equation}\label{mm2}
\widehat{a}_0+\varepsilon\widehat{a}_1+\varepsilon^2\widehat{a}_2+\ldots=
A(\vx)-\varepsilon\widetilde{x}_{1i}\frac{\partial A(\vx)}{\partial x_i} +
\frac{\varepsilon^2}{2}\widetilde{x}_{1i}\widetilde{x}_{1k}\frac{\partial^2 A(\vx)}{\partial x_i\partial x_k} +\ldots
\end{equation}
where $\widehat{a}_n=\widehat{a}_n(\vx,t,\tau)=
\overline{a}_n(\vx,t)+\widetilde{a}_n(\vx,t,\tau)$. The average $\langle\cdot\rangle^{\vx}$ of
this equation yields
\begin{equation}\label{mm3}
\overline{a}_0+\varepsilon\overline{a}_1+\varepsilon^2\overline{a}_2+\ldots=
A(\vx)+
\frac{\varepsilon^2}{2}\langle\widetilde{\xi}_{i}
\widetilde{\xi}_{k}\rangle\frac{\partial^2 A(\vx)}{\partial x_i\partial x_k} +\ldots
\end{equation}
where we have changed $\widetilde{\vx}_1(t,\tau)$ to $\widetilde{\vxi}(t,\tau)$ (\ref{01-appr-5}), which is valid for
the given precision. The $t$-differentiation of (\ref{mm3}) gives
\begin{equation}\label{mm4}
\overline{a}_{0t}+\varepsilon\overline{a}_{1t}+\varepsilon^2\overline{a}_{2t}+\ldots=
\frac{\varepsilon^2}{2}
\langle\widetilde{\xi}_{i}
\widetilde{\xi}_{k}\rangle_t\frac{\partial^2 a_0(\vx)}{\partial x_i\partial x_k}
+\ldots
\end{equation}
where we have used $A=\overline{a}_{0}$ (which follows from (\ref{mm3})). One can see that eqns.
(\ref{4.15})-(\ref{4.17}) taken for zero drift
$\overline{\vV}=\overline{\vV}_0+\varepsilon\overline{\vV}_1+\varepsilon^2\overline{\vV}_2+\ldots\equiv 0$ coincide
with (\ref{mm4}).

It means that in this particular case \emph{PD} represents a diffusion-like term that appears in the averaged equations
as a correction  $\overline{a}_2$ caused by the curvature of the function $\overline{a}_0=\overline{a}_0(\vx)$. In
order to make this explanation clear one can imagine a one-dimensional case of (\ref{osil-rigid}) with $A=A(Z)$ in
(\ref{mm1}) for a single Lagrangian coordinate $Z$. Then the averaging of oscillations of the entire graph $A=A(Z)$ in
$Z$-direction evidently produces a second-order correction (proportional to the curvature $A_{ZZ}$) to the distribution
$\overline{a}_0(z)=A(z)$ which occur in the absence of oscillations. Now one might assume that for general oscillating
flows the nature of pseudo-diffusion is the same:

\underline{\emph{Conjecture:}} Pseudo-diffusion in (\ref{4.17}), (\ref{4.21}) is always caused by the oscillations
of Lagrangian tracer $\widehat{a}(\vX,t,\tau)$ with respect to fixed Eulerian coordinates $\vx$.

According to this conjecture the presence of a drift produces only a logically natural change from $\partial/\partial
t$ (for flows (\ref{mm1}),(\ref{mm4}) without a drift)  to the `material' derivatives (with the drift velocity
$\overline{\vV}$ in (\ref{4.15})-(\ref{4.17})). Hence this conjecture looks reliable, it indicates that
pseudo-diffusion represents a natural as well as necessary term in the averaged equations.

\underline{Remark:} The effects of pseudo-diffusion and diffusion (or anti-diffusion, or
anisotropic diffusion-anti-diffusion) are described by the same equations, hence they are mathematically equivalent to
each other. However, pseudo-diffusion is used only with regular asymptotic procedures.

\subsection{The sub-critical family of oscillations: procedure with $\alpha=1$\label{sect04}}

Let us consider an example of a  sub-critical asymptotic family with $\alpha=1$ and $\beta=0$. For this family
$U=L/T=\const$ in (\ref{path-simple}) which might be seen as an advantage for some applications. Following
(\ref{insp-11c}) we accept that (\ref{insp-1}) is:
\begin{eqnarray}
&& \widehat{\vu}(\vx, t, \tau)=\widetilde{\vu}(\vx,t,\tau),\quad \delta=\omega^{-1},
\label{basic-1sub}
\end{eqnarray}
Then (\ref{insp-6}) yields:
\begin{eqnarray}
&& \mathfrak{D} \widehat{a}=\omega\widehat{a}_\tau+\widehat{a}_t+(\widetilde{\vu}\cdot\nabla)\,\widehat{a}= 0
\label{basic-2sub}
\end{eqnarray}
The small parameter $\varepsilon_{1}= 1/\omega$ (\ref{epsilon}) allows to rewrite it as
\begin{eqnarray}
&& \mathfrak{D}_1
\widehat{a}\equiv\widehat{a}_{\tau}+\varepsilon(\widehat{a}_t+\widetilde{\vu}\cdot\nabla)\,\widehat{a}=0, \quad
\mathfrak{D}_1\equiv\varepsilon\mathfrak{D}=\mathfrak{D}/\omega\label{basic-3sub}
\end{eqnarray}
where the subscript in $\varepsilon_{1}$ has been dropped. We are looking for the solution of (\ref{basic-3sub}) in the
form of regular series (\ref{basic-4aa}) with redifined $\varepsilon$. The substitution of (\ref{basic-4aa}) into
(\ref{basic-3sub}) produces the equations of successive approximations
\begin{eqnarray}
&&\widehat{a}_{0\tau}=0 \label{basic-5sub}\\
&&\widehat{a}_{n\tau}=-(\partial_t+\widetilde{\vu}\cdot\nabla)\,
\widehat{a}_{n-1},\quad\partial_t\equiv\partial/\partial t,\quad n=1,2,3,\dots
\label{basic-7sub}
\end{eqnarray}
The solving of (\ref{basic-5sub}),(\ref{basic-7sub}) yields
\begin{eqnarray}
&&\widetilde{a}_0\equiv 0,\label{4.11asub}\\
&& \widetilde{a}_1= -(\widetilde{\vxi}\cdot\nabla)\,\overline{a}_0,\label{4.11sub}\\
&&\widetilde{a}_{2}=-(\widetilde{\vxi}\cdot\nabla)\, \overline{a}_1
-\{(\widetilde{\vu}\cdot\nabla)\,\widetilde{a}_1\}^\tau-\widetilde{a}_{1t}^\tau, \label{4.12sub}\\
&&\overline{a}_{0t}=0,\label{4.15sub}\\
&&\overline{a}_{1t}+(\overline{\vV}_0\cdot\nabla)\overline{a}_0=0\label{4.16sub}\\
&&\overline{a}_{2t}+ (\overline{\vV}_0\cdot\nabla)\overline{a}_1+(\overline{\vV}_1^+\cdot\nabla)\overline{a}_0=
\frac{\partial}{\partial x_i}\left(
\overline{\chi}_{ik}^\text{sub}\frac{\partial\overline{a}_0}{\partial x_k}\right),
\label{4.17sub}\\
&&\overline{\vV}_1^\text{sub}=\overline{\vV}_1+\frac{1}{2}\langle[\widetilde{\vxi},\widetilde{\vxi}_t]\rangle
+\frac{1}{2}\langle\widetilde{\vxi}\Div\widetilde{\vxi}\rangle_t,\quad
2\overline{\chi}_{ik}^\text{sub}\equiv\langle\widetilde{\xi}_i\widetilde{\xi}_k\rangle_t,\label{4.20sub}
\end{eqnarray}
where $\overline{\vV}_0$ and $\overline{\vV}_1$ are the same as in (\ref{4.18}).  Equations
(\ref{4.15sub})-(\ref{4.17sub}) can be written as a single advection-pseudo-diffusion equation (valid with the error
${O}(\varepsilon^3)$)
\begin{eqnarray}
&&\left(\partial_t+ \overline{\vV}\cdot\nabla\right)\overline{a} =
\frac{\partial}{\partial x_i}\left(\overline{\kappa}_{ik}^\text{sub}\frac{\partial\overline{a}}{\partial x_k}\right)
\label{4.21sub}\\
&&\overline{\vV}=\varepsilon\overline{\vV}_0+\varepsilon^2\overline{\vV}_1^\text{sub},\quad \overline{\kappa}_{ik}^\text{sub}=\varepsilon^2\overline{\chi}_{ik}^\text{sub}\label{4.23sub}\\
&&\overline{a}=\overline{a}_0+\varepsilon\overline{a}_1+\varepsilon^2\overline{a}_2
\label{4.22aasub}
\end{eqnarray}
Eqn. (\ref{4.21sub}) shows that the averaged motion of $\widehat{a}$ represents a \emph{pseudo-drift} with velocity
$\overline{\vV}$ and \emph{pseudo-diffusion} with the matrix-coefficient $\overline{\kappa}_{ik}$. The reason for
introducing the term \emph{pseudo-drift} is the same as we used for \emph{pseudo-diffusion} in (\ref{4.21}): the
drift-like terms in (\ref{4.16sub}),(\ref{4.17sub}),(\ref{4.21sub}) play a part of sources, known from previous
approximations.

\underline{Remarks to \emph{Sect.7.11}:}

1. A simplified (in comparison with (\ref{4.20})) expression for pseudo-diffusivity in (\ref{4.17sub}) complies with
the interpretation of pseudo-diffusivity of previous subsection.

2. The system (\ref{4.15sub})-(\ref{4.20sub}) produces mainly diverging solutions. Indeed, (\ref{4.15sub}) yields
$\overline{a}_0=\overline{a}_0(\vx)$. Let $\overline{a}_0(\vx)\ne\const$, then (\ref{4.16sub}) gives a linear with $t$
growth of $\overline{a}_1$ for any $t$-independent $\overline{\vV}_0$, for example for the Stokes drift (see
\emph{Sect.7.3}). If we choose $\overline{a}_0\equiv\const$, then a similar growth appears for $\overline{a}_2$ \emph{etc.}
A statement that any sub-critical asymptotic procedure with $\overline{a}_t\in \mathbb{O}(1)$ produces diverging
solutions might be considered as a conjecture, but its analysis is beyond the scope of this paper.

3. This equations of this subsection do not produce the transport equation (drift) in any approximation.

4. The arguments of items 2 and 3 emphasize the significance of critical and super-critical asymptotic procedures (of
\emph{Sects.2-7}).

\underline{Remarks to all \emph{Sect.7}:}

1. From the examples of \emph{Sects.7.1-7.9} one can see that the expressions for the drift velocity $\overline{\vV}$
(\ref{4.18}), (\ref{4.19}), (\ref{7.24}) can provide arbitrary functional form and magnitude not higher than $O(1)$. In
order to make further progress one should prescribe some particular oscillating flows (\ref{exact-2}).

2. These examples also show that  pseudo-diffusivity in (\ref{4.21}),(\ref{4.22}) can appear in three qualitatively
different cases: (i) all positive eigenvalues $\overline{\chi}_i$ corresponds to ordinary diffusion; (ii) all negative
$\overline{\chi}_i$ corresponds to anti-diffusion, and (iii) the mixed signs correspond to more complex anisotropic
evolution with diffusion in some directions and anti-diffusion in the others.  The complexity of the problem
(\ref{4.21}) increases if one considers the dependence of $\overline{\chi}_i$ on $\vx$ and $t$. The general theory of
related linear PDEs (\emph{e.g.} (\ref{Example-3-5})) has not been developed yet, see
\cite{Polyanin, Debnath1}.

3. The number of our examples is naturally restricted, therefore \emph{Sect.7} illustrates our studies of
\emph{Sects.3-6} only partially.

\section{Links to Other Theories}

\subsection{TTAM-solution of characteristic equation}

We are not aware about any paper which calculates several approximations for a drift by solving the related ODE by
two-timing method. Therefore we present the \emph{TTAM}-versions of such calculations here. Let us return to the
original dimensional equation (\ref{exact-1}), (\ref{exact-2}) with an oscillating  velocity
\begin{eqnarray}
\widehat{\vu}^*=\widehat{\vu}^*(\widehat{\vx}^*,t^*,\tau) \label{velocity}
\end{eqnarray}
where $\widehat{\vx}^*$ is an upgraded notation for original Eulerian coordinates which is introduced here instead of
$\vx^*$ (see \emph{Sect.2}) since Eulerian coordinates in Lagrangian description  also represent the hat-functions
(\ref{tilde-func-def}). The oscillating trajectories
\begin{eqnarray}
\widehat{\vx}^*=\widehat{\vx}^*(\vX^*,t^*,\tau)\label{traj00}
\end{eqnarray}
(which represent the characteristics for the hyperbolic equation (\ref{exact-1}), (\ref{exact-6}))  can be found by
solving Cauchy's problem for an ODE
\begin{eqnarray}
&&\frac{d\widehat{\vx}^*}{d{s}^*}=\widehat{\vu}^*(\widehat{\vx}^*,t^*,\tau),\quad  \widehat{\vx}^*|_{t=0}=\vX^*\label{Cauchy}\\
&&t^*=s^*,\quad\tau=\omega^* s^*;\quad\frac{d}{d{s}^*}=\frac{\partial}{\partial
t^*}+\omega^*\frac{\partial}{\partial\tau}
\end{eqnarray}
where $\vX^*$ is the Lagrangian coordinate of a fluid particle and $t^*=0$ automatically leads to $\tau=0$. Using the
same procedure as for (\ref{exact-6-L}) we can obtain the dimensionless form of (\ref{Cauchy})
\begin{eqnarray}
\widehat{\vx}=\widehat{\vx}({t,\tau}):
\ \frac{d\widehat{\vx}}{d{s}}=\omega\delta\widehat{\vu}(\widehat{\vx},t,\tau);\quad t=s,\ \tau=\omega s,\
\frac{d}{d{s}}=\frac{\partial}{\partial t}+\omega\frac{\partial}{\partial\tau}\label{CauchyL}
\end{eqnarray}
which represents the two-timing form of a dynamical system for the motion of particles, see \cite{Arnold, Aref, Ottino,
Wiggins}. We accept (\ref{insp-1}) and introduce a test-solution for our inspection procedure
\begin{eqnarray}
&&\widehat{\vx}(t,\tau)=\overline{\vx}_0(t)+\frac{1}{\omega^\alpha}\widehat{\vx}_1(t,\tau),\ \alpha>0;\quad
\delta=\omega^{\beta-1},\ \beta<1\label{test-ODE}
\end{eqnarray}
that is motivated similarly to (\ref{insp-7}) (if one takes $\widetilde{\vx}_0\neq 0$, then $\tau$ ceases to be a fast
variable). The substitution of (\ref{test-ODE}) into (\ref{CauchyL}) yields
\begin{eqnarray}\label{test-ODE-1}
&&\left(\omega\frac{\partial}{\partial\tau}+\frac{\partial}{\partial t}\right)
\left(\overline{\vx}_0+\frac{1}{\omega^\alpha}\widehat{\vx}_1\right)=
\omega^\beta\widetilde{\vu}\left(\overline{\vx}_0+\frac{1}{\omega^\alpha}\widehat{\vx}_1 \right)
\end{eqnarray}
The decomposing of the RHS into Taylor's series with the retaining of two leading terms gives
\begin{eqnarray}\label{test-ODE-2}
\overline{\vx}_{0t}+\omega^{1-\alpha}\widetilde{\vx}_{1\tau}+\omega^{-\alpha}\widehat{\vx}_{1t}=
\omega^\beta\widetilde{\vu}+\omega^{\beta-\alpha}(\widehat{\vx}_1\cdot\nabla)\widetilde{\vu}
\end{eqnarray}
where $\widetilde{\vu}=\widetilde{\vu}(\overline{\vx}(t,\tau),t,\tau)$. The bar- and tilde- parts of this equation are
\begin{eqnarray}
&&\overline{\vx}_{0t}+\omega^{-\alpha}\overline{\vx}_{1t}=
\omega^{\beta-\alpha}\langle(\widetilde{\vx}_1\cdot\nabla)\widetilde{\vu}\rangle\label{test-ODE-3}\\
&&\omega^{1-\alpha}\widetilde{\vx}_{1\tau}+\omega^{-\alpha}\widetilde{\vx}_{1t}=
\omega^\beta\widetilde{\vu}
+\omega^{\beta-\alpha}(\overline{\vx}_1\cdot\nabla)\widetilde{\vu}
+\omega^{\beta-\alpha}\{(\widetilde{\vx}_1\cdot\nabla)\widetilde{\vu}\}
\label{test-ODE-4}
\end{eqnarray}
where we have used the brace notation (\ref{oper-5}). The leading terms in these equations are
\begin{eqnarray}
&&\overline{\vx}_{0t}=\omega^{\beta-\alpha}\langle(\widetilde{\vx}_1\cdot\nabla)\widetilde{\vu}\rangle\label{test-ODE-3L}\\
&&\omega^{1-\alpha}\widetilde{\vx}_{1\tau}=
\omega^\beta\widetilde{\vu}
\label{test-ODE-4L}
\end{eqnarray}
Here the $\tau$-average $\langle\cdot\rangle$ does not carry any superscript, since the spatial independent variable is
absent; at the same time $\widetilde{\vu}$ depends on an unknown function $\overline{\vx}_0(t)$. The equation
(\ref{test-ODE-4L}) gives $\alpha+\beta=1$, after that (\ref{test-ODE-3L}) brings us back to the same notions of
super-critical, critical, and sub-critical asymptotic families (\ref{insp-11a})-(\ref{insp-11c}). For a critical family
$\alpha=\beta=1/2$ and $\varepsilon=\varepsilon_{1/2}=1/\sqrt\omega$, which produces an asymptotic problem
\begin{eqnarray}\label{ODE-crit}
&&\left(\omega\frac{\partial}{\partial\tau}+\frac{\partial}{\partial t}\right)\widehat{\vx}=
\sqrt\omega\widetilde{\vu},\quad \widehat{\vx}=\overline{\vx}_0+
\frac{1}{\sqrt\omega}\widehat{\vx}_1+\frac{1}{\omega}\widehat{\vx}_1+\dots
\end{eqnarray}
or
\begin{eqnarray}\label{ODE-crit1}
&&\widetilde{\vx}_\tau=-\varepsilon^2\widehat{\vx}_t+\varepsilon\widetilde{\vu}(\widehat{\vx},t,\tau),\quad
\widehat{\vx}=\overline{\vx}_0+\varepsilon\widehat{\vx}_1+\varepsilon^2\widehat{\vx}_1+\dots
\end{eqnarray}
The successive approximations are
\begin{eqnarray}
&&\widetilde{\vx}_{0\tau}=0\label{ODE-Appr-0}\\
&&\widetilde{\vx}_{1\tau}=\widetilde{\vu}_0\label{ODE-Appr-1}\\
&&\widetilde{\vx}_{2\tau}+\overline{\vx}_{0t}=(\widehat{\vx}_1\cdot\overline\nabla)\widetilde{\vu}_0
\label{ODE-Appr-2} \\
&&\widetilde{\vx}_{3\tau}+\widetilde{\vx}_{1t}= (\widehat{\vx}_2\cdot\overline\nabla)\widetilde{\vu}_0+
\frac{1}{2}\widehat{x}_{1i}\widehat{x}_{1k}\frac{\partial^2\widetilde{\vu}_0}
{\partial \overline{x}_{0i}\partial \overline{x}_{0k}}\label{ODE-Appr-3}\\
&&\widetilde{\vx}_{4\tau}+\widetilde{\vx}_{2t}= (\widetilde{\vx}_3\cdot\overline\nabla)\widetilde{\vu}_0+
\widetilde{x}_{1i}\widetilde{x}_{2k}\frac{\partial^2\widetilde{\vu}_0}
{\partial \overline{x}_{0i}\partial \overline{x}_{0k}} +
\frac{1}{6}\widetilde{x}_{1i}\widetilde{x}_{1k}\widetilde{x}_{1j}\frac{\partial^3\widetilde{\vu}_0}
{\partial \overline{x}_{0i}\partial \overline{x}_{0k}\partial \overline{x}_{0j}}
\label{ODE-Appr-4}
\end{eqnarray}
where $\widetilde{\vu}_0\equiv\widetilde{\vu}(\overline{\vx}_0,t,\tau)$. The system
(\ref{ODE-Appr-0})-(\ref{ODE-Appr-4}) can be solved using the same method as (\ref{basic-5})-(\ref{basic-7}) and
(\ref{GLM-Appr-1})-(\ref{GLM-Appr-4}). Here we only present the solution for the averaged motion
\begin{eqnarray}
&&\overline{\vx}_{0t}=\overline{\vU}_0\label{ODE-Sol-1}\\
&&\overline{\vx}_{1t}=(\overline{\vx}_1\cdot\nabla)\overline{\vU}_0+\overline{\vU}_1\nonumber\\
&&\overline{\vx}_{2t}=(\overline{\vx}_1\cdot\nabla)\overline{\vU}_1+
(\overline{\vx}_2\cdot\nabla)\overline{\vU}_0+\overline{\vU}_2\nonumber\\
&&\overline{\vU}_0=\langle(\widetilde{\vx}_1\cdot\nabla)\widetilde{\vu}\rangle, \label{GLM-Sol-2}\\
&&\overline{\vU}_1=-\langle(\widetilde{\vx}_1\cdot\nabla)(\widetilde{\vu}\cdot\nabla)\widetilde{\vx}_1\rangle=
\frac{1}{3}\langle[[\widetilde{\vu},\widetilde{\vx}_1],\widetilde{\vx}_1]\rangle\label{GLM-Sol-2a}\\
&&\overline{\vU}_2=\left\langle(\widetilde{\vx}_3\cdot\nabla)\widetilde{\vu}\right\rangle+
\left\langle\widetilde{x}_{1i}\widetilde{x}_{2k}\frac{\partial^2\widetilde{\vu}(\overline{\vx})}
{\partial \overline{x}_i\partial \overline{x}_k}\right\rangle -
\frac{1}{2}\left\langle\widetilde{x}_{1i}\widetilde{x}_{1k}\widetilde{u}_{1j}\frac{\partial^3\widetilde{\vx}(\overline{\vx})}
{\partial \overline{x}_i\partial \overline{x}_k\partial \overline{x}_j}\right\rangle
\label{GLM-Sol-4}
\end{eqnarray}
which can be rewritten as
\begin{eqnarray}
&&\overline{\vx}_t=\overline{\vU}+(\overline{\vl}\cdot\nabla)\overline{\vU}\label{ODE-Sol-2}\\
&&\overline{\vx}=\overline{\vx}_0+\varepsilon\overline{\vx}_1+\varepsilon^2\overline{\vx}_1+\dots,\quad
\overline{\vl}\equiv\overline{\vx}-\overline{\vx}_0\nonumber\\
&&\overline{\vU}=\overline{\vU}_0+\varepsilon\overline{\vU}_1+\varepsilon^2\overline{\vU}_2+\dots\nonumber
\end{eqnarray}
One can see that the system (\ref{ODE-Sol-2}) describes two kinds of motion: in one every material point moves with
velocity $\overline{\vU}$; in the other every small material arc $\delta\overline{\vl}$ is stretched by the same
velocity field according to a standard law, see \cite{Batchelor}.

\subsection{\emph{TTAM}-approach to GLM-kinematics}\label{C}

The most general formula for a drift was obtained in \emph{GLM}-theory by
\cite{McIntyre}, which is referred hereafter as \emph{AM}.
The direct comparison of the results of \emph{Sects.2-7} with those of \emph{AM} is not feasible due to the following
conceptual differences: (i) \emph{GLM}-theory in \emph{AM} is not fully adapted to the two-timing method; (ii)
\emph{AM} uses one small parameter (the amplitude of $\widetilde{\vx}$) while we operate with two small parameters;
(iii) \emph{AM} employs a different averaging operation; and (iv) \emph{AM} expresses a drift in terms of
\emph{generalized Lagrangian displacements} $\widetilde{\vx}$ (\ref{traj-5}), not a given velocity field (as in our
case). We develop here the \emph{TTAM}-version of \emph{GLM}-kinematics with the aim to compare it with the results of
\emph{Sects.2-7}.

At the beginning of this subsection we use dimensional variables, but for brevity we suppress the asterisks; the use of
dimensionless variables (starting from (\ref{GLM-9aL})) will be additionally notified. The core of the whole
\emph{GLM}-theory is a transformation of (\ref{Cauchy}) into a PDE. In order to perform this transformation, we switch
to mutually independent $t$, $\tau$ (see \emph{Comment C} to (\ref{exact-4})) and introduce a $\tau$-averaged
trajectory (\ref{traj00}) as
\begin{eqnarray}
\overline{\vx}=\overline{\vx}(\vX,t)=\langle\widehat{\vx}\rangle^{\vX}
\label{average}
\end{eqnarray}
where $\langle\cdot\rangle^{\vX}$ emphasises that the $\tau$-average (\ref{oper-1}) is performed for fixed $\vX$. From
(\ref{Cauchy}) we get $\overline{\vx}(\vX,0)=\vX$. Equations (\ref{Cauchy}), (\ref{average}) give us three pairs of
main kinematical items:

(i) two kinds of trajectories for a selected particle
\begin{eqnarray}
&& \widehat{\vx}=\widehat{\vx}(\vX,t,\tau)\
\text{- the family of exact trajectories},\label{traj-1}\\
&&\overline{\vx}=\overline{\vx}(\vX,t)\ \text{- the averaged trajectory},
\label{traj-2}\\
&&\text{with}\quad \widehat{\vx}(\vX,0,0)=\overline{\vx}(\vX,0)=\vX,\label{xi=0}
\end{eqnarray}

(ii) two kinds of velocities
\begin{eqnarray}
&& \widehat{\vu}\equiv \partial\widehat{\vx}/\partial {s}|_{\vX} \ \text{-- the family of exact velocities of a
particle},
\label{velocities}\\
&&\overline{\vnu}\equiv\partial\overline{\vx}/\partial {s}|_{\vX} \ \text{-- the averaged velocity of a
particle}\nonumber
\end{eqnarray}

(iii) two Eulerian expressions for the same material derivative
\begin{eqnarray}
&& \partial/\partial {s}|_{\vX}=\partial/\partial {s}|_{\widehat{\vx}}+\widehat{\vu}\cdot\widehat{\nabla}=
\partial/\partial {s}|_{\overline{\vx}}+\overline{\vnu}\cdot\overline{\nabla}
\label{mat-der}
\end{eqnarray}
Relations (\ref{traj-1})-(\ref{mat-der}) show that for Lagrangian description $(\vX,t)$ one can use two different
Eulerian descriptions $(\widehat{\vx},{s})$ and $(\overline{\vx},{s})$ where $\overline{\vx}$ and $\overline{\vnu}$ are
called \emph{the GLM-coordinate} and \emph{GLM-velocity}.  One also should keep in mind that in
(\ref{velocities}),(\ref{mat-der}) according to the chain rule
\begin{eqnarray}
&&{\partial}/{\partial {s}}\vert_{\vX}={\partial}/{\partial t}\vert_{\vX,\tau} +
\omega\,{\partial}/{\partial\tau}\vert_{\vX,t}
\label{exact-5A}
\end{eqnarray}
and similar equalities for fixed $\widehat{\vx}$ or $\overline{\vx}$.  The existence of one-to-one mapping between
$\vX$ and $\widehat{\vx}$ represents a key postulate of classical fluid dynamics; it guarantees the existence of the
unique inverse to (\ref{traj-1}) function
\begin{eqnarray}
&&  \widehat{\vx}=\widehat{\vx}(\vX,t,\tau)\  \Leftrightarrow\ \vX=\vX(\widehat{\vx},t,\tau);\quad 0<J_0<\infty,\
J_0\equiv\frac{\partial(\widehat{x}_1,\widehat{x}_2,\widehat{x}_3)}{\partial(X_1,X_2,X_3)}
\label{traj-4inv}
\end{eqnarray}
with a Jacobian $J_0$. After that the  function
\begin{eqnarray}
&& \overline{\vx}=\overline{\vx}(\widehat{\vx},t,\tau)\label{traj-4aa}
\end{eqnarray}
can be obtained by the completely legal substitution of $\vX=\vX(\widehat{\vx},t,\tau)$ (\ref{traj-4inv}) into
$\overline{\vx}=\overline{\vx}(\vX,t)$ (\ref{average}).

The additional assumption (specific for \emph{GLM}) is an assumption of invertibility of (\ref{traj-2})
\begin{eqnarray}
\overline{\vx}=\overline{\vx}(\vX,t)\Rightarrow \vX=\vX(\overline{\vx},t); \quad
0<J_1<\infty,\ J_1\equiv\frac{\partial(\overline{x}_1,\overline{x}_2,\overline{x}_3)}{\partial(X_1,X_2,X_3)}
\label{traj-inv-aver}
\end{eqnarray}
The coordinate $\overline{\vx}$ is defined by the averaging operation (\ref{average}), hence it does not represent any
physical motion; (\ref{traj-inv-aver}) represents an assumption which can be valid only under some additional
restrictions. For the \emph{GLM}-theory this invertibility is absolutely required; let us also accept it and show how
it leads to \emph{GLM}-kinematics.

The substitution of the inverse function $\vX=\vX(\overline{\vx},t)$ (\ref{traj-inv-aver}) into
$\widehat{\vx}=\widehat{\vx}(\vX,t,\tau)$ (\ref{traj-1}) produces the key (for the
\emph{GLM}-theory) relation between $\widehat{\vx}$ and $\overline{\vx}$
\begin{eqnarray}
&&\widehat{\vx}=\widehat{\vx}(\overline{\vx},t,\tau); \quad 0<J_2<\infty,\
J_2\equiv\frac{\partial(\widehat{x}_1,\widehat{x}_2,\widehat{x}_3)}{(\overline{x}_1,\overline{x}_2,\overline{x}_3)}
 \label{traj-4}
\end{eqnarray}
that is inverse to (\ref{traj-4aa}). The functions (\ref{traj-4aa}), (\ref{traj-4}) give us one-to-one mapping between
two systems of Eulerian coordinates $\widehat{\vx}$ and $\overline{\vx}$;  since $J_2=J_0J_1^{-1}$, the conditions for
the Jacobians in (\ref{traj-inv-aver}) and (\ref{traj-4}) are equivalent to each other, provided (\ref{traj-4inv}) is
valid.

An important property of the averaging operation (\ref{average})
\begin{eqnarray}
&&\langle\cdot\rangle^{\overline{\vx}}=\langle\cdot\rangle^{\vX},
\label{traj-4a}
\end{eqnarray}
follows from the fact that $\tau$ is not involved into transformations (\ref{traj-inv-aver}),(\ref{average}) between
$\vX$ into $\overline{\vx}$. In other words, the $\tau$-averaging operation commutes with the changing of variables
(\ref{traj-inv-aver}). Taking $\langle\cdot\rangle^{\overline{\vx}}$ of (\ref{traj-4}) we get
\begin{eqnarray}
&&\widehat{\vx}(\overline{\vx},t,\tau)=\overline{\vx}+\widetilde{\vx}(\overline{\vx},t,\tau)\quad
\text{with}\quad\langle\widehat{\vx}\rangle^{\overline{\vx}}=\overline{\vx},\
\langle\widetilde{\vx}\rangle^{\overline{\vx}}=0
\label{traj-5}
\end{eqnarray}
where $\langle\widehat{\vx}\rangle^{\overline{\vx}}=\overline{\vx}$ is valid by virtue of (\ref{average}) and
(\ref{traj-4a}). Simultaneously (\ref{traj-5}) represents the definition of $\widetilde{\vx}$ which automatically
possesses zero $\tau$-average; in \emph{AM} $\widetilde{\vx}$ is called \emph{generalized Lagrangian displacement}.

Applying two Eulerian forms of material derivative (\ref{mat-der}) to both sides of (\ref{traj-5}) we obtain
\begin{eqnarray}
&& (\partial/\partial {s}|_{\widehat{\vx}}+\widehat{\vu}\cdot\widehat{\nabla})\widehat{\vx}=
\widehat{\vu}(\widehat{\vx},t,\tau)=(\partial/\partial
s|_{\overline{\vx}}+\overline{\vnu}\cdot\overline{\nabla})(\overline{\vx}+
\widetilde{\vx}(\overline{\vx},t,\tau))
\label{mat-der1}
\end{eqnarray}
which can be rewritten as
\begin{eqnarray}
&&\left(\frac{\partial}{\partial t}+\omega\frac{\partial}{\partial\tau}+\overline{\vnu}\cdot\overline{\nabla}\right)
\left(\overline{\vx}+\widetilde{\vx}(\overline{\vx},t,\tau)\right)=
\widehat{\vu}(\overline{\vx}+\widetilde{\vx}(\overline{\vx},t,\tau),t,\tau)\quad \text{or} \label{GLM-9a}\\
&&\left(\frac{\partial}{\partial t}+\omega\frac{\partial}{\partial\tau}+\overline{\vnu}\cdot\overline{\nabla}\right)
\widetilde{\vx}(\overline{\vx},t,\tau)=\widehat{\vu}(\overline{\vx}+
\widetilde{\vx}(\overline{\vx},t,\tau),t,\tau)-\overline{\vnu}(\overline{\vx},t)
\label{mat-der2}
\end{eqnarray}
where (\ref{GLM-9a}) represents the equation for characteristics (\ref{Cauchy}) written as a PDE in variables
$\overline{\vx},t,\tau$. We call either (\ref{GLM-9a}) or (\ref{mat-der2}) the two-timing form of
\emph{Andrews-McIntyre-Kinematics-Equation}\emph{(AMKE)}; it contains unknown functions $\widetilde{\vx}(\overline{\vx},t,\tau)$ and $\overline{\vnu}(\overline{\vx},t)$. The
description of fluid kinematics by (\ref{GLM-9a}) is rather unusual: \emph{AMKE} describes an effective medium which
moves with the averaged velocity $\overline{\vnu}$ that includes both an advective velocity and a drift. The exact
motion of a material particle in this medium is described by two different fields: the averaged motion is described by
\emph{GLM}-velocity $\overline{\vnu}(\overline{\vx},t)$ and oscillatory displacement (from the \emph{GLM}-position
$\overline{\vx}$) is given by $\widetilde{\vx}(\overline{\vx},t,\tau)$.

The first step in solving (\ref{mat-der2}) is straightforward: its bar-part $\langle\cdot\rangle^{\overline{\vx}}$
gives us \emph{GLM}-velocity $\overline{\vnu}$ expressed in the terms of an original velocity
\begin{eqnarray}
&&\overline{\vnu}(\overline{\vx},t)=
\langle\widehat{\vu}(\overline{\vx}+\widetilde{\vx}(\overline{\vx},t,\tau),t,\tau)\rangle^{\overline{\vx}},
\label{u-v}
\end{eqnarray}
which represents the main kinematical formula by \emph{AM}. The average operation in the RHS of (\ref{u-v}) is called
\emph{GLM-averaging}. \emph{GLM}-averaging of a function $\widehat{a}(\widehat{\vx},t,\tau)$ is denoted by \emph{AM} as
\begin{eqnarray}
&&\bar{a}^L=\bar{a}^L(\vx,t)\equiv
\langle \widehat{a}(\overline{\vx}+\widetilde{\vx}(\overline{\vx},t,\tau),t,\tau)\rangle^{\overline{\vx}}\label{aver_L}
\end{eqnarray}
The advection of a Lagrangian marker is described by the equation (\ref{exact-1})
\begin{eqnarray}
\left({\partial}/{\partial {s}}\vert_{\widehat{\vx}}+\widehat{\vu}(\widehat{\vx},t,\tau)\cdot\widehat{\nabla}\right)
\widehat{a}(\widehat{\vx},t,\tau)=0
\label{exact-1C}
\end{eqnarray}
Due to (\ref{mat-der}), we replace a material derivative in (\ref{exact-1C}) with $\partial/\partial
t|_{\overline{\vx}}+\overline{\vnu}\cdot\overline{\nabla}$ and use (\ref{traj-5})
\begin{equation}\label{aL-}
(\partial/\partial s|_{\overline{\vx}}+\overline{\vnu}\cdot\overline{\nabla})
\widehat{a}(\overline{\vx}+\widetilde{\vx}(\overline{\vx},t,\tau),t,\tau)=0
\end{equation}
Then $\langle\cdot\rangle^{\overline{\vx}}$ yields
\begin{equation}\label{aL}
(\partial/\partial t|_{\overline{\vx}}+\overline{\vnu}\cdot\overline{\nabla})\bar{a}^L=0
\end{equation}
This remarkable equation says that the \emph{GLM}-average $\bar{a}^L$ (\ref{aver_L}) in the \emph{GLM}-coordinates
$\overline{\vx}$ (\ref{average}), (\ref{traj-2}) is purely advected with the \emph{GLM}-velocity $\overline{\vnu}$
(\ref{u-v}).
\emph{The drift} by \emph{AM} is
\begin{eqnarray}
\overline{\vU}(\overline{\vx},t)=\overline{\vnu}(\overline{\vx},t)-
\langle\widehat{\vu}(\overline{\vx},t,\tau)\rangle^{\overline{\vx}}=
\langle\widehat{\vu}(\overline{\vx}+\widetilde{\vx}(\overline{\vx},t,\tau),t,\tau)\rangle^{\overline{\vx}}-
\langle\widehat{\vu}(\overline{\vx},t,\tau)\rangle^{\overline{\vx}}
\label{Stokes-D}
\end{eqnarray}
For a purely oscillating field $\widehat{\vu}=\widetilde{\vu}$ the average
$\langle\widehat{\vu}\rangle^{\overline{\vx}}\equiv 0$, hence
\begin{eqnarray}
&&\overline{\vU}=\overline{\vnu}(\overline{\vx},t)=
\langle\widehat{\vu}(\overline{\vx}+\widetilde{\vx}(\overline{\vx},t,\tau),t,\tau)\rangle^{\overline{\vx}}
\label{V-u-v}
\end{eqnarray}
To follow the route used by \emph{AM} we can take $\widetilde{\vx}$ of a small amplitude  and decompose
(\ref{Stokes-D}) and (\ref{V-u-v}) into Taylor's series; however in the two-timing case this problem contains two small
parameters and is richer in asymptotic procedures. In order to describe them we write the dimensionless form of
\emph{AMKE} (\ref{GLM-9a}) which is obtained by the same procedure as (\ref{exact-6-L})
\begin{eqnarray}
&&\left(\frac{\partial}{\partial t}+\omega\frac{\partial}{\partial\tau}+\overline{\vnu}\cdot\overline{\nabla}\right)
\left(\overline{\vx}+\widetilde{\vx}(\overline{\vx},t,\tau)\right)=
\omega\delta\widetilde{\vu}(\overline{\vx}+\widetilde{\vx}(\overline{\vx},t,\tau),t,\tau) \label{GLM-9aL}
\end{eqnarray}
where the given velocity field is taken as a purely oscillatory one. Starting from (\ref{GLM-9aL}) and further in this
section we use only dimensionless variables and functions, keeping the same notations without asterisks. Using the same
inspection procedure as in
\emph{Sect.3}  we accept (\ref{insp-1}) and introduce a test-solution (similar to (\ref{insp-7}))
\begin{eqnarray}
\widehat{\vx}=\overline{\vx}+\widetilde{\vx}=\overline{\vx}+\frac{1}{\omega^\alpha}\
\widetilde{\vy}(\overline{\vx},t,\tau),\quad \alpha=\const>0
\label{GLM-x-xi}
\end{eqnarray}
where $\overline{\vx}\in \mathbb{B}\cap\mathbb{O}(1),\  \widetilde{\vy}\in \mathbb{T}\cap\mathbb{O}(1)$ and the
amplitude of an oscillating part is given by the small parameter $\varepsilon_\alpha=1/\omega^\alpha$ (\ref{epsilon}).
Hence  we obtain
\begin{eqnarray}
&&\left(\frac{\partial}{\partial t}+\omega\frac{\partial}{\partial\tau}+\overline{\vnu}\cdot\overline{\nabla}\right)
\left(\overline{\vx}+\frac{1}{\omega^\alpha}\ \widetilde{\vy}(\overline{\vx},t,\tau)\right)=
\omega^\beta\widetilde{\vu}\left(\overline{\vx}+\frac{1}{\omega^\alpha}\ \widetilde{\vy}(\overline{\vx},t,\tau)\right)
\label{GLM-9aa}
\end{eqnarray}
Decomposing the right-hand side of (\ref{GLM-9aa}) into Taylor's series and omitting the terms of order
${O}(1/\omega^{2\alpha})$ and above, we have
\begin{eqnarray}
&&\overline{\vnu}+\omega^{1-\alpha}\widetilde{\vy}_\tau+\frac{1}{\omega^\alpha}D_{\overline{\vnu}}\widetilde{\vy}=
\omega^\beta\widetilde{\vu}(\overline{\vx},t,\tau)+
\omega^{\beta-\alpha}(\widetilde{\vy}\cdot\nabla)\widetilde{\vu}(\overline{\vx},t,\tau)\label{GLM-9aaa}
\end{eqnarray}
The bar-part of this equation is
\begin{eqnarray}
\overline{\vnu}=
\omega^{\beta-\alpha}\langle(\widetilde{\vy}\cdot\nabla)\widetilde{\vu}\rangle\label{GLM-9b}
\end{eqnarray}
while the difference between (\ref{GLM-9aaa}) and (\ref{GLM-9b}) produces the tilde-part of the equation. To make a
meaningful equation out of the tilde-part (\emph{cf.} with (\ref{insp-10})) it is necessary to accept that dominating
terms are of the same order:
\begin{equation}
\alpha+\beta=1\quad\text{and}\quad\widetilde{\vy}_\tau=\widetilde{\vu}(\overline{\vx},t,\tau)\label{insp-10a}
\end{equation}
Equations (\ref{GLM-9b}) and (\ref{insp-10a}) brings us back to the same notions  of super-critical, critical, and
sub-critical asymptotic families (\ref{insp-11a})-(\ref{insp-11c}) where a critical family is characterised by
$\alpha=\beta$ (a classical \emph{AM} case corresponds to a sub-critical family with $\alpha=1$ and $\beta=0$). Then
\emph{AMKE} (\ref{GLM-9a}) for the critical asymptotic family $\alpha=\beta=1/2$ takes form
\begin{eqnarray}
&&\left(\omega\frac{\partial}{\partial\tau}+\frac{\partial}{\partial t}+\overline{\vU}\cdot\overline{\nabla}\right)
\left(\overline{\vx}+\widetilde{\vx}(\overline{\vx},t,\tau)\right)=
\sqrt\omega\widehat{\vu}(\overline{\vx}+\widetilde{\vx}(\overline{\vx},t,\tau),t,\tau)\label{GLM-9M}\\
&&\widetilde{\vx}(\overline{\vx},t,\tau)=
\frac{1}{\sqrt\omega}\widetilde{\vx}_1(\overline{\vx},t,\tau)+
\frac{1}{\omega}\widetilde{\vx}_2(\overline{\vx},t,\tau)+\dots\nonumber\\
&&\overline{\vU}(\overline{\vx},t)=\overline{\vU}_0(\overline{\vx},t)+\frac{1}{\sqrt\omega}\overline{\vU}_1(\overline{\vx},t)+
\frac{1}{\omega}\overline{\vU}_2(\overline{\vx},t)+\dots\nonumber
\end{eqnarray}
The use of $\varepsilon=\varepsilon_{1/2}={1}/{\sqrt\omega}$ (\ref{epsilon}) transforms (\ref{GLM-9M}) into
\begin{eqnarray}
&&\left(\frac{\partial}{\partial\tau}+\varepsilon^2\frac{\partial}{\partial
t}+\varepsilon^2(\overline{\vU}_0+\varepsilon\overline{\vU}_1+
\varepsilon^2\overline{\vU}_2+\dots)\cdot\overline{\nabla}\right)
\left(\overline{\vx}+\varepsilon\widetilde{\vx}_1+\varepsilon^2\widetilde{\vx}_2+\dots\right)=\nonumber\\
&&=\varepsilon\widehat{\vu}(\overline{\vx}+\varepsilon\widetilde{\vx}_1+\varepsilon^2\widetilde{\vx}_2+\dots, t, \tau
)\label{GLM-9MM}
\end{eqnarray}
The first four approximations of this equation are:
\begin{eqnarray}
&&\widetilde{\vx}_{1\tau}=\widetilde{\vu}(\overline{\vx},t,\tau)\label{GLM-Appr-1}\\
&&\widetilde{\vx}_{2\tau}+\overline{\vU}_0=(\widetilde{\vx}_1\cdot\nabla)\widetilde{\vu}(\overline{\vx},t,\tau)
\label{GLM-Appr-2} \\
&&\widetilde{\vx}_{3\tau}+\widetilde{\vx}_{1t}+\overline{\vU}_1+(\overline{\vU}_0\cdot\nabla)\widetilde{\vx}_1=
(\widetilde{\vx}_2\cdot\nabla)\widetilde{\vu}(\overline{\vx},t,\tau)+
\frac{1}{2}\widetilde{x}_{1i}\widetilde{x}_{1k}\frac{\partial^2\widetilde{\vu}(\overline{\vx},t,\tau)}
{\partial \overline{x}_i\partial \overline{x}_k}\label{GLM-Appr-3}\\
&&\widetilde{\vx}_{4\tau}+\widetilde{\vx}_{2t}+\overline{\vU}_2+(\overline{\vU}_1\cdot\nabla)\widetilde{\vx}_1
+(\overline{\vU}_0\cdot\nabla)\widetilde{\vx}_2=\label{GLM-Appr-4}\\
&&=(\widetilde{\vx}_3\cdot\nabla)\widetilde{\vu}(\overline{\vx},t,\tau)+
\widetilde{x}_{1i}\widetilde{x}_{2k}\frac{\partial^2\widetilde{\vu}(\overline{\vx},t,\tau)}
{\partial \overline{x}_i\partial \overline{x}_k} +
\frac{1}{6}\widetilde{x}_{1i}\widetilde{x}_{1k}\widetilde{x}_{1j}\frac{\partial^3\widetilde{\vu}(\overline{\vx},t,\tau)}
{\partial \overline{x}_i\partial \overline{x}_k\partial \overline{x}_j}
\nonumber
\end{eqnarray}
These equations can be solved in the same way as the equations in \emph{Sect.4}. The solutions are
\begin{eqnarray}
&&\widetilde{\vx}_1=\widetilde{\vu}^{\tau}\label{GLM-Sol-1}\\
&&\widetilde{\vx}_2=\{(\widetilde{\vx}_1\cdot\nabla)\widetilde{\vu}\}^\tau\label{GLM-Sol-1bb}\\
&&\widetilde{\vx}_{3}=\widetilde{\vx}_{1t}^\tau+(\overline{\vU}_0\cdot\nabla)\widetilde{\vx}_1^\tau+
\{(\widetilde{\vx}_2\cdot\nabla)\widetilde{\vu}\}^\tau+
\frac{1}{2}\left\{\widetilde{x}_{1i}\widetilde{x}_{1k}\frac{\partial^2\widetilde{\vu}(\overline{\vx})}
{\partial \overline{x}_i\partial \overline{x}_k}\right\}^\tau\label{GLM-Sol-3}
\end{eqnarray}
with the same $\overline{\vU}_0, \overline{\vU}_1$, and $\overline{\vU}_2$  as in (\ref{GLM-Sol-2})-(\ref{GLM-Sol-4}).

\underline{Remarks and discussion:}

1. In (\ref{GLM-9aL}) we are looking for the dimensionless drift of order one $\overline{\vnu}={O}(1)$ that (as we know
from \emph{Sects.4} and \emph{5}) is true for critical and super-critical oscillations. The more systematic (but more
cumbersome) approach  is to replace $\overline{\vnu}$ in (\ref{GLM-9aL}) by $\omega\delta\overline{\vnu}$ (which is
required by (\ref{V-u-v})) and to derive rigorously that the first non-vanishing term in $\omega\delta\overline{\vnu}$
is ${O}(1)$.

2. The super-critical versions of the problems considered in \emph{Sects.8.1,8.2} can be solved similarly to the
problems in
\emph{Sects.5,6}.

3. The expression (\ref{ODE-Sol-1}),(\ref{GLM-Sol-2}) was obtained by \cite{Yudovich}.

4. The theories of both \emph{Sect.8.1} and \emph{Sect.8.2} operate with the average taken for a fixed Lagrangian
coordinate $\vX$. Indeed,  it is the only possibility for the problem (\ref{CauchyL}): this problem describes dynamics
of a single particle. As to \emph{AMKE} (\ref{GLM-9a}), it explicitly uses the average for fixed $\overline{\vx}$ which
is the same as for fixed $\vX$, see (\ref{traj-4a}).

5. One can observe a strong resemblance between the system (\ref{ODE-Appr-0})-(\ref{ODE-Appr-4}) and the system
(\ref{GLM-Appr-1})-(\ref{GLM-Appr-4}). In (\ref{ODE-Appr-0})-(\ref{ODE-Appr-4}) terms with $\overline{\vU}_0,
\overline{\vU}_1,\dots$ in LHS are absent and Taylor's series in RHS contain
$\widehat{\vx}=\overline{\vx}+\widetilde{\vx}$ (instead of $\widetilde{\vx}$ in (\ref{GLM-Appr-1})-(\ref{GLM-Appr-4})).
These differences are due to the fact that $\overline{\vx}$ is an independent variable in
(\ref{GLM-Appr-1})-(\ref{GLM-Appr-4}) but in (\ref{ODE-Appr-0})-(\ref{ODE-Appr-4}) it play a part of an unknown
function. However one can show that these two systems are mathematically equivalent to each other, and the drift
velocity $\overline{\vU}$ in \emph{Sect.8.1} and in
\emph{Sect.8.2} is the same. These properties are natural, since \emph{AMKE} (\ref{GLM-9a}) represents a transformed
ODE for characteristics (\ref{Cauchy}) (see (\ref{average})-(\ref{GLM-9a})).

6. Two sets of formulae for drift velocities (\ref{01-appr-5}), (\ref{4.18}) and (\ref{GLM-Sol-1}),
(\ref{GLM-Sol-2}),(\ref{GLM-Sol-2a}) look identical if one makes a correspondence
\begin{eqnarray}
&&\widetilde{\vu}(\vx,t,\tau) \leftrightarrow  \widetilde{\vu}(\overline{\vx},t,\tau),\quad
\widetilde{\vxi}(\vx,t,\tau)\leftrightarrow\widetilde{\vx}(\overline{\vx},t,\tau),\label{analogy}\\
&&\overline{\vV}_0(\vx,t) \leftrightarrow \overline{\vU}_0(\overline{\vx},t),\quad
\overline{\vV}_1(\vx,t) \leftrightarrow \overline{\vU}_1(\overline{\vx},t)\nonumber
\end{eqnarray}
hence one may conclude that our formulae for the zeroth and first approximations of a drift are the same as \emph{AM}.
However this resemblance is misleading, at least partially. The crucial point is the use of the
\emph{GLM}-coordinates $\overline{\vx}$ (\ref{traj-2}) in \emph{GLM}-kinematics \emph{vs.}
the original Eulerian coordinates $\vx\equiv\widehat{\vx}$ (\ref{traj-1}) in
\emph{Sects.2-7}. It leads to the different definitions of the averaging operations with fixed $\overline{\vx}$
\emph{vs.} fixed $\vx$. For example, the expressions $\widetilde{\vx}_1=\widetilde{\vu}^{\tau}$ (\ref{GLM-Sol-1}) and
$\widetilde{\vxi}\equiv\widetilde{\vu}^\tau$ (\ref{01-appr-5}) look identical, however they are different since the
former is expressed as a function of $\overline{\vx}$ and the latter as a function of $\vx\equiv\widehat{\vx}$: these
variables are constant during the $\mathbb{T}$-integrations. The implications of such a difference lay beyond the scope
of this paper. At this stage one can see that: (i) the similarity between the expressions in (\ref{analogy}) indeed
corresponds to the close results for the zeroth and first approximations (due to the smallness of the difference
$\widetilde{\vx}=\widehat{\vx}-\overline{\vx}$); (ii) the expressions for $\overline{\vU}_2$ and $\overline{\vV}_2$ do
not exhibit the similarity shown in (\ref{analogy}).

7. One can see that, in order to express (\ref{aL}) in the spirit of \emph{Sects.2-7}, $\bar{a}^L$ has to be decomposed
into Taylor's series for small $\widetilde{\vx}$. The resulted equations is
\begin{eqnarray}
&&(\partial/\partial t|_{\overline{\vx}}+\overline{\vnu}\cdot\overline{\nabla})\,\overline{R}=0,\label{aL+}\\
&&\overline{R}\equiv\langle
\widehat{a}(\overline{\vx},t,\tau)\rangle^{\overline{\vx}}+\left\langle
\widetilde{x}_i
\frac{\partial  \widehat{a}(\overline{\vx},t,\tau)}{\partial \overline{x}_i}\right\rangle^{\overline{\vx}}
+\frac{1}{2}\left\langle
\widetilde{x}_i
\widetilde{x}_k\frac{\partial^2
\widehat{a}(\overline{\vx},t,\tau)}{\partial \overline{x}_i\partial \overline{x}_k}\right\rangle^{\overline{\vx}}+
\dots\nonumber
\end{eqnarray}
where $\widetilde{\vx}(\overline{\vx},t,\tau)$, $\overline{\vnu}(\overline{\vx},t)$, and
$\widehat{\va}=\overline{\va}+\widetilde{\va}$ must be expressed as in (\ref{GLM-9M})  and (\ref{basic-4aa}). Then the
terms of the type $\langle\widetilde{x}_i\partial\widetilde{a}/\partial \overline{x}_i\rangle$ should be calculated
with the use of the full equation (\ref{aL-}). Such calculations will produce Riemann's invariant $\overline{R}$ which
is constant along the averaged characteristic curves (or $\overline{R}$ is transported with the drift velocity).

8. It is apparent that the establishing of one-to-one correspondence between (\ref{4.21}) and (\ref{aL+}) is not
feasible. Indeed, (\ref{4.21}) has a `dissipative' pseudo-diffusion term which can not be incorporated in the transport
equation (\ref{aL+}). This fact should be expected from \emph{Sect.7.10}, where we have shown that pseudo-diffusion
appears due to oscillations of Lagrangian coordinates with respect to fixed Eulerian coordinates. Hence,
pseudo-diffusion (which is a product of Eulerian averaging) is not compatible with (\ref{aL+}) (which is a product of
Lagrangian averaging, see the
\emph{item 4} above). This incompatibility complies with a general understanding that different averaging operations
preserve and loose original information differently. One can conclude that the averaged fields in
\emph{TTAM}-form of
\emph{GLM}-theory and in the theory of
\emph{Sects. 2-7} contain complementary (or just additional to each other) information.
It opens the opportunity to look for the advantages given by the presence of these two descriptions.

\subsection{Links to other classical theories}

\underline{\emph{The classical drift velocity.}}
The classical expression for a drift velocity is given by the formulae (26) of \cite{LH} or by (5.13.21) of
\cite{Batchelor}. Our $\overline{\vV}_0$ (\ref{4.18}) will be proportional to this expression if it is rewritten with the
use of $\tau$-periodicity and integration by parts. In fact, the conjugated form
$\langle(\widetilde{\vxi}\cdot\nabla)\,\widetilde{\vv}\rangle$ of
$\overline{\vV}_0=-\langle(\widetilde{\vv}\cdot\nabla)\,\widetilde{\vxi}\rangle$ (\ref{2-appr-8}) coincides with the
classical expression. Consequently $\overline{\vV}_0$ gives the correct expression for the drift in a potential
travelling wave (\cite{Stokes}). At the same time, there are four main differences between our results and that of the
quoted classical papers: (i) we  describe all involved variables and fields precisely, within the two-timing framework;
(ii) we employ two small parameters instead of one; (iii) we obtain different magnitudes for a drift while in the
classical paper the leading term is quadratic in small amplitude (which is also obtainable within our approach); (iv)
we obtain two more formulae for drift velocities as well as find pseudo-diffusion.

\underline{\emph{The Krylov-Bogoliubov method.}}
The Krylov-Bogoliubov averaging method \emph{(KBAM)}, see
\cite{Bog,Krylov,Verhulst}, is aimed to solve problems for special classes of ODEs with oscillating coefficients.
This method can be straightforwardly used for the calculating of the averaged equations of characteristic curves
(\ref{CauchyL}). We have exploited this option: the required calculations are rather cumbersome, therefore we formulate
here  the results only. The first
\emph{KBAM}-term for a drift velocity coincides with $\overline{\vU}_0$ (\ref{ODE-Sol-1}), while the
obtaining of the next two terms requires substantially more \emph{KBAM} analytical calculations than
\emph{TTAM} ones. Here we make two comments: (1) \emph{KBAM} as well as
\emph{TTAM} calculations of \emph{Sect.8.1} give the averaged equations for characteristic curves.
Within these two methods applied to an ODE the problem of finding the averaged Riemann invariants cannot be addressed.
(2) The areas of applicability of \emph{TTAM} are much broader than
\emph{KBAM}: for example \emph{TTAM} can operate with PDEs and incorporate molecular diffusivity
to the governing equations (see \emph{Sect.9}).

\underline{\emph{The homogenization theory.}}
\emph{TTAM} has the same methodological roots as the method of
homogenization (see \cite{Lions, Berdichevsky}), which represents a version of the two-scale method. The long-wave
`homogenization' solutions for the transport of scalar admixture obtained by \cite{Papanicolaou, Vergassola1, Frisch}
show the appearance of a `turbulent' diffusion matrix, which is always positive-definite. 
In order to establish further links we consider the case when the given velocity does not contain high frequency
oscillations and can be expressed as $\widetilde{\vu}_1(\vx,t/\omega,t)$ (or just
$\widetilde{\vu}=\widetilde{\vu}_1(\vx,t)$) with the $2\pi$-periodic dependence on $t$. Let us show that in this case
the problem can be transformed to the considered in \emph{Sects.2-7}.

\emph{Problem A:} The dimensionless form of the transport equation (\ref{exact-6-L}), (\ref{insp-6}) for a purely oscillating
velocity
\begin{eqnarray}
\left(\frac{\partial}{\partial {s}}+\omega^{\beta}\widetilde{\vu}\cdot\nabla\right) \widehat{a}=0,\quad
\frac{\partial}{\partial {s}}=\frac{\partial}{\partial t}+\omega\frac{\partial}{\partial\tau}
\label{exact-1R}
\end{eqnarray}
where $\widehat{a}=\widehat{a}(\vx,t,\tau)$, $\widetilde{\vu}=\widetilde{\vu}(\vx,t,\tau)$, $\beta=\const$, the
mutually dependent time-variables are $t={s}$, $\tau\equiv\omega {s}$.

\emph{Problem B:} A similar equation for different time-scales is
\begin{eqnarray}
\left(\frac{\partial}{\partial {s}}+\omega^{\beta_1}\widetilde{\vu}_1\cdot\nabla\right) \widehat{a}_1=0,\quad
\frac{\partial}{\partial {s}}=\frac{1}{\omega}\left(\frac{\partial}{\partial t_1}+\omega\frac{\partial}{\partial\tau_1}\right)
\label{exact-1RR}
\end{eqnarray}
where $\widehat{a}_1=\widehat{a}_1(\vx,t_1,\tau_1)$, $\widetilde{\vu}_1=\widetilde{\vu}(\vx,t_1,\tau_1)$,
$\beta_1=\const$, the mutually dependent time-variables are $t_1={s}/\omega$, $\tau_1\equiv {s}$.

One can see that the equations (\ref{exact-1R}) and (\ref{exact-1RR}) are mathematically identical to each other if one
introduces the link `\emph{Problem A}$\leftrightarrow$\emph{Problem B}' as
\begin{equation}\label{rescale-dict}
    t\leftrightarrow t_1,\quad \tau\leftrightarrow \tau_1,\quad \beta\leftrightarrow\beta_1+1,
    \quad\widehat{a}_t, \widehat{a}_\tau \in\mathbb{O}(1)
    \leftrightarrow\widehat{a}_{t_1}, \widehat{a}_{\tau_1} \in \mathbb{O}(1)
\end{equation}
After such replacements any solution of (\ref{exact-1R}) simultaneously produces a solution of (\ref{exact-1RR}). Hence
this rescaling procedure delivers a counterpart $\widehat{a}(\vx,t_1,\tau_1)$ of $\widehat{a}(\vx,t,\tau)$: in the
\emph{Problem B }(\ref{exact-1RR}) the imposed oscillations have the frequency ${O}(1)$ and the slow-time-scale
${O}(1/\omega)$. The link (\ref{rescale-dict}) allows us to add one more solution for each solution in
\emph{Sects.2-7}. The main motivation behind the rescaling (\ref{rescale-dict}) is: the only solutions
considered in the homogenisation theory, see \cite{Lions, Berdichevsky} are of the type (\ref{exact-1RR}).

\underline{\emph{The presence of molecular diffusivity.}} Let $\mu^*$ and $\mu$ be the dimensional and dimensionless
coefficients of molecular diffusion. Eqn. (\ref{exact-1}) describes the advection-diffusion of a scalar admixture in an
incompressible fluid after adding  the term $\mu^*\nabla^{*2}a$ to the RHS.  All results of \emph{Sects.2-6} can be
straightforwardly generalized to the flows with $\mu=O(1)$. Such generalization is achieved by replacing operators
$\partial/\partial{s}^*\rightarrow
\partial/\partial {s}^*-\mu^*\nabla^{*2}$ in (\ref{exact-1}) and $\partial/\partial t\rightarrow \partial/\partial
t-\mu\nabla^{2}$ in (\ref{exact-6}) and in all subsequent formulae  The limit $\mu\to 0$ is a regular one for our
strictly regular asymptotic procedures (\ref{basic-4aa}), so all the problems with $\mu\neq 0$ have the  cases studied
in \emph{Sects.2-6} with $\mu\equiv 0$ as their limits.

\underline{\emph{Magnetohydrodynamics (MHD)}.}
One can also develop the same asymptotic procedures and to derive the averaged equations similar to
(\ref{4.15})-(\ref{4.20a}) for a vectorial passive admixture such as  magnetic field $\vh^*(\vx^*,{s}^*)$ in the
kinematic MHD-dynamo problem for a given oscillating velocity field, see
\cite{Moffatt}. In this case one can start with equations
\begin{eqnarray}
{\partial\vh^*}/{\partial {s}^*}+ [\vh^*,\vu^*] =0,\quad \Div \vh^*=0,\quad \Div \vu^*=0
\label{h-1}
\end{eqnarray}
which replace (\ref{exact-1}). Taking the same velocity field (\ref{basic-1}) and repeating the same steps as in
\emph{Sect.4} lead to the averaged equation (valid with the error ${O}(\varepsilon^3)$)
\begin{eqnarray}
&&\overline{\vh}_t+ (\overline{\vV}\cdot\nabla)\overline{\vh}-(\overline{\vh}\cdot\nabla)\overline{\vV}
=\frac{\partial}{\partial x_i}\left(\overline{\kappa}_{ik}\frac{\partial\overline{\vh}}{\partial x_k}\right),\quad \Div
\overline{\vh}=0
\label{beauty-Eqn-h}\\
&&\overline{\vh}=\overline{\vh}_0+\varepsilon\overline{\vh}_1+\varepsilon^2\overline{\vh}_2,\quad
\overline{\vV}=\overline{\vV}_0+\varepsilon\overline{\vV}_1+\varepsilon^2\overline{\vV}_2, \quad
\overline{\kappa}_{ik}=\epsilon^2\overline{\chi}_{ik}
\nonumber
\end{eqnarray}
where all coefficients  are the same as in (\ref{4.15})-(\ref{4.20a}). For this problem we have performed the detailed
calculations of $\overline{\vV}_0$ and $\overline{\vV}_1$, while the expressions for $\overline{\vV}_2$ and
$\overline{\kappa}_{ik}$ represent reliable conjectures. In these calculations one should essentially use
(\ref{oper-13}) and (\ref{oper-15}). Notice that eqn. (\ref{beauty-Eqn-h}) describes evolution of
$\overline{\vh}(\vx,t)$ for an arbitrary spatial scale $L$, while the homogenisation approach to the same problem (see
\emph{e.g.} \cite{Frisch, Zheligovsky}) operates with long-wave solutions. However, the rescaling
of spatial variables (similar to one for time variables in (\ref{exact-1R}),(\ref{exact-1RR})) is applicable.

\section{Discussion and Conclusions}


1. The motivation behind this paper is to study in full a variety of asymptotic solutions and procedures for the
advection of a scalar or vector field in an oscillating flow. The main result of the paper is the general understanding
of the differences between different classes of asymptotic solutions.

2. The author is not aware about any research paper entirely devoted to the general analysis of motions of a scalar (or
vector) admixture in high-frequency oscillating flows. At the same time this topic is exploited in many papers in the
solutions of particular applied problems (\emph{e.g.} \cite{Pedley}). We hope, that our general study can underpin
further applied studies.

3. Several well-known papers are devoted to the motion of a scalar admixture in spatially oscillating flows,
\emph{e.g.} \cite{Papanicolaou, Vergassola1}, and a review in \cite{Frisch}. These authors have considered only
one asymptotic family of solutions (following to \cite{Lions}). Therefore it can be useful to study both high-frequency
and short-wave problems at a greater level of generality and then to consider different asymptotic procedures for flows
that oscillate both in space and time.

4. In our paper we are dealing with the systematic calculations of solutions. A few restrictions have been accepted at
the very beginning, the rest of the paper does not contain any physical input. Hence, our results require physical
interpretations, explanations, and applications.

5. The asymptotic validity of our results is provided by the fact that we build only the solutions for high frequency
$\omega$. In our approach $\omega$ is not linked to any dynamical equations. Instead, we require that dimensional
frequency $\omega^*$ is high in comparison with the inverted slow-time-scale $1/T$ (\ref{scales}) of an arbitrary
prescribed velocity field (\ref{exact-2}). If velocity field (\ref{exact-2}) is purely oscillatory, then the choice of
$T$ is not unique, which is explained in (\ref{eqn-lambda}),(\ref{eqn-lambda-1}).

6. All results of this paper have been obtained for the $\tau$-periodic functions from the class $\mathbb{H}$
(\ref{tilde-func-def}), which is closed with respect to  all used operations. One can try to consider more complex
classes of quasi-periodic, non-periodic, or chaotic solutions. The generalization of our results to quasi-periodic
solutions (containing several $\tau$-periods) looks rather straightforward, provided one can deal with resonances. At
the same time, chaotic oscillations (containing low $\omega$ in their spectrum) can be hardly treated by our method
since $\omega$ enters the denominators of asymptotic series (the apparent requirement for the applicability of our
method is separating the frequency spectrum from zero). Some generalizations of our results can be achieved after
replacing  the integration over $\tau_0<\tau<\tau_0+2\pi$  by integration over $-\infty<\tau<+\infty$ in the definition
of average (\ref{oper-1}) (as it has been widely accepted for the spatial average in the homogenization theory, see
\cite{Berdichevsky}).

7. In principle, \emph{TTAM} allows to produce the approximate asymptotic solutions with as small an error
(RHS-residual) as needed. However, the next logical step is more challenging: one has to prove that a solution with a
small RHS-residual is close to the exact one. This problem is equivalent to the presence of a small additional force.
Such proofs had been performed by
\cite{Simonenko, Levenshtam} for vibrational convection. Similar justifications of
\emph{TTAM} for other oscillating flows are not available yet.

8. Different asymptotic procedures $\varepsilon_{1/2}$, $\varepsilon_{1/3}$, $\varepsilon_{1/4}$, \emph{etc.}
correspond to different asymptotic paths on the plane of two scaling parameters (\ref{exact-6a}). Two additional
asymptotic paths represent two successive limits: (i) first $\delta\to 0$ and then $1/\omega\to 0$ or (ii) first
$1/\omega\to 0$ and then $\delta\to 0$ (these paths are not considered in this paper but they are worth studying). The
relevance of different available paths to particular physical situations represents a major problem for further
studies. At this stage we can mention only that different asymptotic paths correspond to different relations between
physical parameters. For example, $\delta=\omega^{-1/2}$ (\ref{basic-1}) and $\delta=\omega^{-1/3}$ (\ref{5.1}) in
their dimensional form (\ref{exact-6aa}) give
\begin{eqnarray}\label{U_}
U=LT^{-1/2}\omega^{*1/2}\quad \text{and}\quad U=LT^{-1/3}\omega^{*2/3}
\end{eqnarray}
correspondingly. For the asymptotic procedures  with large $\omega^*$ (\ref{path-simple}) it means that the imposed
oscillatory velocity (\ref{exact-2}) is higher in the latter (super-critical)  case.

9. To be able to compare different flows physically one might introduce a
\emph{drift-efficiency} of an oscillatory flow as the ratio
\begin{eqnarray}
E_{\alpha}\equiv(\text{amplitude of drift velocity})/(\text{amplitude of velocity oscillations})\nonumber
\end{eqnarray}
with $\alpha$ defined in (\ref{epsilon}). The dimensionless drifts in both cases (\ref{U_}) are $O(1)$ and in
dimensional form they are of order $L/T$. Then one might conclude that the efficiencies of (\ref{U_}) are
$E_{1/2}=(\omega^* T)^{-1/2}$ and $E_{1/3}=(\omega^* T)^{-2/3}$ correspondingly, which indicates that the critical
family  solution (\ref{basic-1}) produces stronger drift than the super-critical one (\ref{5.1}). However, this
comparison is not fair since the solution (\ref{5.1}) is valid only when the leading term in (\ref{basic-1}) vanishes
(\ref{5.0}). Taking the degeneration into account one obtains the efficiencies of (\ref{U_}) as
\begin{eqnarray}\nonumber\label{E_*}
E^d_{1/2}=(\omega^* T)^{-1}\quad \text{and}\quad E_{1/3}=(\omega^* T)^{-2/3}
\end{eqnarray}
where $E^d_{1/2}$ represents the drift-efficiency of the degenerated critical solution (\ref{basic-1}),(\ref{5.0}).
Now, one can conclude that a super-critical solution is more drift-efficient than a critical one.

10. The appearance of small pseudo-diffusion (\emph{PD}) in the high-frequency asymptotic problem (\ref{basic-2}),
(\ref{4.15})-(\ref{4.23}) is an accurate and qualitatively new result. One can make two conjectures: (i) a solution can
slowly self-concentrate (due to small negative pseudo-diffusivity) or it can undergo unusual anisotropic evolution (due
to sign-indefinite pseudo-diffusivity); and (ii) the maximum principle can be violated (since the original equation
(\ref{exact-1}) expresses the conservation of $\widehat{a}$ in each fluid particle, hence the values of
$\sup\widehat{a}$ and $\inf\widehat{a}$ do not change with time). One can argue that the conjecture (ii) is not valid
due to the explanations given in \emph{Sect.7.10}, while (i) requires additional studies.

11. It is worth introducing the finite molecular diffusivity $\mu=O(1)$ (see \emph{Sect.8}) that can improve the
convergence of all used asymptotic procedures. The introducing of asymptotically small or large diffusivity (\emph{e.g.
}$\mu=O(1/\omega)$ or $\mu=O(\omega)$) will generally lead to different asymptotic theories corresponding to three
independent small parameters from the extended list (\ref{exact-6aa}).

12. In \emph{Sect.7.9} we have shown that the drift, caused by a relatively simple oscillatory velocity, produces
chaotic dynamics of particles. This result leads to numerous new questions (see the end of \emph{Sect.7.9}) and
deserves a serious elaboration.

13. \emph{In Sect.2.3} we have demonstrated that an infinite and continuous range of slow-time scales
$1/\omega^*<T<\infty$ can be used in an important case where slow-time $t$ is absent in the expression for velocity:
$\widehat{\vu}=\widehat{\vu}(\vx,\tau)$. We have formulated this result with the aim to clarify the existence of many
scales (which often causes confusions) and to show that an infinite number of similar (to the considered in
\emph{Sects.3-7}) solutions are available. At the same time, this result can lead to studying the motions with
simultaneous presence of several different scales $T$, and to developing multi-scale (triple-scale,
\emph{etc.}) theories.

14. It is worth completing the calculations of Riemann's invariant and (\ref{aL+}) and to compare the related solutions
for $\widehat{a}$ with that of (\ref{4.21}). The aim is to find the advantages given by two complementary averaged
solutions of the same problem.


15. The version of \emph{TTAM}-theory for a vectorial passive admixture (with the averaged equations
(\ref{beauty-Eqn-h})) is linked with the problem of kinematic $MHD$-dynamo (see \cite{Moffatt}) and can bring new
results. It is apparent, that for the majority of drift velocities $\overline{\vV}(\vx,t)$ the stretching of material
elements will produce linear growth $|\overline{\vh}|\sim t$. At the same time, there are known examples with an
exponential stretching of material lines with time, say, in the flows near stagnation points or in chaotic flows. These
examples will provide the exponential growth of $|\overline{\vh}|$.

16. Our theory brings up  a new possibility: one can find an oscillatory  velocity $\widetilde{\vu}$ which produces any
required drift field (for example, a drift in the form of ABC-flow). The results of
\emph{Sects.4-7} (including (\ref{bi-linear}),(\ref{7.19a})) indicate that the solution of this problem is not
unique.

17. \emph{TTAM}-method and results have potential applications in the studies of a broad variety of flows. The
discovery of drift motions with the velocities $\overline{\vV}_1={O}(1)$ and $\overline{\vV}_2={O}(1)$ can lead to new
applications. The other advantages of our method, that can be exploited in applications, are: (i) the mathematical
generality of the imposed oscillatory velocity fields (\ref{exact-2}),(\ref{7.1}); (ii) the wide class of scalings
(\ref{exact-7}); and (iii) the most straightforward Eulerian average.  In particular, our method can be useful for
applications to biologically motivated flows possessing complex geometry (\emph{e.g.} \cite{Pedley1, Pedley,
Goldstein}), for microhydrodynamics (\emph{e.g.}
\cite{Leal1, Kim, Hinch, Moffatt2}); for flow mixing (\emph{e.g.} {\cite{Ottino, Carlsson, Carlsson1, Leal}), for
the flow of blood and flows in lungs, see \cite {Pedley2}, and for the pipes with oscillating walls, see
\cite{Pedley3}. The  spreading of ash from recently erupted Eyjaffjallaj\"{o}kull volcano has made the research on the
general area of the
 transport of an admixture in a fluid (which can oscillate due to various reasons) even more important.

\begin{acknowledgments}
This research was partially supported by EPSRC grants GR/S96616/01, GR/S96616/02, and EP/D035635/1. The author thanks
the Department of Mathematics of the University of York for the research-stimulating environment. The author is
grateful to Profs. A.D.D.Craik, R.Grimshaw, K.I.Ilin, M.E.McIntyre, H.K.Moffatt, A.B.Morgulis, T.J.Pedley, and
V.A.Zheligovsky  for helpful discussions.
\end{acknowledgments}

\appendix

\section{Calculations for $\alpha=1/2$ of Sect.4}\label{A}

\underline{\emph{The zero-order equation}}  (\ref{basic-5})  is:
\begin{eqnarray}
&&\widehat{a}_{0\tau}=0\label{01-appr-1}
\end{eqnarray}
The substitution of $\widehat{a}_{0}=\overline{a}_0(\vx,t)+\widetilde{a}_0(\vx,t,\tau)$ into (\ref{01-appr-1}) gives
$\widetilde{a}_{0\tau}= 0$. Its $\mathbb{T}$-integration (\ref{oper-7}) produces a unique (inside the
$\mathbb{T}$-class) solution $\widetilde{a}_0\equiv 0$. At the same time (\ref{01-appr-1}) does not impose any
restrictions on $\overline{a}_0(\vx,t)$, which must be determined from the next approximations. Thus the results
derivable from (\ref{01-appr-1}) are:
\begin{eqnarray} &&\widetilde{a}_0(\vx,t,\tau)\equiv 0,\quad \forall\overline{a}_{0}=\overline{a}_0(\vx,t);\quad
\widehat{a}^{[0]}\equiv\overline{a}_0+\widetilde{a}_0=\overline{a}_0(\vx,t) \label{01-appr-1a}
\end{eqnarray}
The substitution of the truncated solution $\widehat{a}^{[0]}$ into the governing equation (\ref{basic-3}) produces the
RHS-residual of order $\varepsilon$:
\begin{eqnarray}
\mathfrak{D}_2 \widehat{a}^{[0]}=\mathrm{Res}{[0]}\equiv\varepsilon(\widetilde{\vu}\cdot\nabla)\,\overline{a}_0+
\varepsilon^2\overline{a}_{0t}={O}(\varepsilon)
\label{01-appr-2}
\end{eqnarray}
\underline{\emph{The first-order equation} } (\ref{basic-6})  is
\begin{eqnarray}
&&\widehat{a}_{1\tau}=-(\widetilde{\vu}\cdot\nabla)\,\widehat{a}_0 \label{01-appr-3}
\end{eqnarray}
The use of $\widetilde{a}_0\equiv 0$ (\ref{01-appr-1a}) and $\overline{a}_{1\tau}\equiv 0$ reduces (\ref{01-appr-3}) to
the equation $\widetilde{a}_{1\tau}=-(\widetilde{\vu}\cdot\nabla)\,
\overline{a}_0$. Its $\mathbb{T}$-integration (\ref{oper-7}) gives the unique solution for $\widetilde{a}_{1}$
\begin{eqnarray}
&&\widetilde{a}_{1}=-(\widetilde{\vxi}\cdot\nabla)\,\overline{a}_0\label{01-appr-7}
\end{eqnarray}
where $\widetilde{\vxi}\equiv \widetilde{\vu}^\tau$ (\ref{01-appr-5}). Hence, $\widehat{a}_1$ and $\widehat{a}^{[1]}$
are
\begin{eqnarray}
&&\widehat{a}_1=\overline{a}_1 -(\widetilde{\vxi}\cdot\nabla)\,\overline{a}_0,\quad
\widehat{a}^{[1]}=\overline{a}_0+\varepsilon\widehat{a}_1;\quad \forall\overline{a}_0(\vx,t)\ \text{and}\
\overline{a}_1(\vx,t)
\label{01-appr-8}
\end{eqnarray}
The substitution of $\widehat{a}^{[1]}$ into (\ref{basic-3}) produces the RHS-residual of order $\varepsilon^2$:
\begin{eqnarray}
\mathfrak{D}_2 \widehat{a}^{[1]}=\mathrm{Res}{[1]}\equiv\varepsilon^2\left((\widetilde{\vu}\cdot\nabla)\,\widehat{a}_1+
\partial_t\,\widehat{a}^{[1]}\right)={O}(\varepsilon^2)\label{01-appr-9}
\end{eqnarray}
Formulae (\ref{01-appr-1a}) and (\ref{01-appr-7}) give (\ref{4.11a}) and (\ref{4.11}).

\underline{\emph{The second-order equation}} ((\ref{basic-7}) for $n=2$) is
\begin{eqnarray}
&&\widehat{a}_{2\tau}=-(\widetilde{\vu}\cdot\nabla)\,\widehat{a}_1-
\widehat{a}_{0t}\label{2-appr-1}
\end{eqnarray}
The use of (\ref{01-appr-1a}) and $\overline{a}_{2\tau}\equiv 0$ transforms (\ref{2-appr-1}) to
\begin{eqnarray}
&&\widetilde{a}_{2\tau}=-(\widetilde{\vu}\cdot\nabla)\,\overline{a}_1- (\widetilde{\vu}\cdot\nabla)\,
\widetilde{a}_1-
\overline{a}_{0t}\label{2-appr-2}
\end{eqnarray}
Its $\mathbb{B}$-part is
\begin{eqnarray}
&&\overline{a}_{0t}=-\langle(\widetilde{\vu}\cdot\nabla)\,\widetilde{a}_1\rangle
\label{2-appr-4}
\end{eqnarray}
where we have used $\langle{\widetilde{a}_{2\tau}}\rangle=0$,
$\langle(\widetilde{\vu}\cdot\nabla)\,\overline{a}_1\rangle=0$, and $\langle
\overline{a}_{0t}\rangle= \overline{a}_{0t}$.
The substitution of (\ref{01-appr-7}) into (\ref{2-appr-4}) produces the equation
\begin{eqnarray}\label{2-appr-5}
&&\overline{a}_{0t}=
\langle(\widetilde{\vu}\cdot\nabla)(\widetilde{\vxi}\cdot\nabla)\rangle\,\overline{a}_0
\end{eqnarray}
which represents a  version of inspection equation (\ref{insp-12b}) systematically derived. One may expect that the RHS
of (\ref{2-appr-5}) contains both first and second spatial derivatives of $\overline{a}_0$, however \emph{all second
derivatives vanish}. In order to prove it we introduce a commutator (\ref{oper-11})
\begin{eqnarray}
&&\widehat{\vK}\equiv[\widetilde{\vxi},\widetilde{\vu}]=
(\widetilde{\vu}\cdot\nabla)\widetilde{\vxi}-(\widetilde{\vxi}\cdot\nabla)\widetilde{\vu},\label{App1-2}\\
&&(\widetilde{\vu}\cdot\nabla)(\widetilde{\vxi}\cdot\nabla)-(\widetilde{\vxi}\cdot\nabla)(\widetilde{\vu}\cdot\nabla)=
\widehat{\vK}\cdot\nabla.\label{App1-1}
\end{eqnarray}
The $\mathbb{B}$-part of (\ref{App1-1}) is
\begin{eqnarray}
&&\langle(\widetilde{\vu}\cdot\nabla)(\widetilde{\vxi}\cdot\nabla)\rangle
=\langle(\widetilde{\vxi}\cdot\nabla)(\widetilde{\vu}\cdot\nabla)\rangle+\overline{\vK}\cdot\nabla\label{App1-3}
\end{eqnarray}
At the same time the integration by parts (\ref{oper-9}) gives
\begin{eqnarray}
&&\langle(\widetilde{\vu}\cdot\nabla)(\widetilde{\vxi}\cdot\nabla)\rangle
=-\langle(\widetilde{\vxi}\cdot\nabla)(\widetilde{\vu}\cdot\nabla)\rangle,\quad\widetilde{\vu}\equiv\widetilde{\vxi}_\tau
\label{App1-4}
\end{eqnarray}
Combining (\ref{App1-3}) and (\ref{App1-4}) we obtain
\begin{eqnarray}
&&\langle(\widetilde{\vu}\cdot\nabla)(\widetilde{\vxi}\cdot\nabla)\rangle
=\frac{1}{2}\overline{\vK}\cdot\nabla\label{App1-5}
\end{eqnarray}
which reduces (\ref{2-appr-5}) to the advection equation (\ref{4.15}) with
\begin{eqnarray}
&&\overline{\vV}_0\equiv-\langle(\widetilde{\vu}\cdot\nabla)\,\widetilde{\vxi}\rangle=
-\frac{1}{2}\langle[\widetilde{\vxi},\widetilde{\vu}]\rangle=-\frac{1}{2}\overline{\vK}
\label{2-appr-8}
\end{eqnarray}
which coincides with (\ref{4.18}). The $\mathbb{T}$-part of (\ref{2-appr-2}) appears after subtracting (\ref{2-appr-4})
from (\ref{2-appr-2}) and the use of notation (\ref{oper-5}):
\begin{eqnarray}
&&\widetilde{a}_{2\tau}= -(\widetilde{\vu}\cdot\nabla)\, \overline{a}_1 -
\{(\widetilde{\vu}\cdot\nabla)\,\widetilde{a}_1\}.
\label{2-appr-10}
\end{eqnarray}
Its $\mathbb{T}$-integration (\ref{oper-7}) with the use of (\ref{01-appr-7}) gives (\ref{4.12}):
\begin{eqnarray}
&&\widetilde{a}_{2}=-(\widetilde{\vxi}\cdot\nabla)\, \overline{a}_1 +
\{(\widetilde{\vu}\cdot\nabla)(\widetilde{\vxi}\cdot\nabla)\}^\tau\overline{a}_0,\quad \forall\overline{a}_1
\label{2-appr-11}
\end{eqnarray}
Hence, $\widehat{a}_2$ and $\widehat{a}^{[2]}$ can be written as
\begin{eqnarray}
&&\widehat{a}_2=\overline{a}_2+\widetilde{a}_2,\quad \widehat{a}^{[2]}=\overline{a}_0+\varepsilon
\left(\overline{a}_1 - (\widetilde{\vxi}\cdot\nabla)\,\overline{a}_0\right)+\varepsilon^2\widehat{a}_2,
\quad \forall\overline{a}_1,\overline{a}_2
\label{2-appr-12}
\end{eqnarray}
where $\overline{a}_0$ and $\widetilde{a}_{2}$ are given by (\ref{4.15}), (\ref{2-appr-11}). The substitution of
$\widehat{a}^{[2]}$ into (\ref{basic-3}) produces the RHS-residual of order $\varepsilon^3$
\begin{eqnarray}
\mathfrak{D}_2\widehat{a}^{[2]}=\mathrm{Res}{[2]}\equiv\varepsilon^3\left((\widetilde{\vu}\cdot\nabla)\,\widehat{a}_2+
\partial_t\,(\widehat{a}_1+\varepsilon\widehat{a}_2)\right) ={O}(\varepsilon^3).\label{2-appr-13}
\end{eqnarray}
\underline{\emph{The third-order equation}} ((\ref{basic-7}) for $n=3$) is:
\begin{eqnarray}
&&\widetilde{a}_{3\tau}=-(\widetilde{\vu}\cdot\nabla)\,\widehat{a}_2-\widehat{a}_{1t}\label{3-appr-1}
\end{eqnarray}
Its $\mathbb{B}$-part is
\begin{eqnarray}
&&\overline{a}_{1t}= -\langle(\widetilde{\vu}\cdot\nabla)\,\widetilde{a}_2\rangle.
\label{3-appr-2}
\end{eqnarray}
The substitution of (\ref{2-appr-11}) into (\ref{3-appr-2}), the use of  $\widetilde{\vu}\equiv\widetilde{\vxi}_\tau$,
and the integration by parts (\ref{oper-9}) yield
\begin{eqnarray}
&&\overline{a}_{1t}=
\langle(\widetilde{\vu}\cdot\nabla)(\widetilde{\vxi}\cdot\nabla)\rangle\overline{a}_1+
\langle(\widetilde{\vxi}\cdot\nabla)(\widetilde{\vu}\cdot\nabla)(\widetilde{\vxi}\cdot\nabla)\rangle\overline{a}_0
\label{3-appr-3}
\end{eqnarray}
where $\langle(\widetilde{\vu}\cdot\nabla)(\widetilde{\vxi}\cdot\nabla)\rangle$ has been already simplified in
(\ref{App1-5}). The second term in the RHS of (\ref{3-appr-3}) formally contains the third, second, and first spatial
derivatives of $\overline{a}_0$; however \emph{all the third  and  second derivatives vanish}. To prove it, first, we
use (\ref{oper-9a}):
\begin{eqnarray}
&&\langle(\widetilde{\vxi}\cdot\nabla)(\widetilde{\vu}\cdot\nabla)(\widetilde{\vxi}\cdot\nabla)\rangle=
-\langle(\widetilde{\vu}\cdot\nabla)(\widetilde{\vxi}\cdot\nabla)(\widetilde{\vxi}\cdot\nabla)\rangle-
\langle(\widetilde{\vxi}\cdot\nabla)(\widetilde{\vxi}\cdot\nabla)(\widetilde{\vu}\cdot\nabla)\rangle\label{App2-1}
\end{eqnarray}
Then we use (\ref{App1-2}), (\ref{App1-1}) to transform the sequence of operators $(\widetilde{\vxi}\cdot\nabla)$ and
$(\widetilde{\vu}\cdot\nabla)$ in each term in the RHS of (\ref{App2-1}) into their sequence in the LHS. The result is
\begin{eqnarray}
&&\langle(\widetilde{\vxi}\cdot\nabla)(\widetilde{\vu}\cdot\nabla)(\widetilde{\vxi}\cdot\nabla)\rangle=
\frac{1}{3}\overline{\vK'}\cdot\nabla,\quad \widehat{\vK'}\equiv[\widehat{\vK},\widetilde{\vxi}]\label{App2-2}
\end{eqnarray}
As the result (\ref{3-appr-3}) takes form (\ref{4.16})  with $\overline{\vV}_0$ (\ref{2-appr-8}) and
\begin{eqnarray}
&&\overline{\vV}_1\equiv-\langle(\widetilde{\vxi}\cdot\nabla)(\widetilde{\vu}\cdot\nabla)\widetilde{\vxi})\rangle=
-\frac{1}{3}\langle[[\widetilde{\vxi},\widetilde{\vu}],\widetilde{\vxi}]\rangle=-\frac{1}{3}\overline{\vK'}
\label{3-appr-4a}
\end{eqnarray}
which gives (\ref{4.18}). The $\mathbb{T}$-part of (\ref{3-appr-1}) after its $\mathbb{T}$-integration gives
(\ref{4.13})
\begin{eqnarray}
&&\widetilde{a}_{3}=-(\widetilde{\vxi}\cdot\nabla)\,\overline{a}_2-
\{(\widetilde{\vu}\cdot\nabla)\,\widetilde{a}_2\}^\tau-\widetilde{a}_{1t}^\tau,
\quad\widetilde{a}_1^\tau=-(\widetilde{\vxi}^\tau\cdot\nabla)\, \overline{a}_0 \label{3-appr-5}
\end{eqnarray}
where $\widetilde{a}_2$ is given by (\ref{2-appr-11}).

Hence, $\widehat{a}_3$ and $\widehat{a}^{[3]}$ can be written as
\begin{eqnarray}
&&\widehat{a}_3=\overline{a}_3+\widetilde{a}_3,\quad \widehat{a}^{[3]}=\overline{a}_0+\varepsilon (\overline{a}_1 -
(\widetilde{\vxi}\cdot\nabla)\,\overline{a}_0)+\varepsilon^2\widehat{a}_2+\varepsilon^3\widehat{a}_3,
\quad\forall\overline{a}_2,\overline{a}_3
\label{3-appr-6}
\end{eqnarray}
where $\overline{a}_0$, $\overline{a}_1$, $\widetilde{a}_{2}$, and $\widetilde{a}_{3}$ are given by (\ref{4.15}),
(\ref{4.16}), (\ref{2-appr-11}), and (\ref{3-appr-5}). The substitution of $\widehat{a}^{[3]}$ into (\ref{basic-3})
produces the RHS-residual of order $\varepsilon^4$:
\begin{eqnarray}
\mathfrak{D}_2\widehat{a}^{[3]}=\mathrm{Res}{[3]}={O}(\varepsilon^4)\label{3-appr-7}
\end{eqnarray}
Explicit formulae for residuals (similar to (\ref{01-appr-2}), (\ref{01-appr-9}), (\ref{2-appr-13})) for
(\ref{3-appr-7}),  (\ref{4-appr-7}) and for all other considered cases can be calculated straightforwardly; however for
brevity we do not present them.

\underline{\emph{The fourth-order equation}} ((\ref{basic-7}) for $n=4$) is:
\begin{eqnarray}
&&\widetilde{a}_{4\tau}=-(\widetilde{\vu}\cdot\nabla)\,\widehat{a}_3-\widehat{a}_{2t}\label{4-appr-1}
\end{eqnarray}
Its $\mathbb{B}$-part is
\begin{eqnarray}
&&\overline{a}_{2t}= -\langle(\widetilde{\vu}\cdot\nabla)\,\widetilde{a}_3\rangle
\label{4-appr-2}
\end{eqnarray}
The substitution of $\widetilde{a}_3$ (\ref{3-appr-5})  into (\ref{4-appr-2}), $\widetilde{a}_2$ (\ref{2-appr-11}) into
$\widetilde{a}_3$ (\ref{3-appr-5}), the integration by parts (\ref{oper-9}), and the use of (\ref{01-appr-5}),
(\ref{2-appr-8}), (\ref{3-appr-4a}) yield
\begin{eqnarray}
&&\langle(\widetilde{\vu}\cdot\nabla)\,\widetilde{a}_3\rangle=(\overline{\vV}_0\cdot\nabla)\overline{a}_2+
(\overline{\vV}_1\cdot\nabla)\overline{a}_1+\langle(\widetilde{\vxi}\cdot\nabla)(\widetilde{\vxi}\cdot\nabla)\rangle
(\overline{\vV}_0\cdot\nabla)\overline{a}_0-\label{App3-1}\\
&&-\langle(\widetilde{\vxi}\cdot\nabla)(\widetilde{\vxi}_t\cdot\nabla)\rangle\overline{a}_0+
\overline{\mathfrak{X}}\overline{a}_0,\ \text{where}\
\overline{\mathfrak{X}}\equiv\langle(\widetilde{\vxi}\cdot\nabla)(\widetilde{\vu}\cdot\nabla)
\{(\widetilde{\vu}\cdot\nabla)(\widetilde{\vxi}\cdot\nabla)\}^\tau\rangle
\nonumber
\end{eqnarray}
The shorthand operator $\overline{\mathfrak{X}}$ (as well as the operators $\widehat{\mathfrak{Y}}$,
$\mathfrak{\overline{A}}$, $\mathfrak{\overline{B}}$, $\overline{\mathfrak{C}}$, and $\overline{\mathfrak{F}}$ below)
acts on $\overline{a}_0$. The RHS of (\ref{App3-1}) formally contains the fourth, third, second, and first spatial
derivatives of $\overline{a}_0$; however \emph{all the fourth and third derivatives vanish}. In order to prove it we
first rewrite $\overline{\mathfrak{X}}$ as
\begin{eqnarray}
&&\overline{\mathfrak{X}}=\langle(\widetilde{\vxi}\cdot\nabla)(\widetilde{\vu}\cdot\nabla)
\widetilde{\mathfrak{Y}}^\tau\rangle\quad\text{where}\quad
\widehat{\mathfrak{Y}}\equiv(\widetilde{\vu}\cdot\nabla)(\widetilde{\vxi}\cdot\nabla)
\label{App3-2}
\end{eqnarray}
The use of (\ref{oper-9a}) and (\ref{App1-2}), (\ref{App1-1}) transforms (\ref{App3-2}) to
\begin{eqnarray}
&&2\overline{\mathfrak{X}}=-\overline{\mathfrak{A}}+\overline{\mathfrak{B}}+
\frac{1}{2}\langle(\widetilde{\vxi}\cdot\nabla)(\widetilde{\vxi}\cdot\nabla)\rangle(\overline{\vK}\cdot\nabla)
\label{App3-3}\\
&&\overline{\mathfrak{A}}\equiv\langle(\widetilde{\vxi}\cdot\nabla)(\widetilde{\vxi}\cdot\nabla)
(\widetilde{\vu}\cdot\nabla)(\widetilde{\vxi}\cdot\nabla)\rangle,\quad
\overline{\mathfrak{B}}\equiv\langle(\widetilde{\vK}^\tau\cdot\nabla)(\widetilde{\vu}\cdot\nabla)
(\widetilde{\vxi}\cdot\nabla)\rangle
\nonumber
\end{eqnarray}
Let us now simplify $\overline{\mathfrak{A}}$ and $\overline{\mathfrak{B}}$.  For $\overline{\mathfrak{B}}$ we use
(\ref{oper-9a})
\begin{eqnarray}
&&\overline{\mathfrak{B}}\equiv
\langle(\widetilde{\vK}^\tau\cdot\nabla)(\widetilde{\vu}\cdot\nabla)(\widetilde{\vxi}\cdot\nabla)\rangle=-
\langle(\widetilde{\vK}\cdot\nabla)(\widetilde{\vxi}\cdot\nabla)(\widetilde{\vxi}\cdot\nabla)\rangle-
\langle(\widetilde{\vK}^\tau\cdot\nabla)(\widetilde{\vxi}\cdot\nabla)(\widetilde{\vu}\cdot\nabla)\rangle\nonumber
\end{eqnarray}
To change $(\widetilde{\vxi}\cdot\nabla)(\widetilde{\vu}\cdot\nabla)$ into
$(\widetilde{\vu}\cdot\nabla)(\widetilde{\vxi}\cdot\nabla)$ in the last term we use (\ref{App1-2}), (\ref{App1-1}) that
yields:
\begin{eqnarray}
&&\overline{\mathfrak{B}}=-\frac{1}{2}\langle(\widetilde{\vK}\cdot\nabla)(\widetilde{\vxi}\cdot\nabla)
(\widetilde{\vxi}\cdot\nabla)\rangle-\frac{1}{4}\overline{\vkappa}\cdot\nabla,\quad
\widehat{\vkappa}\equiv[\widetilde{\vK}^\tau,\widetilde{\vK}]\label{App3-4}
\end{eqnarray}
The operator $\overline{\mathfrak{A}}$ is simplified by the version of (\ref{oper-9a}) with four multipliers
\begin{eqnarray}
&&\mathfrak{\overline{A}}\equiv\langle(\widetilde{\vxi}\cdot\nabla)(\widetilde{\vxi}\cdot\nabla)
(\widetilde{\vu}\cdot\nabla)(\widetilde{\vxi}\cdot\nabla)\rangle=
-\langle(\widetilde{\vu}\cdot\nabla)(\widetilde{\vxi}\cdot\nabla)(\widetilde{\vxi}\cdot\nabla)
(\widetilde{\vxi}\cdot\nabla)\rangle-\label{App3-5}\\
&&-\langle(\widetilde{\vxi}\cdot\nabla)(\widetilde{\vu}\cdot\nabla)(\widetilde{\vxi}\cdot\nabla)
(\widetilde{\vxi}\cdot\nabla)\rangle-
\langle(\widetilde{\vxi}\cdot\nabla)(\widetilde{\vxi}\cdot\nabla)
(\widetilde{\vxi}\cdot\nabla)(\widetilde{\vu}\cdot\nabla)\rangle
\nonumber
\end{eqnarray}
The multiple use of commutator (\ref{App1-2}), (\ref{App1-1})  allows us to transform the sequence of operators
$(\widetilde{\vxi}\cdot\nabla)$ and $(\widetilde{\vu}\cdot\nabla)$ in each term in the RHS of (\ref{App3-5}) to the
sequence in its LHS. The result is
\begin{eqnarray}
&&\overline{\mathfrak{A}}=-\frac{1}{2}\langle(\widehat{\vK}\cdot\nabla)(\widetilde{\vxi}\cdot\nabla)
(\widetilde{\vxi}\cdot\nabla)\rangle+\frac{1}{4}\overline{\vK''}\cdot\nabla,\quad
\widehat{\vK''}\equiv[\widehat{\vK'},\widetilde{\vxi}]\label{App3-6}
\end{eqnarray}
Now, (\ref{App3-3}), (\ref{App3-4}), and (\ref{App3-6}) yield
\begin{eqnarray}
&&\overline{\mathfrak{X}}=
\frac{1}{4}(\overline{\vK}\cdot\nabla)\langle(\widetilde{\vxi}\cdot\nabla)(\widetilde{\vxi}\cdot\nabla)\rangle+
\frac{1}{4}\langle(\widetilde{\vxi}\cdot\nabla)(\widetilde{\vxi}\cdot\nabla)\rangle(\overline{\vK}\cdot\nabla)-
\frac{1}{8}(\overline{\vkappa}+\overline{\vK''})\cdot\nabla\nonumber
\label{App3-7}
\end{eqnarray}
The substitution of this expression into (\ref{App3-1}), (\ref{4-appr-2})  gives
\begin{eqnarray}
&&\overline{a}_{2t}+(\overline{\vV}_0\cdot\nabla)\overline{a}_2+
(\overline{\vV}_1\cdot\nabla)\overline{a}_1-\frac{1}{8}(\overline{\vkappa}+\overline{\vK''})\cdot\nabla \overline{a}_0
+\frac{1}{4}\overline{\mathfrak{C}}\overline{a}_0-\overline{\mathfrak{F}}\overline{a}_0=0,\label{App3-8}\\
&&\overline{\mathfrak{C}}\equiv(\overline{\vK}\cdot\nabla)\langle(\widetilde{\vxi}\cdot\nabla)(\widetilde{\vxi}\cdot\nabla)
\rangle-\langle(\widetilde{\vxi}\cdot\nabla)(\widetilde{\vxi}\cdot\nabla)\rangle(\overline{\vK}\cdot\nabla),\
\overline{\mathfrak{F}}\equiv\langle(\widetilde{\vxi}\cdot\nabla)(\widetilde{\vxi}_t\cdot\nabla)\rangle
\nonumber
\end{eqnarray}
Additional transformations of the last two operators in (\ref{App3-8}) yield
\begin{eqnarray}
&&\frac{1}{4}\overline{\mathfrak{C}}-\overline{\mathfrak{F}}=
\frac{1}{2}\langle[\widetilde{\vxi},\widetilde{\vxi}_t]\rangle\cdot\nabla
-\frac{1}{2}\langle(\widetilde{\vu}'\cdot\nabla)\widetilde{\vxi}+
(\widetilde{\vxi}\cdot\nabla)\widetilde{\vu}'\rangle\cdot\nabla - \frac{1}{2}\langle
\widetilde{u}'_i\widetilde{\xi}_k+\widetilde{u}'_k\widetilde{\xi}_i\rangle\frac{\partial^2}{\partial x_i\partial x_k}=\nonumber\\
&&=\frac{1}{2}\langle[\widetilde{\vxi},\widetilde{\vxi}_t]\rangle\cdot\nabla-\frac{\partial}{\partial
x_k}\left(\overline{\chi}_{ik}\frac{\partial}{\partial
x_i}\right)+\frac{1}{2}\langle\widetilde{\vxi}\Div\widetilde{\vu}'+\widetilde{\vu}'\Div\widetilde{\vxi}\rangle
\label{Ap-trans}\\
&&\widetilde{\vu}'\equiv\widetilde{\vxi}_t-[\overline{\vV}_0,\widetilde{\vxi}],\quad
\overline{\chi}_{ik}\equiv\frac{1}{2}\langle \widetilde{u}'_i\widetilde{\xi}_k+\widetilde{u}'_k\widetilde{\xi}_i\rangle
 \label{v-prime}
\end{eqnarray}
The  substitution of (\ref{Ap-trans}) into (\ref{App3-8}) leads to the equation for $\overline{a}_2$ (\ref{4.17}) where
the formula (\ref{4.20}) for $\overline{\chi}_{ik}$ is obtained from (\ref{v-prime}) by the use of definition
$\widetilde{\vu}'$.

The $\mathbb{T}$-part of (\ref{4-appr-1}) after its $\mathbb{T}$-integration gives (\ref{4.14})
\begin{eqnarray}
&&\widetilde{a}_{4}=-(\widetilde{\vxi}\cdot\nabla)\,\overline{a}_3-
\{(\widetilde{\vu}\cdot\nabla)\,\widetilde{a}_3\}^\tau-\widetilde{a}_{2t}^\tau \label{4-appr-5}
\end{eqnarray}
where $\widetilde{a}_3$ is given by (\ref{3-appr-5}).

\section{Calculations for Super-critical Families}\label{B}

\subsection{Calculations for $\alpha={1/3}$ of Sect.5}

The equations of the \emph{zeroth and first approximations} (\ref{5.5}),(\ref{5.6}) are the same as
(\ref{basic-5}),(\ref{basic-6}) hence they have the same solutions
\begin{eqnarray}
&&\widehat{a}_0=\overline{a}_0(\vx,t),\quad  \widehat{a}_1=\overline{a}_1
-(\widetilde{\vxi}\cdot\nabla)\overline{a}_0;\quad
\forall \,\overline{a}_0,\ \overline{a}_1\in \mathbb{B}\cap \mathbb{O}(1)\label{0-1-1/3}
\end{eqnarray}
An essential difference from the $\omega^{1/2}$-procedure emerges for the \emph{second approximation} (\ref{5.7}) where
instead of an advection equation for $\overline{a}_0$ (\ref{4.15}) we obtain a compatibility condition
\begin{eqnarray}
&&(\overline{\vV}_0\cdot\nabla)\,\overline{a}_0=0\label{altern-10}
\end{eqnarray}
which represents an identity due to (\ref{5.0}). The $\mathbb{T}$-part of (\ref{5.7}) produces the same
$\widetilde{a}_{2}$ as in (\ref{2-appr-11}). Hence
\begin{eqnarray}
&&\widehat{a}_2=\overline{a}_2+\widetilde{a}_2,\quad \widetilde{a}_{2}=-(\widetilde{\vxi}\cdot\nabla)\, \overline{a}_1
+\{(\widetilde{\vu}\cdot\nabla)(\widetilde{\vxi}\cdot\nabla)\}^\tau\overline{a}_0,\quad
\forall \overline{a}_2\in \mathbb{B}\cap \mathbb{O}(1)\label{altern-10a}
\end{eqnarray}
The equation for the \emph{third approximation} ($n=3$ in (\ref{5.8})) jointly with (\ref{5.0}) produces the
$\mathbb{B}$-part of the equation
\begin{eqnarray}\label{altern-13}
\overline{a}_{0t}=
\langle(\widetilde{\vxi}\cdot\nabla)(\widetilde{\vu}\cdot\nabla)(\widetilde{\vxi}\cdot\nabla)\rangle\overline{a}_0
\end{eqnarray}
with the same triple-correlation as in (\ref{App2-2}), (\ref{3-appr-4a}); hence
\begin{eqnarray}
\left(\frac{\partial}{\partial t}+ \overline{\vV}_1\cdot\nabla\right)\overline{a}_0=0,\quad
\text{where}\quad\overline{\vV}_1=
-\frac{1}{3}\langle[[\widetilde{\vxi},\widetilde{\vu}],\widetilde{\vxi}]\rangle
\label{altern-14}
\end{eqnarray}
The $\mathbb{T}$-part of (\ref{5.8}) after its $\mathbb{T}$-integration gives a simpler expression
\begin{eqnarray}
&&\widetilde{a}_{3}=-(\widetilde{\vxi}\cdot\nabla)\,\overline{a}_2-
\{(\widetilde{\vu}\cdot\nabla)\,\widetilde{a}_2\}^\tau \label{3-appr-5b}
\end{eqnarray}
with $\widetilde{a}_2$ (\ref{altern-10a}). Hence, $\widehat{a}_3$ and $\widehat{a}^{[3]}$ can be written as
\begin{eqnarray}
&&\widehat{a}_3=\overline{a}_3+\widetilde{a}_3,\quad \widehat{a}^{[3]}=\overline{a}_0+\varepsilon (\overline{a}_1 -
(\widetilde{\vxi}\cdot\nabla)\,\overline{a}_0)+\varepsilon^2\widehat{a}_2+\varepsilon^3\widehat{a}_3
\label{3-appr-6}
\end{eqnarray}
where $\overline{a}_0$,  $\widetilde{a}_{2}$, and $\widetilde{a}_{3}$ are given by (\ref{altern-14}),
(\ref{altern-10a}), and (\ref{3-appr-5b}), while $\overline{a}_1$, $\overline{a}_2$ and $\overline{a}_3$ are to be
found from further approximations. Hence the $\omega^{1/3}$-procedure shows that in the presence of degeneration
(\ref{5.0}) a drift velocity remains ${O}(1)$ and is given by the expression $\overline{\vV}_1$ (\ref{altern-14}).

The equation  for \emph{the fourth approximation} ((\ref{5.8}) with $n=4$) is:
\begin{eqnarray}
&&\widetilde{a}_{4\tau}=-(\widetilde{\vu}\cdot\nabla)\,\widehat{a}_3-\widehat{a}_{1t}\label{4-appr-4}
\end{eqnarray}
After the use of the same steps as in the $\omega^{1/2}$-procedure we have $\mathbb{B}$-part
\begin{eqnarray}
&&\overline{a}_{1t}+ (\overline{\vV}_1\cdot\nabla)\overline{a}_1+(\overline{\vV}_2\cdot\nabla)\overline{a}_0=0,
\label{App3-8ba}
\end{eqnarray}
with the same notations as in (\ref{altern-14}), (\ref{5.17}), (\ref{5.17a}). The $\mathbb{T}$-part of (\ref{4-appr-4})
after its $\mathbb{T}$-integration gives
\begin{eqnarray}
&&\widetilde{a}_{4}=-(\widetilde{\vxi}\cdot\nabla)\,\overline{a}_3-
\{(\widetilde{\vu}\cdot\nabla)\,\widetilde{a}_3\}^\tau-\widetilde{a}_{1t}^\tau \label{4-appr-5a}
\end{eqnarray}
with $\widetilde{a}_3$ (\ref{3-appr-5b}).

\subsection{Calculations for $\alpha={1/4}$ of Sect.5}

The equation of the \emph{zeroth, first, and second approximations} (\ref{5.24}),(\ref{5.25}), and (\ref{5.26}) are the
same as (\ref{5.5}),(\ref{5.6}), and (\ref{5.7}) so they have the same solutions (\ref{0-1-1/3})-(\ref{altern-10a})
\begin{eqnarray}
&&\widehat{a}_0\equiv\overline{a}_0+\widetilde{a}_0=\overline{a}_0(\vx,t),\quad
\widehat{a}_1=\overline{a}_1 -(\widetilde{\vxi}\cdot\nabla)\,\overline{a}_0,\quad
\widehat{a}_2=\overline{a}_2+\widetilde{a}_2
\label{altern-8-4}
\end{eqnarray}
An essential difference from the $\omega^{1/3}$-procedure emerges for the \emph{third approximation} (\ref{5.26}) where
instead of an advection equation for $\overline{a}_0$ (\ref{altern-14}) we obtain a compatibility condition
\begin{eqnarray}
&&(\overline{\vV}_1\cdot\nabla)\,\overline{a}_0=0\label{altern-10-4}
\end{eqnarray}
which represents an identity due to (\ref{5.19}). The $\mathbb{T}$-part of (\ref{5.27}) produces the same
$\widetilde{a}_{3}$ as in (\ref{3-appr-5b}). The equation  for \emph{the fourth approximation} ((\ref{5.28}) with
$n=4$) is:
\begin{eqnarray}
&&\widetilde{a}_{4\tau}=-(\widetilde{\vu}\cdot\nabla)\,\widehat{a}_3-\widehat{a}_{0t}\label{4-appr-4-4}
\end{eqnarray}
After performing the same steps as in the $\omega^{1/3}$-procedure we obtain $\mathbb{B}$-part
\begin{eqnarray}
&&\overline{a}_{0t}+(\overline{\vV}_2\cdot\nabla)\overline{a}_0=0,
\label{App3-8ba-4}
\end{eqnarray}
with the same notations as in (\ref{5.17}), (\ref{5.17a}). The $\mathbb{T}$-part of (\ref{4-appr-4-4}) after its
$\mathbb{T}$-integration gives
\begin{eqnarray}
&&\widetilde{a}_{4}=-(\widetilde{\vxi}\cdot\nabla)\,\overline{a}_3-
\{(\widetilde{\vu}\cdot\nabla)\,\widetilde{a}_3\}^\tau \label{4-appr-5a-4}
\end{eqnarray}
where $\widetilde{a}_3$ is given in (\ref{3-appr-5b}).

\end{document}